\RequirePackage{fixltx2e}
\documentclass[11pt,a4paper]{article}
\pdfoutput=1
\usepackage[utf8]{inputenc}
\usepackage{jheppub}
\hypersetup{unicode=true,bookmarksopen=true}

\usepackage{amsthm}
\usepackage{mathtools}
\usepackage{amsthm}
\usepackage{lmodern}
\usepackage[T1]{fontenc}
\usepackage{bbm}
\DeclareMathAlphabet{\mathbfi}{OML}{cmm}{b}{it}

\let\originalleft\left
\let\originalright\right
\renewcommand{\left}{\mathopen{}\mathclose\bgroup\originalleft}
\renewcommand{\right}{\aftergroup\egroup\originalright}
\makeatletter
\newcommand{\biggg}{\bBigg@\thr@@}
\newcommand{\Biggg}{\bBigg@{3.5}}
\makeatother

\makeatletter
\newenvironment{equations}[1][]{\subequations\ifx\relax#1\relax\else\label{#1}\fi\align\ignorespaces}{\endalign\ignorespacesafterend\endsubequations}
\def\@spliteq#1{\begin{equation}\begin{split}#1\end{split}\end{equation}}
\def\@spliteqstar#1{\begin{equation*}\begin{split}#1\end{split}\end{equation*}}
\def\splitequation{\collect@body\@spliteq}
\expandafter\def\csname splitequation*\endcsname{\collect@body\@spliteqstar}

\expandafter\def\csname endsplitequation*\endcsname{\ignorespacesafterend}
\makeatother

\renewcommand{\vec}[1]{{\ifnum9<1#1\mathbf{#1}\else\ifcat\noexpand#1\relax\boldsymbol{#1}\else\mathbfi{#1}\fi\fi}}

\newcommand{\mathi}{\mathrm{i}}
\let\oldre\Re
\let\oldim\Im
\renewcommand{\Re}{\oldre\mathfrak{e}\,}
\renewcommand{\Im}{\oldim\mathfrak{m}\,}
\newcommand{\total}{\mathop{}\!\mathrm{d}}
\newcommand{\laplace}{\mathop{}\!\bigtriangleup}
\newcommand{\abs}[1]{{\left\lvert{#1}\right\rvert}}

\newcommand{\unitmatrix}{\mathbbm{1}}
\newcommand{\1}{\mathbbm{1}}

\newcommand{\eqend}[1]{\,#1}
\newcommand{\bigo}[1]{\mathcal{O}\left({#1}\right)}

\newcommand{\expect}[1]{\left\langle{#1}\right\rangle}

\newcommand{\op}{\mathcal{O}}
\newcommand{\normord}[1]{\mathopen{:}{#1}\mathclose{:}}

\theoremstyle{definition}

\theoremstyle{plain}

\theoremstyle{remark}

\allowdisplaybreaks

\begin{document}

\title{Recursive construction of the operator product expansion in curved space}

\author{Markus B. Fröb}
\affiliation{Institut f{\"u}r Theoretische Physik, Universit{\"a}t Leipzig, Br{\"u}derstra{\ss}e 16, 04103 Leipzig, Germany}

\emailAdd{mfroeb@itp.uni-leipzig.de}

\abstract{I derive a formula for the coupling-constant derivative of the coefficients of the operator product expansion (Wilson OPE coefficients) in an arbitrary curved space, as the natural extension of the quantum action principle. Expanding the coefficients themselves in powers of the coupling constants, this formula allows to compute them recursively to arbitrary order. As input, only the OPE coefficients in the free theory are needed, which are easily obtained using Wick's theorem. I illustrate the method by computing the OPE of two scalars $\phi$ in hyperbolic space (Euclidean Anti-de Sitter space) up to terms vanishing faster than the square of their separation to first order in the quartic interaction $g \phi^4$, as well as the OPE coefficient $\mathcal{C}^\1_{\phi \phi}$ at second order in $g$.}

%\keywords{TODO}

\maketitle

\section{Introduction}
\label{sec_intro}

The operator product expansion, first conceived by Wilson~\cite{wilson1969}, has found many applications in quantum field theory, especially conformal theories and in the context of the AdS/CFT correspondence~\cite{maldacena1997,witten1998}. The OPE is the statement that~\cite{wilson1969,wilsonzimmermann1972,zimmermann1973b}
\begin{equation}
\label{sec_intro_ope_general}
\expect{ \op_{A_1}(x_1) \cdots \op_{A_k}(x_k) }_\Omega \sim \sum_B \mathcal{C}^B_{A_1 \cdots A_k}(x_1,\ldots,x_k;y) \expect{ \op_B(y) }_\Omega \eqend{,}
\end{equation}
where $\op_A$ denotes a composite operator, the expectation value $\expect{ \cdot }_\Omega$ is taken in any suitable (interacting) state $\Omega$ which may include other (spectator) fields, where the sum on the right-hand side runs over all composite operators in the theory, and where the OPE coefficients $\mathcal{C}^B_{A_1 \cdots A_k}(x_1,\ldots,x_k;y)$ are independent of the state $\Omega$, such that one really has an ``expansion of operators''. The concrete meaning of $\sim$ depends on the theory: while for conformal field theories it has been shown that the OPE is convergent~\cite{mack1977,pappadopuloetal2012,rychkovynernay2016,gilliozetal2020}, such that one can replace $\sim$ by $=$, in general it is only supposed to hold as an asymptotic expansion. That is, if one scales the insertion points $x_i$ to the expansion point $y$ and includes in the sum on the right-hand side of~\eqref{sec_intro_ope_general} all operators $\op_B$ up to a fixed dimension $[\op_B] \leq \Delta$, the difference between left- and right-hand side vanishes like $d^{\Delta-\sum_{i=1}^k [\op_{A_i}]}$, where $d$ is the largest distance between $y$ and any of the $x_i$.

On the other hand, in recent work by Holland, Hollands and Kopper~\cite{hollandskopper2012,hollandetal2014} it has been proven that the OPE~\eqref{sec_intro_ope_general} actually also converges for (Euclidean) scalar $\lambda \phi^4$ theory, to all orders in perturbation theory. Since the rate of convergence depends on the order of perturbation theory (and becomes worse at higher orders), this result is not as far-reaching as the non-perturbative convergence in conformal field theories, but shows that the OPE is much better behaved as originally expected. It seems thus possible to define a quantum field theory not by its correlation functions (obtained, e.g., from a classical Lagrangian in perturbation theory), but instead by the OPE coefficients which encode the algebraic properties of the theory and the one-point expectation values (form factors) $\expect{ \op_B(y) }_\Omega$, which determine the quantum state $\Omega$. For this, one needs in addition certain factorisation and associativity conditions on the OPE coefficients. The factorisation conditions are obtained by performing the OPE~\eqref{sec_intro_ope_general} for a subset of the operators first, and then a second one for the remaining operators, which leads to
\begin{equation}
\label{sec_intro_ope_factor}
\mathcal{C}^B_{A_1 \cdots A_k}(x_1,\ldots,x_k;z) \sim \sum_C \mathcal{C}^C_{A_1 \cdots A_m}(x_1,\ldots,x_m;y) \, \mathcal{C}^B_{C A_{m+1} \cdots A_k}(y,x_{m+1},\ldots,x_k;z) \eqend{,}
\end{equation}
while the associativity conditions come from demanding that two different ways of performing such a factorisation agree. These conditions hold again as equalities in conformal field theories~\cite{pappadopuloetal2012,gilliozetal2020} for suitable configurations of the points $x_i$, $y$ and $z$ (where the associativity conditions lead to crossing symmetry constraints), and have also been proven for scalar $\lambda \phi^4$ theory to all orders in perturbation theory~\cite{hollandhollands2015a,hollandhollands2015b}.

It remains to give an algorithm to determine the OPE coefficients themselves. In free theories, they can be obtained directly by defining composite operators as the normal-ordered ones, using Wick's theorem and expanding the normal-ordered products around the expansion point~\cite{hollandhollands2015a,collins1986}. For example, in flat space one has
\begin{splitequation}
\label{sec_intro_ope_free_phiphi}
\phi(x) \phi(y) &= \normord{ \phi(x) \phi(y) }_G + G(x,y) \1 \\
&= G(x,y) \1 + \normord{ \phi^2(z) }_G + \left[ (x-z)^\mu + (y-z)^\mu \right] \normord{ ( \phi \partial_\mu \phi )(z) }_G + \cdots \eqend{,}
\end{splitequation}
where $G(x,y) = \expect{ \phi(x) \phi(y) }_0$ is the free two-point function and normal-ordering is performed with respect to $G$, and can read off the two-point OPE coefficients
\begin{equation}
\label{sec_intro_ope_free_coeffs}
\mathcal{C}^\1_{\phi \phi}(x,y;z) = G(x,y) \eqend{,} \qquad \mathcal{C}^{\phi^2}_{\phi \phi}(x,y;z) = 1 \eqend{,} \qquad \mathcal{C}^{\phi \partial_\mu \phi}_{\phi \phi}(x,y;z) = (x-z)^\mu + (y-z)^\mu \eqend{.}
\end{equation}
In general interacting theories, the textbook way to compute the coefficients is to take certain high-momenta limits of correlation functions~\cite{collins1986}, which then gives the most singular parts of the OPE coefficients, the ones that contribute most to the expansion~\eqref{sec_intro_ope_general} for small separations. Obviously, this contradicts the idea to determine quantum field theories by OPE coefficients and form factors without taking recourse to correlation functions, and other constructions are needed.

For conformal field theories, conformal symmetry and associativity place severe restrictions on the OPE coefficients. This is especially fruitful in two dimensions (where the above conditions can be formulated using vertex algebras~\cite{borcherds1983,kac1996,hollandsolbermann2009}), and one can non-perturbatively construct a large class of models (see, e.g.,~\cite{belavinetal1984,verlinde1988,fredenhagenjoerss1994,lechner2008,bostelmanncadamuro2012} for very general results). In higher dimensions, these conditions are less restrictive, but can still be used to constrain and even numerically construct conformal theories, an approch known as conformal bootstrap~\cite{ferraragrillogatto1973,polyakov1974,luescher1976,mack1977,belavinetal1984,elshowketal2012,elshowkpaulos2013,elshowketal2014,elshowkpaulos2018}. For non-conformal field theories, no such strong restrictions on the OPE coefficients exist, and other ideas are needed.

One natural condition that can be imposed in a general field theory on the OPE are the (interacting) equations of motion, e.g.,
\begin{equation}
\mathcal{C}^B_{(\partial^2 \phi) A_1 \cdots A_k}(x,x_1,\ldots,x_k;y) = \partial_x^2 \mathcal{C}^B_{\phi A_1 \cdots A_k}(x,x_1,\ldots,x_k;y) = \lambda \mathcal{C}^B_{\phi^3 A_1 \cdots A_k}(x,x_1,\ldots,x_k;y)
\end{equation}
for a massless scalar $\lambda \phi^4$ interaction in flat space. Using associativity and this condition, an explicit recursive construction of the OPE coefficients as a power series in $\lambda$ was given by Holland~\cite{holland2009}, which involves an inversion of $\partial^2$ at each step and is technically involved. Much more useful would be a formula for the dependence of OPE coefficients on coupling constants that only depends on the coefficients themselves, involving the composite operator conjugate to the coupling constant. Namely, expanding such a formula in perturbation theory one could compute the OPE coefficients recursively in perturbation theory, using as input only the OPE coefficients of lower orders. Such formulas are closely related to Schwinger's quantum action principle expressing the change in correlation functions due to a change in couplings, which was shown to hold in renormalised perturbation theory (with an appropriate choice of renormalisation conditions) for both massive and massless fields in flat space by Zimmermann, Lowenstein, Schroer, Gomes and Lam~\cite{lowenstein1971,lowensteinschroer1972,gomeslowenstein1972,lam1972,zimmermann1973a}. The possibility of extending the quantum action principle to OPE coefficients was mentioned by Wilson~\cite{wilson1969}. That the counterterms in a renormalized action principle involve the OPE coefficients was shown by Sonoda~\cite{sonoda1992}, who then derived from this a variational formula for the dependence of the OPE coefficients on coupling constants~\cite{sonoda1993a}. A similar formula was also obtained by Guida and Magnoli~\cite{guidamagnoli1996} and Bochicchio and Becchetti~\cite{bochicchio2017,becchettibochicchio2019}, but all these formulas either still need to be renormalised or depend on correlation functions (in addition to OPE coefficients). The first fully renormalised variational formula that only involves the OPE coefficients themselves was derived (in perturbation theory) for scalar field theory in flat Euclidean space by Holland and Hollands~\cite{hollandhollands2015a,hollandhollands2015b} and generalised to gauge theories by Holland and the author~\cite{froebholland2016}.

In this paper, I derive an analogous formula valid also for the case of more than one coupling, and in curved space. Consider an interaction Lagrangian
\begin{equation}
L = \sum_A \int g^A(x) \op_A(x) \total x \eqend{,}
\end{equation}
where $\total x \equiv \sqrt{\det g} \, \total^n x$ in an $n$-dimensional Riemannian space with metric $g_{\mu\nu}$, and the couplings $g^A$ are in general position-dependent. The main result of this paper is the following: there exists a renormalisation scheme such that the OPE coefficients fulfill~\eqref{sec_ope_var_redef_opeder}
\begin{splitequation}
\label{sec_intro_ope_recursive}
&\frac{\delta}{\delta g^D(z)} \mathcal{C}^B_{A_1 \cdots A_k}(x_1,\ldots,x_k;y) = - \mathcal{C}^B_{D A_1 \cdots A_k}(z,x_1,\ldots,x_k;y) \\
&\hspace{4em}+ \sum_{m=1}^k \sum_{E\colon [\op_E] \leq [\op_{A_m}]} \mathcal{C}^E_{D A_m}(z,x_m;x_m) \, \mathcal{C}^B_{A_1 \cdots E \cdots A_k}(x_1,\ldots,x_k;y) \\
&\hspace{4em}+ \sum_{C\colon [\op_C] < [\op_B]} \mathcal{C}^B_{DC}(z,y;y) \, \mathcal{C}^C_{A_1 \cdots A_k}(x_1,\ldots,x_k;y) \\
&\hspace{4em}+ \int \sum_{C,E\colon [\op_C] \leq [\op_E]} \mathcal{C}^C_{DE}(z,u;u) g^E(u) \frac{\delta}{\delta g^C(u)} \mathcal{C}^B_{A_1 \cdots A_k}(x_1,\ldots,x_k;y) \total u \eqend{.}
\end{splitequation}
In this formula, the first term involving the OPE coefficient $\mathcal{C}^B_{D A_1 \cdots A_k}$ with an additional insertion of the interaction operator $\op_D$ corresponding to the coupling $g^D$ is the naively expected one which would also follow from a formal path-integral treatment, with the minus sign arising because I work in a Riemannian setting where the action enters in the form $\exp(-S)$ instead of $\exp(\mathi S)$. The terms in the second line are UV counterterms: if $z$ is close to one of the $x_i$, the OPE coefficient $\mathcal{C}^B_{D A_1 \cdots A_k}(z,x_1,\ldots,x_k;y)$ factorises according to the condition~\eqref{sec_intro_ope_factor} and one obtains
\begin{splitequation}
&\mathcal{C}^B_{D A_1 \cdots A_k}(z,x_1,\ldots,x_k;y) - \sum_{E\colon [\op_E] \leq [\op_{A_i}]} \mathcal{C}^E_{D A_i}(z,x_i;x_i) \, \mathcal{C}^B_{A_1 \cdots E \cdots A_k}(x_1,\ldots,x_k;y) \\
&\quad\sim \sum_{E\colon [\op_E] > [\op_{A_i}]} \mathcal{C}^E_{D A_i}(z,x_i;x_i) \, \mathcal{C}^B_{A_1 \cdots E \cdots A_k}(x_1,\ldots,x_k;y) \eqend{.}
\end{splitequation}
If the interaction is renormalisable, $[\op_D] \leq n$ for all interaction operators $[\op_D]$, the most singular contribution to the coefficient $\mathcal{C}^E_{D A_i}(z,x_i;x_i)$ scales like $d(x_i,z)^{[\op_E] - [\op_D] - [\op_{A_i}]}$ up to logarithmic corrections in perturbation theory. For $[\op_E] > [\op_{A_i}]$, this is less singular than $d(x_i,z)^{-n}$ and thus integrable over $z$. Similarly, the terms in the third line of~\eqref{sec_intro_ope_recursive} are IR subtractions, ensuring the decay of the right-hand side if $z$ is far from the expansion point $y$ (and thus all points). Finally, the terms in the last line arise from the specific choice of renormalisation scheme, and in special cases can be absorbed into redefinitions of the couplings (e.g., in flat space and if there is only one coupling as in $\lambda \phi^4$ theory~\cite{hollandhollands2015a,hollandhollands2015b}).

I derive formula~\eqref{sec_intro_ope_recursive} in the algebraic approch to quantum field theory on curved space(-time)s~\cite{hollandswald2015,fewsterverch2015,fredenhagenrejzner2016}, which has the advantage of cleanly separating the algebraic issues in quantum field theory (including renormalisation and possible renormalisation ambiguities) from the choice of states. Section~\ref{sec_aqft} gives a short overview of the algebraic approach, including the exact definition of composite operators in curved space that is used and the definition of (physically acceptable) Hadamard quantum states. In section~\ref{sec_ope}, I give the definition of the OPE coefficients, derive formula~\eqref{sec_intro_ope_recursive}, and show that with this definition the OPE exists in the asymptotic sense~\eqref{sec_intro_ope_general} and fulfills the factorisation and associativity conditions~\eqref{sec_intro_ope_factor} in any Hadamard state. Finally, in section~\ref{sec_example} I use the formula~\eqref{sec_intro_ope_recursive} to compute the OPE of $\phi$ with itself in hyperbolic space (Euclidean anti-de Sitter space) to first order in $g$ for a $g \phi^4$ interaction, and conclude in section~\ref{sec_outlook}. As usual, $\hbar = c = 1$.

\section{Quantum field theory in curved space from an algebraic point of view}
\label{sec_aqft}

Consider a complete smooth Riemannian manifold $M$ of dimension $\operatorname{dim} M = n \geq 2$, with smooth metric $g_{\mu\nu}$ and covariant derivative $\nabla$. An important concept in Riemannian geometry is the exponential map $\exp_x$ around a point $x$, which determines points $y$ in a neighbourhood of $x$ as endpoints of geodesics passing through $x$. That is, given the geodesic from $x$ to $y$ and the tangent vector $\xi = \xi(x,y)$ along this geodesic at the point $x$, we have $\exp_x(\xi) = y$. This is well-defined in a sufficiently small neighbourhood of $x$, called normal geodesic neighbourhood, where a unique geodesic between $x$ and $y$ exists, and the length of the tangent vector $\abs{\xi} \equiv \sqrt{ g_{\mu\nu}(x) \xi^\mu \xi^\nu }$ is equal to the geodesic length $d(x,y)$. In the following, we work exclusively in such a normal geodesic neighbourhood, which is not a restriction since the OPE is anyway a local expansion. Nevertheless, if $M$ is complete and has non-positive sectional curvature (e.g., flat Euclidean space $\mathbb{R}^n$ or hyperbolic space $\mathbb{H}^n$), the exponential map is even globally well-defined and the normal geodesic neighbourhood of any point is equal to the whole $M$, or if $M$ is not simply connected to its universal cover~\cite[Ch.~1]{cheegerebin1975}.

A useful coordinate system are Riemann normal coordinates (RNC), where geodesics are straight lines and consequently $y = \exp_x(\xi) = x + \xi$. Since RNC are uniquely defined, partial derivatives in RNC can be expressed using covariant derivatives, for which we use the short-hand notations $\nabla^{(\mu_1} \cdots \nabla^{\mu_k)} \equiv \nabla^{\mu_1 \cdots \mu_k} \equiv \nabla^k$. Concretely, the tensor $U$ defined by $U(x) = \partial^k T(x) = \partial_\xi^k T(y) \rvert_{y = x}$ in RNC at $x$ and symmetrised over all indices is equal to $\nabla^k T$, symmetrised over all indices, which is shown as follows: In RNC around $x$, geodesics have the form $z(\tau) = x + \tau \xi$, such that $T(y) = T(x+\xi)$ and $\partial_x^k T(x) = \partial_\xi^k T(x+\xi) \rvert_{\xi = 0}$. Differentiating the geodesic equation $\ddot z^\mu(\tau) + \Gamma^\mu_{\alpha\beta}(z(\tau)) \dot z^\alpha(\tau) \dot z^\beta(\tau) = 0$, one obtains
\begin{equation}
\left( \partial_{\rho_1} \cdots \partial_{\rho_n} \Gamma^\mu_{\alpha\beta} \right)(z) \xi^\alpha \xi^\beta \xi^{\rho_1} \cdots \xi^{\rho_n} = 0 \eqend{.}
\end{equation}
In particular, since $\xi$ is an arbitrary vector, it follows for $\tau = 0$ (thus $z = x$) that
\begin{equation}
\label{sec_aqft_lemma_covd_gamma0}
\left[ \partial_{(\rho_1} \cdots \partial_{\rho_n} \Gamma^\mu_{\alpha\beta)} \right](x) = 0 \eqend{.}
\end{equation}
Consider now in RNC the totally symmetrised expression
\begin{equation}
\left[ \partial_{(\alpha_1} \cdots \partial_{\alpha_m} T_{\beta_1 \cdots \beta_n)} \right](x)
\end{equation}
for some tensor $T$. Replacing a partial by a covariant derivative, this is equal to
\begin{equation}
\left[ \partial_{(\alpha_1} \cdots \partial_{\alpha_{m-1}} \nabla_{\alpha_m} T_{\beta_1 \cdots \beta_n)} \right](x) + \sum_{k=1}^n \left[ \partial_{(\alpha_1} \cdots \partial_{\alpha_{m-1}} \left( \Gamma^\mu_{\alpha_m \beta_k} T_{\beta_1 \cdots \beta_{k-1} |\mu| \beta_{k+1} \cdots \beta_n)} \right) \right](x) \eqend{.}
\end{equation}
Distributing the partial derivatives in the second term, one obtains a sum of products of totally symmetrised derivatives acting on the Christoffel symbol, and derivatives acting on $T$. Since however the totally symmetrised derivatives acting on the Christoffel symbol vanish at $x$~\eqref{sec_aqft_lemma_covd_gamma0}, all these terms vanish. Inductively one thus obtains
\begin{equation}
\left[ \partial_{(\alpha_1} \cdots \partial_{\alpha_m} T_{\beta_1 \cdots \beta_n)} \right](x) = \left[ \nabla_{(\alpha_1} \cdots \nabla_{\alpha_m} T_{\beta_1 \cdots \beta_n)} \right](x) \eqend{.}
\end{equation}
In particular, for a scalar $\phi$ one has $\partial_\xi^k \phi = \nabla^k \phi$.

\subsection{The free theory}
\label{sec_aqft_free}

For simplicity, we restrict to the case of a real scalar field $\phi$. The free theory depends on the equation-of-motion operator $P \equiv - \laplace + m^2 + \xi R$, where $m$ is the mass of the scalar field and $\xi$ denotes a non-minimal coupling to the scalar curvature $R$. We assume that there exists a Green's function $G(x,y)$ satisfying $P_x G(x,y) = \delta(x,y) = P_y G(x,y)$ with the covariant $\delta$ distribution $\delta(x,y) \equiv \delta^n(x-y)/\sqrt{\det g}$. We further assume that $G(x,y)$ is of Hadamard form $G(x,y) = H(x,y) + W(x,y)$, where $H(x,y)$ is the Hadamard parametrix that is (up to a choice of reference scale) uniquely determined by the local geometry and singular as $x \to y$, while $W(x,y)$ is a smooth function depending on the boundary conditions. Analogously to the Lorentzian case~\cite{baerginouxpfaeffle2007,hadamard1932}, the parametrix has the expansion
\begin{equation}
\label{sec_aqft_free_hadamard_parametrix}
H(x,y) = c_n \left[ \frac{U(x,y)}{\sigma(x,y)^\frac{n-2}{2}} + V(x,y) \ln\left( \mu^2 \sigma(x,y) \right) \right] \eqend{,}
\end{equation}
where $c_n$ is a normalisation constant depending on the dimension $n$, $\sigma(x,y) = \frac{1}{2} d(x,y)^2$ is Synge's world function equal to one half of the geodesic distance squared between the points $x$ and $y$, $\mu$ is a reference scale needed to make the argument of the logarithm dimensionless, and $U(x,y)$ and $V(x,y)$ are smooth symmetric functions with a computable expansion in powers of $\sigma$, see, e.g.~\cite{hollandswald2005,decaninifolacci2008,hackmoretti2012} and references therein. In even dimensions, the expansion of $U$ is finite and $V$ has an infinite expansion (convergent in analytic spaces), while for odd dimensions $V = 0$,\footnote{In odd-dimensional spaces, there is no need for a reference scale and $H$ is thus uniquely determined.} and $U$ has an infinite expansion (again convergent in analytic spaces). The expansion coefficients fulfill recursion relations that are obtained by imposing the equation of motion for $G(x,y)$ and comparing powers of $\sigma$ (expanding also $W(x,y)$), but their concrete form is irrelevant for the purposes of this paper, where it is only important that they are completely determined by the local geometry, concretely by the metric and curvature tensors along the geodesic from $x$ to $y$. On the other hand, $W(x,y)$ determines the state-dependence of the correlation functions and depends on global boundary conditions.

In the Lorentzian case, the parametrix $H$ is constrained by the fact that one needs to obtain the correct commutator $\left[ \phi(x), \phi(y) \right] = \mathi \left[ G_\text{adv}(x,y) - G_\text{ret}(x,y) \right]$, where $G_\text{adv/ret}$ are the advanced and retarded propagators for $P$. This is ensured by a proper $\mathi \epsilon$ prescription in the parametrix~\eqref{sec_aqft_free_hadamard_parametrix}, which leads to retarded, advanced, Feynman, etc. parametrices. Furthermore, $W(x,y)$ is constrained by the requirement of positivity, i.e., one must have $G(f^*,f) = \iint G(x,y) f^*(x) f(y) \total x \total y \geq 0$ for all test functions $f$\footnote{In flat Euclidean space $\mathbb{R}^n$, this is equivalent to the statement that the Fourier transform of $G$ must be positive.}, which is necessary for the probability interpretation. Contrary to the Lorentzian case, in the Euclidean case the fields $\phi$ commute everywhere, and the positivity requirement is replaced by reflection positivity, which ensures that correlation functions can be Wick-rotated back to Lorentzian signature and can then be interpreted as the correlation functions of a unitary theory~\cite{osterwalderschrader1973,osterwalderschrader1975}. However, for a general Riemannian manifold $M$ there is no corresponding Lorentzian manifold $M'$ obtained by any kind of Wick rotation, such that in the following I do not impose any condition on $W(x,y)$ (other than it being smooth).

In the algebraic approach~\cite{brunettifredenhagenkoehler1996,brunettifredenhagen2000,hollandswald2001}, one first constructs the algebra of non-interacting fields $\mathfrak{A}_0(M,g)$, which in the Riemannian case is just the commutative algebra generated by the $\phi$'s and the identity $\1$. Non-interacting quantum states $\expect{ \cdot }_\omega$ are then linear functionals on $\mathfrak{A}_0$, and if $\omega$ is a free state with two-point function $G$ one has the usual Wick theorem
\begin{splitequation}
\label{sec_aqft_free_propagator_wick}
\expect{ \phi(x_1) \cdots \phi(x_k) }_\omega = \left[ \frac{\partial^k}{\partial t_1 \cdots \partial t_k} \exp\left( \frac{1}{2} \sum_{i,j=1}^k t_i t_j G(x_i,x_j) \right) \right]_{t_i = 0} \eqend{.}
\end{splitequation}
As is well known, one has to perform normal ordering to define powers of $\phi$ at the same point because of the singularities of $G(x,y)$ as $y \to x$. There are essentially two ways: one is to perform normal ordering with respect to the full two-point function, which gives
\begin{equation}
\label{sec_aqft_free_propagator_normord}
\normord{ \phi(x_1) \cdots \phi(x_k) }_G \equiv \left[ \frac{\partial^k}{\partial t_1 \cdots \partial t_k} \exp\left( \sum_{j=1}^k t_j \phi(x_j) - \frac{1}{2} \sum_{i,j=1}^k t_i t_j G(x_i,x_j) \unitmatrix \right) \right]_{t_i = 0}
\end{equation}
and $\expect{ \normord{ \phi(x_1) \cdots \phi(x_k) }_G }_\omega = 0$. However, since the state-dependent part $W(x,y)$ of $G$ depends on global boundary conditions, the so-defined normal-ordered products do not transform locally and covariantly under diffeomorphisms, or more generally diffeomorphic embeddings. That is, consider an embedding $\chi\colon M \to M'$ of $M$ into another complete smooth Riemannian manifold $M'$ which is isometric (such that the metric $g'$ of $M'$ is given by $g' = \chi^* g$), the two associated free-field algebras $\mathfrak{A}_0(M,g)$ and $\mathfrak{A}_0'(M',g')$ and the two associated Green's functions $G$ and $G'$. While the Hadamard parametrix is completely determined by the local geometry, such that one has $H'(x,y) = H(\chi(x), \chi(y))$, this is in general not the case for the state-dependent part $W$ since it depends on boundary conditions: $W'(x,y) \neq W(\chi(x), \chi(y))$. It follows immediately that the induced map $\alpha_\chi$ between the free-field algebras $\alpha_\chi\colon \mathfrak{A}_0 \to \mathfrak{A}_0'$, defined by $\alpha_\chi(\phi(x_1) \cdots \phi(x_k)) = \phi(\chi(x_1)) \cdots \phi(\chi(x_k))$, that realizes the local covariant transformation of free fields does not map the normal products into themselves: $\alpha_\chi( \normord{ \phi(x_1) \cdots \phi(x_k) }_G ) \neq \normord{ \phi(\chi(x_1)) \cdots \phi(\chi(x_k)) }_{G'}$.

To remedy this situation, one has to perform normal ordering with respect to the Hadamard parametrix only~\cite{hollandswald2001,hollandswald2002}:
\begin{equation}
\label{sec_aqft_free_hadamard_normord}
\normord{ \phi(x_1) \cdots \phi(x_k) }_H \equiv \left[ \frac{\partial^k}{\partial t_1 \cdots \partial t_k} \exp\left( \sum_{j=1}^k t_j \phi(x_j) - \frac{1}{2} \sum_{i,j=1}^k t_i t_j H(x_i,x_j) \unitmatrix \right) \right]_{t_i = 0} \eqend{.}
\end{equation}
The Wick theorem for the Hadamard-normal-ordered products then reads
\begin{splitequation}
\label{sec_aqft_free_hadamard_wick}
\expect{ \normord{ \phi(x_1) \cdots \phi(x_k) }_H }_\omega = \left[ \frac{\partial^k}{\partial t_1 \cdots \partial t_k} \exp\left( \frac{1}{2} \sum_{i,j=1}^k t_i t_j W(x_i,x_j) \right) \right]_{t_i = 0} \eqend{,}
\end{splitequation}
and we see that the expectation values of normal-ordered products are smooth functions, such that the limit of coinciding points $x_i \to x$ can be taken. Since by construction the Hadamard parametrix only depends on the local geometry, one sees immediately that $\alpha_\chi( \normord{ \phi(x_1) \cdots \phi(x_k) }_H ) = \normord{ \phi(\chi(x_1)) \cdots \phi(\chi(x_k)) }_{H'}$, and the Hadamard-normal-ordered products transform in a locally covariant way. Composite operators that transform covariantly are then obtained by taking the limit $x_i \to x$, see the next subsection. For the product of two normal-ordered quantities, one obtains the formula
\begin{splitequation}
\label{sec_aqft_free_hadamard_product}
&\normord{ \phi(x_1) \cdots \phi(x_n) }_H \, \normord{ \phi(y_1) \cdots \phi(y_m) }_H \\
&\quad= \normord{ \phi(x_1) \cdots \phi(x_n) \exp\left( \iint \overleftarrow{\frac{\delta}{\delta \phi(u)}} H(u,v) \overrightarrow{\frac{\delta}{\delta \phi(v)}} \total u \total v \right) \phi(y_1) \cdots \phi(y_m) }_H \eqend{,}
\end{splitequation}
where the variational derivatives formally act to the left and right like on classical fields, as indicated by the arrows. This formula follows directly from the definition~\eqref{sec_aqft_free_hadamard_normord}, and an analogous formula with all $H$'s replaced by $G$'s holds for products normal ordered with respect to the full two-point function.

Lastly, we note that one can also consider more general (non-free) Hadamard states, which are linear functionals on $\mathfrak{A}_0$ whose connected two-point function has the Hadamard form $H(x,y) + W(x,y)$ as above, and whose connected $n$-point functions for $n \neq 2$ are smooth functions $W(x_1,\ldots,x_n)$.\footnote{A proof that all well-defined states on $\mathfrak{A}_0$ are Hadamard states has been given in the Lorentzian case by Hollands and Ruan~\cite{hollandsruan2002} and Sanders~\cite{sanders2010}, but the proof makes essential use of the positivity condition.} This is especially important because unlike in the Lorentzian case where there are plenty of free Hadamard states~\cite{fullingnarcowichwald1981,junkerschrohe2001}, on a large class of Riemannian manifolds there is a unique fundamental solution for the Green's function~\cite{malgrange1956,litam1987,urakawa1993}, and thus a unique free (``ground'' or ``vacuum'') state. However, from a given Hadamard state one can easily construct a new one by adding a spectator field and defining
\begin{equation}
\label{sec_aqft_free_omegaf_def}
\expect{ A }_{\omega,f} \equiv \frac{\expect{ \phi(f^*) A \, \phi(f) }_\omega}{\expect{ \phi(f^*) \phi(f) }_\omega}
\end{equation}
for any smooth function $f$ with $\phi(f) = \int \phi(x) f(x) \total x$ and $\expect{ \phi(f^*) \phi(f) }_\omega \neq 0$, which gives another (non-free) Hadamard state.

\subsection{Composite operators}
\label{sec_aqft_ops}

A composite operator is given by products of derivatives of $\phi$, and I use an abstract multi-index notation, where a label $A = ( a_1, \ldots, a_n )$ with $a_i \in \mathbb{N}_0$ is associated to the operator
\begin{equation}
\label{sec_aqft_ops_def}
\op_A = \nabla^{a_1} \phi \cdots \nabla^{a_n} \phi \quad \Leftrightarrow \quad \op_{\mu_1 \cdots \nu_{a_n}} = \nabla_{(\mu_1} \cdots \nabla_{\mu_{a_1})} \phi \cdots \nabla_{(\nu_1} \cdots \nabla_{\nu_{a_n})} \phi \eqend{.}
\end{equation}
Setting
\begin{equation}
\abs{A}_k = \abs{ \{ i \colon a_i = k \} } \eqend{,} \quad \abs{A} = \sum_{k=0}^\infty \abs{A}_k = n \eqend{,} \quad [\op_A] \equiv n [\phi] + \sum_{k=1}^n a_k = \sum_{k=0}^\infty \left( [\phi] + k \right) \abs{A}_k \eqend{,}
\end{equation}
we seen that $\abs{A}_k$ counts the number of fields $\phi$ with $k$ derivatives appearing in $\op_A$, $\abs{A}$ the total number of fields $\phi$, and $[\op_A]$ is the engineering dimension. The (countable) set of composite operators can then be ordered first by $[\op_A]$, and by some fixed but unspecified order for each given $[\op_A]$. For later use, define
\begin{equation}
\partial^B \op_A = \prod_{k=0}^\infty \frac{1}{\abs{B}_k!} \left( \frac{\partial}{\partial \nabla^k \phi} \right)^{\abs{B}_k} \op_A \eqend{,}
\end{equation}
such that $\partial^A \op_A = \unitmatrix$. Two operators are identified if their labels only differ by a permutation, which corresponds to the fact that the fields $\phi$ commute; the number of possible permutations for a given $A$ is given by
\begin{equation}
\label{sec_aqft_ops_calp_def}
\mathcal{P}_A \equiv \frac{\abs{A}!}{\prod_{k=0}^\infty \abs{A}_k!} \eqend{.}
\end{equation}
The product $\op_A(x) \op_B(x)$ of two composite operators is the composite operator $\op_C(x)$, where the multi-index $C = A \cdot B$ is given by $C = (a_1, \ldots, a_\abs{A}, b_1, \ldots, b_\abs{B})$, or any permutation thereof. In particular, when in later formulas one sums over multi-indices $A$ and $B$ with the restriction that $A \cdot B = C$ for some given multi-index $C$, this sum extends over multi-indices $A$ and $B$ corresponding to distinct composite operators $\op_A$ and $\op_B$ (i.e., counting $A$ and $A'$ that differ by a permutation only once), and such that $A \cdot B$ is some permutation of the given multi-index $C$. In other words, the sum extends over composite operators $\op_A$ and $\op_B$ such that $\op_A(x) \op_B(x) = \op_C(x)$. Since $B$ is (up to permutation) uniquely defined by this condition, instead of the sum over $A$ and $B$, one can also just sum over $A$ and denote the corresponding $B$ by $C/A$; in other words, $\op_{C/A}$ is the unique composite operator such that $\op_A(x) \op_{C/A}(x) = \op_C(x)$ holds (if a solution exists), and the corresponding sums range only over those $A$ such that $\op_{C/A}$ exists. It follows immediately that $\cdot$ and $/$ behave in the same way as the usual arithmetic operations, and in particular $A \cdot B = B \cdot A$ and $(C/A)/B = C/(A \cdot B)$.

The corresponding (Hadamard) normal-ordered composite operators $\normord{ \op_A(x) }_G$ and $\normord{ \op_A(x) }_H$ are defined in the obvious way by normal-ordering $\nabla^{a_1} \phi(x) \cdots \nabla^{a_n} \phi(x)$. In the following, I concentrate on the Hadamard-normal-ordered quantities, but analogous formulas (with $H$ replaced everywhere by $G$) hold for normal ordering with respect to the full two-point function. Since as explained above in normal-ordered expressions one can take the limit of coinciding points, and since symmetrised covariant derivatives are equal to partial derivatives in RNC, we have
\begin{equation}
\label{sec_aqft_ops_normord_op_def}
\normord{ \op_A(x) }_H = \lim_{\xi_i \to 0} \left[ \partial^{a_1}_{\xi_1} \cdots \partial^{a_n}_{\xi_n} \, \normord{ \phi(\exp_x(\xi_1)) \cdots \phi(\exp_x(\xi_n)) }_H \right] \eqend{,}
\end{equation}
and similar expressions for the product of more than one composite operator. Since expectation values of normal-ordered quantities~\eqref{sec_aqft_free_hadamard_wick} are smooth functions of the points, it follows that also the expectation value of composite operators $\expect{ \normord{ \op_{A_1}(x_1) \cdots \op_{A_k}(x_k) }_H }_\omega$ is a smooth function of the points $x_i$.

For the OPE, we need to extract the term with a given composite operator $\op_A$ from a sum of normal-ordered expressions, which I do as in~\cite{gorishnylarintkachov1983,gorishnylarin1987,bostelmann2005,hollands2007} by defining a functional $\mathcal{D}^B_x$ in such a way that
\begin{equation}
\label{sec_aqft_ops_dx_delta}
\mathcal{D}^B_x \normord{ \op_A(x) }_H = \delta_A^B
\end{equation}
holds, which the exact definition of $\delta_A^B$ given later in equation~\eqref{sec_aqft_ops_deltadef}. $\mathcal{D}^B_x$ is defined to act on a normal-ordered product of $m$ fields $\phi$ by
\begin{equation}
\label{sec_aqft_ops_dx_def}
\mathcal{D}^B_x \normord{ \phi(x_1) \cdots \phi(x_m) }_H \equiv \frac{\delta_{m,\abs{B}}}{b_1! \cdots b_\abs{B}!} \mathcal{P}_B \, \mathbb{S}_{\{b_i\}} \, \xi_1^{\otimes b_1} \cdots \xi_\abs{B}^{\otimes b_\abs{B}} \eqend{,}
\end{equation}
where the $\xi_i$ are the tangent vectors along the geodesic from $x$ to $x_i$, and where $\mathbb{S}$ denotes the idempotent symmetrisation with respect to all possible permutations of the $b_i$. This ensures that the definition does not depend on the choice of an ordering among the $b_i$ nor among the $x_i$ (since the normal-ordered product is symmetric under an exchange of its arguments). This definition is extended to all normal-ordered products by linearity and by postulating that $\mathcal{D}$ commutes with taking derivatives in RNC and taking limits. Taking derivatives one obtains
\begin{equation}
\mathcal{D}^B_x \normord{ \nabla^{a_1} \phi(x_1) \cdots \nabla^{a_m} \phi(x_m) }_H = \mathcal{P}_B \, \delta_{m,\abs{B}} \, \mathbb{S}_{\{b_i\}} \, \frac{\xi_1^{\otimes (b_1-a_1)} \cdots \xi_m^{\otimes (b_m-a_m)}}{(b_1-a_1)! \cdots (b_m-a_m)!} \eqend{,}
\end{equation}
and taking the limit $x_i \to x$ ($\xi_i \to 0$) it follows that
\begin{equation}
\label{sec_aqft_ops_deltadef}
\mathcal{D}^B_x \normord{ \op_A(x) }_H = \delta_{\abs{A},\abs{B}} \mathcal{P}_B \, \mathbb{S}_{\{b_i\}} \delta_{a_1,b_1} \cdots \delta_{a_\abs{A},b_\abs{B}} \equiv \delta_A^B \eqend{.}
\end{equation}
This corresponds to a formal Taylor expansion of the $\phi(x_i)$ around $x$, picking the coefficient which has the correct number of derivatives to correspond to $\op_B$. As one checks easily, the inclusion of the combinatorial factor $\mathcal{P}_B$ in the definition~\eqref{sec_aqft_ops_dx_def} of $\mathcal{D}^B$ ensures that the definition of $\delta_A^B$ is consistent. Namely, we have
\begin{equation}
\sum_B \delta^B_A \op_B = \op_A
\end{equation}
without any combinatorial factors when summing only over the multi-indices $B$ corresponding to distinct operators $\op_B$.

\subsection{Renormalised products}
\label{sec_aqft_renorm}

In the Riemannian setting, (free) renormalised products (called E-products in~\cite{dappiaggidragorinaldi2019}) are the analogue of the (free) time-ordered products in the Lorentzian case. In the algebraic approach, they are constructed recursively in the number of arguments, imposing certain conditions that are detailed in the following~\cite{brunettifredenhagenkoehler1996,brunettifredenhagen2000,hollandswald2001,hollandswald2002,brunettifredenhagenverch2003,hollandswald2005,dappiaggidragorinaldi2019}. The renormalised product of a single argument is just the Hadamard-normal-ordered product
\begin{equation}
\label{sec_aqft_renorm_r_1arg}
\mathcal{R}\left[ \op_A(x) \right] \equiv \normord{ \op_A(x) }_H \eqend{,}
\end{equation}
which was defined in the previous subsections. The recursive construction then proceeds by imposing the conditions of symmetry, factorisation, scaling, field independence, and local covariance, together with some other technical conditions which are not needed for the present paper. Since all fields commute, the symmetry condition is just the obvious one
\begin{equation}
\label{sec_aqft_renorm_r_symm}
\mathcal{R}\left[ \cdots \op_A(x) \cdots \op_B(y) \cdots \right] = \mathcal{R}\left[ \cdots \op_B(y) \cdots \op_A(x) \cdots \right] \eqend{,}
\end{equation}
where the dots stand for an arbitrary number of other operators. The Riemannian analogue of the well-known causal factorisation of time-ordered products in the Lorentzian case
\begin{equation}
\mathcal{T}\left[ \op_A(x) \op_B(y) \right] = \begin{cases} \mathcal{T}\left[ \op_A(x) \right] \mathcal{T}\left[ \op_B(y) \right] & x \text{ in the future of } y \\ \mathcal{T}\left[ \op_B(y) \right] \mathcal{T}\left[ \op_A(x) \right] & y \text{ in the future of } x \\ \text{either of the above} & x \text{ and } y \text{ space-like related } \end{cases}
\end{equation}
(for two arguments, and similar expressions in the general case) is given by
\begin{splitequation}
\label{sec_aqft_renorm_r_factor}
&\mathcal{R}\left[ \op_{A_1}(x_1) \cdots \op_{A_k}(x_k) \op_{B_1}(y_1) \cdots \op_{B_\ell}(x_\ell) \right] \\
&\quad= \mathcal{R}\left[ \op_{A_1}(x_1) \cdots \op_{A_k}(x_k) \right] \mathcal{R}\left[ \op_{B_1}(y_1) \cdots \op_{B_\ell}(x_\ell) \right] \\
&\quad= \mathcal{R}\left[ \op_{B_1}(y_1) \cdots \op_{B_\ell}(x_\ell) \right] \mathcal{R}\left[ \op_{A_1}(x_1) \cdots \op_{A_k}(x_k) \right] \eqend{,}
\end{splitequation}
if all the points $x_i$ are distinct from all the $y_j$. That is, since in the Riemannian case all points are space-like related, we have factorisation in either order if not all points coincide. Field independence is the condition
\begin{splitequation}
\label{sec_aqft_renorm_r_fieldindep}
&\frac{\delta}{\delta \phi(y)} \mathcal{R}\left[ \op_{A_1}(x_1) \cdots \op_{A_k}(x_k) \right] \\
&\quad= \sum_{i=1}^k \mathcal{R}\left[ \op_{A_1}(x_1) \cdots \op_{A_{i-1}}(x_{i-1}) \frac{\delta \op_{A_i}(x_i)}{\delta \phi(y)} \op_{A_{i+1}}(x_{i+1}) \cdots \op_{A_k}(x_k) \right] \eqend{,}
\end{splitequation}
and is clearly satisfied for the renormalised products of a single argument~\eqref{sec_aqft_renorm_r_1arg}. Local covariance requires that the renormalised products transform locally and covariantly under diffeomorphisms, which is the reason that one needs to use the Hadamard-normal-ordered product in~\eqref{sec_aqft_renorm_r_1arg}. Field independence and local covariance ensure the existence of a local Wick expansion (shown in appendix~\ref{appendix_wick})
\begin{splitequation}
\label{sec_aqft_renorm_r_wick}
\mathcal{R}\left[ \op_{A_1}(x_1) \cdots \op_{A_k}(x_k) \right] = \sum_{B_1,\ldots,B_k} r\left[ \partial^{B_1} \op_{A_1}(x_1) \cdots \partial^{B_k} \op_{A_k}(x_k) \right] \normord{ \op_{B_1}(x_1) \cdots \op_{B_k}(x_k) }_H \eqend{,}
\end{splitequation}
where the $r$ are numerical (locally covariant) distributions which are smooth functions for non-coinciding points. Using the local Wick expansion for each of the renormalised products in the factorisation condition~\eqref{sec_aqft_renorm_r_factor}, and formula~\eqref{sec_aqft_free_hadamard_product} for the product of two Hadamard-normal-ordered products, the distributions $r$ are already fully determined except for points which all coincide. The extension of $r$ to those points corresponds to renormalisation, and the renormalisation freedom is given by local terms ($\delta$ distributions and their derivatives, which correspond to polynomials in momenta in flat space). Furthermore, the condition of local covariance severely restricts the coefficients of these local terms, which must be polynomials of curvature tensors and their covariant derivatives of the correct dimension, together with powers of mass and other parameters of the theory~\cite{hollandswald2001,hollandswald2002,hollandswald2005}. Lastly, the scaling condition is a bound on the degree of divergence as the points are scaled together, and can be given both for the renormalised products $\mathcal{R}$ and the numerical distributions $r$, and the latter will be important for this paper. The scaling of points is defined using RNC: writing $y = \exp_x(\xi)$ with the tangent vector $\xi$ along the geodesic from $x$ to $y$, the rescaled point is given by $y^\tau \equiv \exp_x(\tau \xi)$ for $0 \leq \tau \leq 1$. The scaling condition then reads
\begin{equation}
\label{sec_aqft_renorm_r_scaling}
r\left[ \op_{A_1}(x_1^\tau) \cdots \op_{A_k}(x_k^\tau) \right] = \bigo{\tau^{- \sum_{i=1}^k [\op_{A_i}]}} \eqend{,}
\end{equation}
where I write
\begin{equation}
\label{sec_aqft_renorm_ordertau}
f(\tau) = \bigo{\tau^\alpha} \quad \Leftrightarrow \quad \lim_{\tau \to 0} \tau^{-\alpha+\delta} f(\tau) = 0 \quad\text{for all } \delta > 0 \eqend{,}
\end{equation}

In the Lorentzian case, one then includes the interaction $L$ by the Bogoliubov formula~\cite{bogoliubovshirkov1959} and defines the interacting time-ordered products $\mathcal{T}_L$ as
\begin{equation}
\mathcal{T}_L\left[ \op_{A_1}(x_1) \cdots \op_{A_k}(x_k) \right] \equiv \mathcal{T}\left[ \exp\left( \frac{\mathi}{\hbar} L \right) \right]^{(-1)} \mathcal{T}\left[ \op_{A_1}(x_1) \cdots \op_{A_k}(x_k) \exp\left( \frac{\mathi}{\hbar} L \right) \right] \eqend{.}
\end{equation}
Since the time-ordered products factorise only if its arguments are causally separated, this gives a non-trivial result. For example, to first order one obtains for the interacting observables/composite operators the expression
\begin{equation}
\mathcal{T}_L\left[ \op_A(x) \right] = \mathcal{T}\left[ \op_A(x) \right] + \frac{\mathi}{\hbar} \left( \mathcal{T}\left[ \op_A(x) L \right] - \mathcal{T}\left[ L \right] \mathcal{T}\left[ \op_A(x) \right] \right) + \bigo{L^2} \eqend{,}
\end{equation}
and the second term is non-vanishing for (the part of) $L$ in the past light cone of $x$. In the Riemannian case, this does not work for the simple reason that the renormalised products factorise everywhere (except for coinciding points). Instead, one has to define expectation values in the interacting theory (the Schwinger functions) by the Gell-Mann--Low formula~\cite{gellmannlow1951}
\begin{equation}
\label{sec_aqft_renorm_omegal}
\expect{ \op_{A_1}(x_1) \cdots \op_{A_k}(x_k) }_\Omega \equiv \frac{\expect{ \mathcal{R}\left[ \op_{A_1}(x_1) \cdots \op_{A_k}(x_k) \exp\left( - L \right) \right] }_\omega}{\expect{ \mathcal{R}\left[ \exp\left( - L \right) \right] }_\omega}
\end{equation}
for some state $\omega$ of the non-interacting theory.\footnote{This is nevertheless no real difference since one can show~\cite{duch2017} (under some reasonable technical conditions) that the expectation value of the interacting time-ordered product $\expect{ \mathcal{T}_L }_\omega$ agrees with the Gell-Mann--Low formula in the adiabatic limit.}

\section{The operator product expansion}
\label{sec_ope}

One would like to define the OPE coefficients $\mathcal{C}^B_{A_1 \cdots A_k}(x_1,\ldots,x_k;y)$ such that the OPE holds for all Schwinger functions~\eqref{sec_aqft_renorm_omegal} as an asymptotic equality:
\begin{equation}
\label{sec_ope_schwinger}
\expect{ \op_{A_1}(x_1) \cdots \op_{A_k}(x_k) }_\Omega \sim \sum_B \mathcal{C}^B_{A_1 \cdots A_k}(x_1,\ldots,x_k;y) \, \expect{ \op_B(y) }_\Omega \eqend{.}
\end{equation}
In perturbation theory, multiplying both sides with $\expect{ \mathcal{R}\left[ \exp\left( - L \right) \right] }_\omega$ this is equivalent to
\begin{equation}
\label{sec_ope_schwinger_numerator}
\expect{ \mathcal{R}\left[ \op_{A_1}(x_1) \cdots \op_{A_k}(x_k) \exp\left( - L \right) \right] }_\omega \sim \sum_B \mathcal{C}^B_{A_1 \cdots A_k}(x_1,\ldots,x_k;y) \, \expect{ \mathcal{R}\left[ \op_B(y) \exp\left( - L \right) \right] }_\omega \eqend{,}
\end{equation}
which is easier to work with. Instead of performing the OPE for all operators appearing in the expectation value $\expect{ \mathcal{R}\left[ \op_{A_1}(x_1) \cdots \op_{A_k}(x_k) \exp\left( - L \right) \right] }_\omega$, one can also perform an expansion of the first $m$ operators, and then an expansion of the remaining ones. Demanding that this agrees with the expansion for all operators at once, one obtains the factorisation condition
\begin{equation}
\label{sec_ope_factorisation}
\tilde{\mathcal{C}}^B_{A_1 \cdots A_k}(x_1,\ldots,x_k;z) \sim \sum_C \tilde{\mathcal{C}}^C_{A_1 \cdots A_m}(x_1,\ldots,x_m;y) \, \tilde{\mathcal{C}}^B_{C A_{m+1} \cdots A_k}(y,x_{m+1},\ldots,x_k;z) \eqend{.}
\end{equation}
By demanding that two different factorisations agree, the associativity condition follows automatically:
\begin{splitequation}
\label{sec_ope_associativity}
&\sum_B \tilde{\mathcal{C}}^B_{A_1 \cdots A_m}(x_1,\ldots,x_m;y) \, \tilde{\mathcal{C}}^C_{B A_{m+1} \cdots A_k}(y,x_{m+1},\ldots,x_k;z) \\
&\quad\sim \sum_B \tilde{\mathcal{C}}^B_{A_1 \cdots A_n}(x_1,\ldots,x_n;y) \, \tilde{\mathcal{C}}^C_{B A_{n+1} \cdots A_k}(y,x_{n+1},\ldots,x_k;z) \quad (m \neq n) \eqend{,}
\end{splitequation}
and other associativity conditions are obtained by permuting the operators $\op_{A_i}$ before performing the OPE, using the symmetry condition~\eqref{sec_aqft_renorm_r_symm} for the renormalised products.

In subsections~\ref{sec_ope_coeffs} to~\ref{sec_ope_var_redef} I will first define the OPE coefficients and derive a recursion formula for them, which then allows a quite short proof of the OPE~\eqref{sec_ope_schwinger_numerator}, as well as the factorisation~\eqref{sec_ope_factorisation} and associativity~\eqref{sec_ope_associativity} conditions for the coefficients as asymptotic equalities.

\subsection{Definition of OPE coefficients}
\label{sec_ope_coeffs}

Assuming that an OPE of the form~\eqref{sec_ope_schwinger_numerator} exists, one can obtain the OPE coefficients by extracting the coefficient of the corresponding operator. That is, one would like to define
\begin{equation}
\mathcal{C}^B_{A_1 \cdots A_k}(x_1,\ldots,x_k;y) \equiv \mathcal{D}^B_{L,y} \mathcal{R}\left[ \op_{A_1}(x_1) \cdots \op_{A_k}(x_k) \exp\left( - L \right) \right] \eqend{,}
\end{equation}
where $\mathcal{D}^B_{L,y}$ is a linear functional fulfilling
\begin{equation}
\label{sec_ope_coeffs_hatd_cond}
\mathcal{D}^B_{L,x} \mathcal{R}\left[ \op_A(x) \exp\left( - L \right) \right] = \delta_A^B \eqend{,}
\end{equation}
with $\delta_A^B$ defined by~\eqref{sec_aqft_ops_deltadef}. This ``projection'' of the renormalized product onto a given operator was first proposed by Gorishny, Larin and Tkachov~\cite{gorishnylarintkachov1983,gorishnylarin1987}, and independently by Bostelmann~\cite{bostelmann2005} (in the algebraic approach to QFT); the generalisation to curved spacetimes is due to Hollands~\cite{hollands2007}. In the free theory where $L = 0$, we have $\mathcal{R}\left[ \op_A(x) \right] = \normord{ \op_A(x) }_H$~\eqref{sec_aqft_renorm_r_1arg}, and the corresponding condition~\eqref{sec_aqft_ops_dx_delta} is then fufilled by the free functionals $\mathcal{D}^B_x$~\eqref{sec_aqft_ops_dx_delta}. The condition~\eqref{sec_ope_coeffs_hatd_cond} is thus equivalent to
\begin{equation}
\label{sec_ope_coeffs_hatd_cond2}
\left( \mathcal{D}^B_{L,x} - \mathcal{D}^B_x \right) \mathcal{R}\left[ \op_A(x) \exp\left( - L \right) \right] = - \mathcal{D}^B_x \Big( \mathcal{R}\left[ \op_A(x) \exp\left( - L \right) \right] - \mathcal{R}\left[ \op_A(x) \right] \Big) \eqend{,}
\end{equation}
from which an expression for the difference $\mathcal{D}^B_{L,x} - \mathcal{D}^B_x$ can be obtained to arbitrary order in perturbation theory. Namely, consider the Neumann series formed by the right-hand side
\begin{equation}
D^B_x{}_A \equiv - \mathcal{D}^B_x \Big( \mathcal{R}\left[ \op_A(x) \exp\left( - L \right) \right] - \mathcal{R}\left[ \op_A(x) \right] \Big) \eqend{,}
\end{equation}
interpreted as multiplication operators from the set of composite operators or multi-indices to itself. The formal sum
\begin{equation}
\label{sec_ope_coeffs_hatdba_def}
\hat{D}^B_x{}_A \equiv \sum_{k=0}^\infty \left( D_x^k \right)^B{}_A = \delta_A^B + D^B_x{}_A + \sum_C D^B_x{}_C D^C_x{}_A + \cdots
\end{equation}
is then equal to $\left( \delta_A^B - D^B_x{}_A \right)^{-1}$, i.e., we have
\begin{equation}
\delta_A^B = \sum_C \hat{D}^B_x{}_C \left( \delta_A^C - D^C_x{}_A \right) = \sum_C \hat{D}^B_x{}_C \mathcal{D}^C_x \mathcal{R}\left[ \op_A(x) \exp\left( - L \right) \right]
\end{equation}
and thus
\begin{equation}
\mathcal{D}^B_{L,x} \mathcal{R}\left[ \op_A(x) \exp\left( - L \right) \right] = \delta_A^B = \sum_C \hat{D}^B_x{}_C \mathcal{D}^C_x \mathcal{R}\left[ \op_A(x) \exp\left( - L \right) \right] \eqend{,}
\end{equation}
which is solved by
\begin{equation}
\mathcal{D}^B_{L,x} = \sum_C \hat{D}^B_x{}_C \mathcal{D}^C_x \eqend{.}
\end{equation}
Since the sum over $C$, and also the sums implicit in the definition~\eqref{sec_ope_coeffs_hatdba_def} of $\hat{D}^B_x{}_C$ are infinite, this is only a formal solution. Nevertheless, I will now proceed and see later on how to turn this into a well-defined solution.

Using that $\hat{D}^B_x{}_A = \delta^B_A + \sum_C D^B_x{}_C \hat{D}^C_x{}_A$, it follows that
\begin{splitequation}
\mathcal{C}^B_{A_1 \cdots A_k}(x_1,\ldots,x_k;y) &= \sum_C \hat{D}^B_y{}_C \mathcal{D}^C_y \mathcal{R}\left[ \op_{A_1}(x_1) \cdots \op_{A_k}(x_k) \exp\left( - L \right) \right] \\
&= \mathcal{D}^B_y \mathcal{R}\left[ \op_{A_1}(x_1) \cdots \op_{A_k}(x_k) \exp\left( - L \right) \right] \\
&\quad+ \sum_{C,E} D^B_y{}_E \hat{D}^E_y{}_C \mathcal{D}^C_y \mathcal{R}\left[ \op_{A_1}(x_1) \cdots \op_{A_k}(x_k) \exp\left( - L \right) \right] \\
&= \mathcal{D}^B_y \mathcal{R}\left[ \op_{A_1}(x_1) \cdots \op_{A_k}(x_k) \exp\left( - L \right) \right] \\
&\quad+ \sum_E \mathcal{C}^E_{A_1 \cdots A_k}(x_1,\ldots,x_k;y) D^B_y{}_E \eqend{,}
\end{splitequation}
which is a recursive definition for the OPE coefficients in perturbation theory since $D^B_y{}_E$ is at least of first order. By restricting the sum over $E$ to operators with $[\op_E] < \Delta$ for some $\Delta > 0$, this becomes well-defined in perturbation theory. That is, the OPE coefficients with cutoff $\Delta$ and expansion point $y$ are defined as a formal power series in perturbation theory by the recursion
\begin{splitequation}
\label{sec_ope_coeffs_def_eq}
&\mathcal{C}^{B,\Delta}_{A_1 \cdots A_k}(x_1,\ldots,x_k;y) \equiv \mathcal{D}^B_y \mathcal{R}\left[ \op_{A_1}(x_1) \cdots \op_{A_k}(x_k) \exp\left( - L \right) \right] \\
&\qquad- \sum_{C\colon [\op_C] < \Delta} \mathcal{C}^{C,\Delta}_{A_1 \cdots A_k}(x_1,\ldots,x_k;y) \mathcal{D}^B_y \Big( \mathcal{R}\left[ \op_C(y) \exp\left( - L \right) \right] - \mathcal{R}\left[ \op_C(y) \right] \Big) \eqend{.}
\end{splitequation}
One checks that the same recursion formula can be obtained from the definition of Hollands~\cite{hollands2007}.

To show existence of the OPE in any form it is obviously important that no artificial cutoffs are in place. In the free theory, the definition~\ref{sec_ope_coeffs_def_eq} of the OPE coefficients does not depend on $\Delta$, but including interactions one has to perform a redefinition of composite operators to be able to take the limit $\Delta \to \infty$. In general, such a redefinition is of the form
\begin{equation}
\tilde{\op}_A(x) = \sum_B \mathcal{Z}_A{}^B(x) \op_B(x) \eqend{,}
\end{equation}
where the $\mathcal{Z}_A{}^B(x)$ are in general functions of $x$. In order for this redefinition to correspond to an allowed mixing of composite operators, $\mathcal{Z}$ must be lower triangular, i.e. $\mathcal{Z}_A{}^B = 0$ if $[\op_B] > [\op_A]$, which ensures that a composite operator only mixes with operators of equal or lower dimension such that $[\tilde{\op}_A] = [\op_A]$, and that the sum over $B$ is finite. For the purposes of removing the cutoff $\Delta$, the redefinition will be at least of first order in $L$ and depend on $\Delta$, such that
\begin{equation}
\mathcal{Z}^\Delta_A{}^B(x) = \delta_A^B + \bigo{L} \eqend{.}
\end{equation}
Under such a redefinition, the OPE coefficients $\mathcal{C}^B_{A_1 \cdots A_k}(x_1,\ldots,x_k;y)$ should obviously transform covariantly, i.e., with $\mathcal{Z}_{A_k}{}^{C_k}(x_k)$ for the operators $\op_{A_k}$ and with $\left( \mathcal{Z}^{-1} \right)_D{}^B(y)$ for $\op_B$, where the inverse is given as a formal power series in perturbation theory by the geometric series
\begin{equation}
\left( \mathcal{Z}^\Delta \right)^{-1}_A{}^B = \delta_A{}^B + \sum_{k=\mathi1}^\infty (-1)^k \left[ \left( \mathcal{Z}^\Delta - \delta \right)^k \right]_A{}^B \eqend{,}
\end{equation}
and is again lower triangular. The OPE coefficients should also transform covariantly, which means that (in the sense of formal perturbation theory)
\begin{splitequation}
&\tilde{\mathcal{C}}^B_{A_1 \cdots A_k}(x_1,\ldots,x_k;y) \\
&\quad= \sum_{C_1, \ldots, C_k, D} \left( \mathcal{Z}^{-1} \right)_D{}^B(y) \mathcal{Z}_{A_1}{}^{C_1}(x_1) \cdots \mathcal{Z}_{A_k}{}^{C_k}(x_k) \, \mathcal{C}^D_{C_1 \cdots C_k}(x_1,\ldots,x_k;y) \eqend{.}
\end{splitequation}
However, since there is a cutoff $\Delta$ on operator dimensions it is only necessary to demand this equality for operators of low dimension; concretely I set
\begin{splitequation}
\label{sec_ope_coeffs_redef_transform}
\tilde{\mathcal{C}}^{B,\Delta}_{A_1 \cdots A_k}(x_1,\ldots,x_k;y) &\equiv \sum_{C_1, \ldots, C_k} \mathcal{Z}^\Delta_{A_1}{}^{C_1}(x_1) \cdots \mathcal{Z}^\Delta_{A_k}{}^{C_k}(x_k) \\
&\quad\times \begin{cases} \displaystyle\sum_{D\colon [\op_D] < \Delta} \left( \mathcal{Z}^\Delta \right)^{-1}_D{}^B(y) \, \mathcal{C}^{D,\Delta}_{C_1 \cdots C_k}(x_1,\ldots,x_k) & [\op_B] < \Delta \\ \mathcal{C}^{B,\Delta}_{C_1 \cdots C_k}(x_1,\ldots,x_k;y) & [\op_B] \geq \Delta \eqend{.} \end{cases}
\end{splitequation}
The mixing matrix $\mathcal{Z}^\Delta$ should be determined such that the recursion formula for the redefined coefficients reads
\begin{splitequation}
\label{sec_ope_coeffs_redef_recursion1}
&\tilde{\mathcal{C}}^{B,\Delta}_{A_1 \cdots A_k}(x_1,\ldots,x_k;y) \equiv \mathcal{D}^B_y \mathcal{R}\left[ \tilde{\op}_{A_1}(x_1) \cdots \tilde{\op}_{A_k}(x_k) \exp\left( - L \right) \right] \\
&\qquad- \sum_{C\colon [\op_C] < \min([\op_B],\Delta)} \tilde{\mathcal{C}}^{C,\Delta}_{A_1 \cdots A_k}(x_1,\ldots,x_k;y) \mathcal{D}^B_y \Big( \mathcal{R}\left[ \tilde{\op}_C(y) \exp\left( - L \right) \right] - \mathcal{R}\left[ \tilde{\op}_C(y) \right] \Big) \eqend{,}
\end{splitequation}
since then the limit $\Delta \to \infty$ can be taken without problems. Combining all of the above, $\mathcal{Z}^\Delta$ needs to be determined such that
\begin{splitequation}
&\sum_{D\colon [\op_D] < \Delta} \left[ \left( \mathcal{Z}^\Delta \right)^{-1}_D{}^B(y) - \delta_D^B \right] \mathcal{C}^{D,\Delta}_{A_1 \cdots A_k}(x_1,\ldots,x_k;y) = \sum_{C,D,E\colon [\op_C] \geq [\op_B], [\op_D] < \Delta} \left( \mathcal{Z}^\Delta \right)^{-1}_D{}^C(y) \\
&\hspace{8em}\times \mathcal{Z}^\Delta_C{}^E(y) \, \mathcal{C}^{D,\Delta}_{A_1 \cdots A_k}(x_1,\ldots,x_k;y) \, \mathcal{D}^B_y \Big( \mathcal{R}\left[ \op_E(y) \exp\left( - L \right) \right] - \mathcal{R}\left[ \op_E(y) \right] \Big)
\end{splitequation}
holds for $[\op_B] < \Delta$, while no condition arises for $[\op_B] \geq \Delta$. By ``cancelling'' the OPE coefficients on both sides, one thus obtains an explicit recursion formula
\begin{splitequation}
\label{sec_ope_coeffs_zinv_recursive}
\left( \mathcal{Z}^\Delta \right)^{-1}_D{}^B(x) = \delta_D^B &+ \sum_{C,E\colon [\op_C] \geq [\op_B]} \left( \mathcal{Z}^\Delta \right)^{-1}_D{}^C(x) \mathcal{Z}^\Delta_C{}^E(x) \\
&\qquad\times \mathcal{D}^B_x \Big( \mathcal{R}_{>0}\left[ \op_E(x) \exp\left( - L \right) \right] - \mathcal{R}_{>0}\left[ \op_E(x) \right] \Big)
\end{splitequation}
valid for $[\op_B],[\op_D] < \Delta$, while if either $[\op_B] \geq \Delta$ or $[\op_D] \geq \Delta$ we simply have $\mathcal{Z}^\Delta_D{}^B(x) = \delta_D^B$. Since both $[\op_B]$ and $[\op_D]$ are smaller than $\Delta$, in the sum also $[\op_C] < \Delta$ and then $[\op_E] < \Delta$, and the formula closes on the finite submatrix with operator dimensions less than $\Delta$. Moreover, the solution of the recursion is explicitly lower triangular: if $[\op_B] > [\op_D]$, we have $[\op_C] \geq [\op_B] > [\op_D]$ and the inverse $\left( \mathcal{Z}^\Delta \right)^{-1}_D{}^C$ vanishes by induction. Multiplying by $\mathcal{Z}^\Delta_A{}^D$ and summing over $D$, one also obtains the recursive formula
\begin{equation}
\label{sec_ope_coeffs_z_recursive}
\mathcal{Z}^\Delta_A{}^B(x) = \delta_A^B - \begin{cases} \displaystyle \sum_E \mathcal{Z}^\Delta_A{}^E(x) \, \mathcal{D}^B_x \Big( \mathcal{R}\left[ \op_E(x) \exp\left( - L \right) \right] - \mathcal{R}\left[ \op_E(x) \right] \Big) & [\op_B] \leq [\op_A] < \Delta \\ 0 & \text{else}\eqend{,} \end{cases}
\end{equation}
where being lower triangular is obvious. Since $\mathcal{D}^B_y \mathcal{R}\left[ \op_C(y) \right] = \delta^B_C$~\eqref{sec_aqft_ops_dx_delta} and the mixing matrix is lower triangular, the very last term $\mathcal{R}\left[ \tilde{\op}_C(y) \right] = \sum_E \mathcal{Z}^\Delta_C{}^E(y) \mathcal{R}\left[ \op_E(y) \right]$ in~\eqref{sec_ope_coeffs_redef_recursion1} does not contribute, and it follows that for each $\Delta > 0$, there exists an admissible redefinition of composite operators $\tilde{\op}_A$, given by equation~\eqref{sec_ope_coeffs_z_recursive}, such that the redefined OPE coefficients $\tilde{C}^{B,\Delta}_{A_1 \cdots A_k}$ satisfy the recursion~\eqref{sec_ope_coeffs_redef_recursion1} without the very last term. Furthermore, from the explicit formula~\eqref{sec_ope_coeffs_z_recursive} for the mixing matrix $\mathcal{Z}^\Delta$, it follows that the redefinition is stable in the sense that for all composite operators $\op_A$ with $[\op_A] < \Delta$ the redefined operators $\tilde{\op}_A$ do not depend on $\Delta$, that is, they do not change if one considers a different $\Delta' > \Delta$. Consequently, if $[\op_B] < \Delta$ and $[\op_{A_i}] < \Delta$ for all $i$ we have $\tilde{\mathcal{C}}^{B,\Delta'}_{A_1 \cdots A_k}(x_1,\ldots,x_k;y) = \tilde{\mathcal{C}}^{B,\Delta}_{A_1 \cdots A_k}(x_1,\ldots,x_k;y)$ for all $\Delta' > \Delta$ by the recursion formula~\eqref{sec_ope_coeffs_redef_recursion1}, and the limit $\Delta \to \infty$ can be taken. The resulting OPE coefficients $\tilde{\mathcal{C}}^B_{A_1 \cdots A_k}(x_1,\ldots,x_k;y)$ are then obtained recursively as
\begin{splitequation}
\label{sec_ope_coeffs_redef_recursion}
&\tilde{\mathcal{C}}^B_{A_1 \cdots A_k}(x_1,\ldots,x_k;y) = \mathcal{D}^B_y \mathcal{R}\left[ \tilde{\op}_{A_1}(x_1) \cdots \tilde{\op}_{A_k}(x_k) \exp\left( - L \right) \right] \\
&\hspace{6em}- \sum_{C\colon [\op_C] < [\op_B]} \tilde{\mathcal{C}}^C_{A_1 \cdots A_k}(x_1,\ldots,x_k;y) \mathcal{D}^B_y \mathcal{R}\left[ \tilde{\op}_C(y) \exp\left( - L \right) \right] \eqend{.}
\end{splitequation}

\subsection{Variational formula for the original OPE coefficients}
\label{sec_ope_var_orig}

I start with the simpler case of the original definition~\eqref{sec_ope_coeffs_def_eq} with cutoff $\Delta$. Since the renormalised products are defined in the free theory, the renormalisation procedure is in particular independent of any coupling constant. Since also the definition of composite operators~\eqref{sec_aqft_ops_def} and the Hadamard-normal ordering~\eqref{sec_aqft_ops_normord_op_def},~\eqref{sec_aqft_free_hadamard_normord} are independent of the coupling, one simply has
\begin{equation}
\label{sec_ope_var_orig_dgrenorm}
\frac{\delta}{\delta g(z)} \mathcal{R}\left[ \op_{A_1}(x_1) \cdots \op_{A_k}(x_k) \exp\left( - L \right) \right] = - \mathcal{R}\left[ \op_{A_1}(x_1) \cdots \op_{A_k}(x_k) \op_\mathcal{I}(z) \exp\left( - L \right) \right]
\end{equation}
for $L = \int g(z) \, \op_\mathcal{I}(z) \total z$. This formula is basically the naive one that one expects from a path integral definition of correlation functions, with the small (but important) difference that both sides are fully renormalised. Since also the free functional $\mathcal{D}^B_y$~\eqref{sec_aqft_ops_dx_def} is independent of the coupling, it is straightforward to take a derivative of the recursion~\eqref{sec_ope_coeffs_def_eq} and use equation~\eqref{sec_ope_var_orig_dgrenorm} to obtain
\begin{splitequation}
&\frac{\delta}{\delta g(z)} \mathcal{C}^{B,\Delta}_{A_1 \cdots A_k}(x_1,\ldots,x_k;y) = - \mathcal{C}^{B,\Delta}_{A_1 \cdots A_k \mathcal{I}}(x_1,\ldots,x_k,z;y) \\
&\qquad+ \sum_{C\colon [\op_C] < \Delta} \, \mathcal{C}^{C,\Delta}_{A_1 \cdots A_k}(x_1,\ldots,x_k;y) \, \mathcal{C}^{B,\Delta}_{C \mathcal{I}}(y,z;y) \\
&\qquad- \sum_{C\colon [\op_C] < \Delta} \frac{\delta}{\delta g(z)} \mathcal{C}^{C,\Delta}_{A_1 \cdots A_k}(x_1,\ldots,x_k;y) \mathcal{D}^B_y \Big( \mathcal{R}\left[ \op_C(y) \exp\left( - L \right) \right] - \mathcal{R}\left[ \op_C(y) \right] \Big) \\
&\qquad- \sum_{C\colon [\op_C] < \Delta} \mathcal{C}^{C,\Delta}_{A_1 \cdots A_k \mathcal{I}}(x_1,\ldots,x_k,z;y) \mathcal{D}^B_y \Big( \mathcal{R}\left[ \op_C(y) \exp\left( - L \right) \right] - \mathcal{R}\left[ \op_C(y) \right] \Big) \\
&\qquad+ \sum_{C\colon [\op_C] < \Delta} \mathcal{C}^{C,\Delta}_{A_1 \cdots A_k}(x_1,\ldots,x_k;y) \\
&\qquad\qquad\times \sum_{D\colon [\op_D] < \Delta} \mathcal{C}^{D,\Delta}_{C \mathcal{I}}(y,z;y) \mathcal{D}^B_y \Big( \mathcal{R}\left[ \op_D(y) \exp\left( - L \right) \right] - \mathcal{R}\left[ \op_D(y) \right] \Big) \eqend{,}
\end{splitequation}
where I used equation~\eqref{sec_ope_coeffs_def_eq} to express $\mathcal{D}^B_y$ acting on a renormalised product again by OPE coefficients. Renaming $C \leftrightarrow D$ in the last line, it is seen that the three last lines are of higher order in perturbation theory, and that the unique solution of this equation reads
\begin{splitequation}
\frac{\delta}{\delta g(z)} \mathcal{C}^{B,\Delta}_{A_1 \cdots A_k}(x_1,\ldots,x_k;y) &= - \mathcal{C}^{B,\Delta}_{A_1 \cdots A_k \mathcal{I}}(x_1,\ldots,x_k,z;y) \\
&\quad+ \sum_{C\colon [\op_C] < \Delta} \mathcal{C}^{C,\Delta}_{A_1 \cdots A_k}(x_1,\ldots,x_k;y) \, \mathcal{C}^{B,\Delta}_{C \mathcal{I}}(y,z;y) \eqend{.}
\end{splitequation}

That is, the variational derivative of an OPE coefficient with respect to a coupling constant is given by the (naively expected) coefficient with an additional insertion of the corresponding interaction operator $\op_\mathcal{I}$, and a correction term that involves a product of two OPE coefficients. Since a variational derivative was taken, to obtain the correction to the OPE coefficient itself one has to integrate this equation over $z$; the correction term ensures (at least formally) that this integral is well defined in the IR. Namely, if the point $z$ is far away from the $x_i$, the OPE coefficient $\mathcal{C}^{B,\Delta}_{A_1 \cdots A_k \mathcal{I}}(x_1,\ldots,x_k,z;y)$ should factorise exactly in the way given by the sum, such that the whole expression vanishes at least in the formal limit $\Delta \to \infty$. On the other hand, since the OPE coefficients are by definition well-defined (renormalised) distributions, there is no UV problem when $z$ is close to any of the $x_i$. In the next subsection, we will see how this works when the limit $\Delta \to \infty$ can actually be taken.

\subsection{Variational formula for the redefined OPE coefficients}
\label{sec_ope_var_redef}

I now assume that the interaction $L$ has the generic form
\begin{equation}
L = \int \sum_A g^A(x) \op_A(x) \total x = \int \sum_A \tilde{g}^A(x) \tilde{\op}_A(x) \total x
\end{equation}
with
\begin{equation}
\label{sec_ope_var_redef_tildegdef}
\tilde{g}^A(x) \equiv \sum_B \mathcal{Z}^{-1}_B{}^A(x) g^B(x) \eqend{,} \quad g^A(x) = \sum_B \mathcal{Z}_B{}^A(x) \tilde{g}^B(x) \eqend{.}
\end{equation}
Since the inverse mixing matrix is lower triangular, if there is a finite number of couplings $g^A$ before the redefinition, there is also a finite number of redefined couplings $\tilde{g}^A$ after it. Consider the quantity
\begin{equation}
\label{sec_ope_var_redef_gamma_def}
\Gamma^\Delta_{E A}{}^B(y,x) \equiv \sum_C \left[ \frac{\delta}{\delta g^E(y)} \mathcal{Z}^\Delta_A{}^C(x) \right] \left( \mathcal{Z}^\Delta \right)^{-1}_C{}^B(x) = - \sum_C \mathcal{Z}^\Delta_A{}^C(x) \frac{\delta}{\delta g^E(y)} \left( \mathcal{Z}^\Delta \right)^{-1}_C{}^B(x) \eqend{,}
\end{equation}
which vanishes for $[\op_B] > [\op_A]$ since then at least one of the (inverse) mixing matrices vanishes, or if $[\op_A] \geq \Delta$ since then $\mathcal{Z}^\Delta_A{}^C(x) = \delta_A^C$. Using the second form and the recursive formula~\eqref{sec_ope_coeffs_zinv_recursive} for the inverse mixing matrix, one derives
\begin{splitequation}
\Gamma^\Delta_{E A}{}^B(y,x) &= \sum_{C,D\colon [\op_C] \geq [\op_B]} \Gamma^\Delta_{E A}{}^C(y,x) \mathcal{Z}^\Delta_C{}^D(x) \, \mathcal{D}^B_x \Big( \mathcal{R}\left[ \op_D(x) \exp\left( - L \right) \right] - \mathcal{R}\left[ \op_D(x) \right] \Big) \\
&+ \begin{cases} \displaystyle \begin{split}
&\sum_F \mathcal{Z}^\Delta_A{}^F(x) \, \mathcal{D}^B_x \mathcal{R}\left[ \op_E(y) \op_F(x) \exp\left( - L \right) \right] \\
&- \sum_{D,F} \Gamma^\Delta_{E A}{}^F(y,x) \mathcal{Z}^\Delta_F{}^D(x) \, \mathcal{D}^B_x \mathcal{R}\left[ \op_D(x) \exp\left( - L \right) \right] \\
&+ \sum_{F} \Gamma^\Delta_{E A}{}^F(y,x) \mathcal{Z}^\Delta_F{}^B(x) \\
\end{split} & [\op_B] \leq [\op_A] < \Delta \\ 0 & \text{else}\eqend{.} \end{cases} \raisetag{2em}
\end{splitequation}
It is a long but straightforward computation to check that
\begin{equation}
\label{sec_ope_var_redef_gammasol}
\Gamma^\Delta_{E A}{}^B(y,x) = \begin{cases} \displaystyle \sum_{D,F} \mathcal{Z}^\Delta_A{}^D(x) \left( \mathcal{Z}^\Delta \right)^{-1}_F{}^B(x) \, \mathcal{C}^{F,\Delta}_{E D}(y,x;x) & [\op_B] \leq [\op_A] < \Delta \\ 0 & \text{else} \end{cases}
\end{equation}
is a solution of this recursive equation (uniquely defined in perturbation theory), using the recursive definition of the OPE coefficients~\eqref{sec_ope_coeffs_def_eq} in the form
\begin{splitequation}
&\mathcal{D}^B_x \mathcal{R}\left[ \op_E(y) \op_F(x) \exp\left( - L \right) \right] \\
&\quad= \mathcal{C}^{B,\Delta}_{EF}(y,x;x) + \sum_{D\colon [\op_D] < \Delta} \mathcal{C}^{D,\Delta}_{EF}(y,x;x) \, \mathcal{D}^B_x \Big( \mathcal{R}\left[ \op_D(x) \exp\left( - L \right) \right] - \mathcal{R}\left[ \op_D(x) \right] \Big)
\end{splitequation}
and the recursive formula~\eqref{sec_ope_coeffs_zinv_recursive} for the inverse mixing matrix. Note that the sum over $F$ in equation~\eqref{sec_ope_var_redef_gammasol} is restricted to $[\op_F] < \Delta$, since for $[\op_F] \geq \Delta$ we have $\left( \mathcal{Z}^\Delta \right)^{-1}_F{}^B = \delta_F^B$ which does not contribute to the sum because $[\op_B] < \Delta$. However, this implies that the limit $\Delta \to \infty$ of equation~\eqref{sec_ope_var_redef_gammasol} is not well-defined. Instead, using the definition of the redefined OPE coefficients~\eqref{sec_ope_coeffs_redef_transform} one obtains from this equation
\begin{equation}
\label{sec_ope_var_redef_zgammasum}
\sum_E \mathcal{Z}^\Delta_D{}^E(y) \Gamma^\Delta_{E A}{}^B(y,x) = \begin{cases} \tilde{\mathcal{C}}^{B,\Delta}_{D A}(y,x;x) & [\op_B] \leq [\op_A] < \Delta \\ 0 & \text{else} \eqend{.} \end{cases}
\end{equation}
Since the limit $\Delta \to \infty$ exists for the redefined OPE coefficients, the mixing matrix and $\Gamma^\Delta_{EA}{}^B$ (from the definition~\eqref{sec_ope_var_redef_gamma_def}), and the sum over $E$ is finite if the number of couplings $g^E$ is finite, it is possible to take the limit $\Delta \to \infty$ of this last equation.

In this limit, I now compute
\begin{splitequation}
\label{sec_ope_var_redef_renorm_abl}
&\frac{\delta}{\delta g^E(y)} \mathcal{R}\left[ \tilde{\op}_{A_1}(x_1) \cdots \tilde{\op}_{A_k}(x_k) \exp\left( - L \right) \right] \\
&\quad= - \sum_D \mathcal{Z}^{-1}_E{}^D(y) \mathcal{R}\left[ \tilde{\op}_D(y) \tilde{\op}_{A_1}(x_1) \cdots \tilde{\op}_{A_k}(x_k) \exp\left( - L \right) \right] \\
&\qquad+ \sum_{m=1}^k \sum_D \Gamma_{E A_m}{}^D(y,x_m) \mathcal{R}\left[ \tilde{\op}_{A_1}(x_1) \cdots \tilde{\op}_D(x_m) \cdots \tilde{\op}_{A_k}(x_k) \exp\left( - L \right) \right] \eqend{,}
\end{splitequation}
and thus for the redefined OPE coefficients~\eqref{sec_ope_coeffs_redef_recursion} and using the result~\eqref{sec_ope_var_redef_zgammasum} (in the limit $\Delta \to \infty$) it follows that
\begin{splitequation}
&\sum_E \mathcal{Z}_D{}^E(z) \frac{\delta}{\delta g^E(z)} \tilde{\mathcal{C}}^B_{A_1 \cdots A_k}(x_1,\ldots,x_k;y) = - \tilde{\mathcal{C}}^B_{D A_1 \cdots A_k}(z,x_1,\ldots,x_k;y) \\
&\hspace{8em}+ \sum_{m=1}^k \sum_{E\colon [\op_E] \leq [\op_{A_m}]} \tilde{\mathcal{C}}^E_{D A_m}(z,x_m;x_m) \, \tilde{\mathcal{C}}^B_{A_1 \cdots E \cdots A_k}(x_1,\ldots,x_k;y) \\
&\hspace{8em}+ \sum_{C\colon [\op_C] < [\op_B]} \tilde{\mathcal{C}}^C_{A_1 \cdots A_k}(x_1,\ldots,x_k;y) \, \tilde{\mathcal{C}}^B_{DC}(z,y;y) \\
&\quad- \sum_{C\colon [\op_C] < [\op_B]} \Bigg[ \sum_E \mathcal{Z}_D{}^E(z) \frac{\delta}{\delta g^E(z)} \tilde{\mathcal{C}}^C_{A_1 \cdots A_k}(x_1,\ldots,x_k;y) + \tilde{\mathcal{C}}^C_{D A_1 \cdots A_k}(z,x_1,\ldots,x_k;y) \\
&\hspace{4em}- \sum_{m=1}^k \sum_{E\colon [\op_E] \leq [\op_{A_m}]} \tilde{\mathcal{C}}^E_{D A_m}(z,x_m;x_m) \, \tilde{\mathcal{C}}^C_{A_1 \cdots E \cdots A_k}(x_1,\ldots,x_k;y) \\
&\hspace{4em}- \sum_{E\colon [\op_E] < [\op_C]} \tilde{\mathcal{C}}^E_{A_1 \cdots A_k}(x_1,\ldots,x_k;y) \, \tilde{\mathcal{C}}^C_{DE}(z,y;y) \Bigg] \mathcal{D}^B_y \mathcal{R}\left[ \tilde{\op}_C(y) \exp\left( - L \right) \right] \eqend{.}
\end{splitequation}
The unique solution (in perturbation theory) is obviously given by the first three lines, but one still has to express the variational derivative using the new couplings $\tilde{g}^E$~\eqref{sec_ope_var_redef_tildegdef}. Using the definition of $\Gamma_{EA}{}^C$~\eqref{sec_ope_var_redef_gamma_def} and the sum~\eqref{sec_ope_var_redef_zgammasum} (in the limit $\Delta \to \infty$), I compute
\begin{splitequation}
\label{sec_ope_var_redef_zdg_in_tildeg}
\sum_E \mathcal{Z}_D{}^E(z) \frac{\delta}{\delta g^E(z)} &= \sum_E \mathcal{Z}_D{}^E(z) \int \sum_C \frac{\delta \tilde{g}^C(y)}{\delta g^E(z)} \frac{\delta}{\delta \tilde{g}^C(y)} \total y \\
&= \frac{\delta}{\delta \tilde{g}^D(z)} - \int \sum_{C,E\colon [\op_C] \leq [\op_E]} \tilde{\mathcal{C}}^C_{DE}(z,y;y) \tilde{g}^E(y) \frac{\delta}{\delta \tilde{g}^C(y)} \total y
\end{splitequation}
and thus
\begin{splitequation}
\label{sec_ope_var_redef_opeder}
&\frac{\delta}{\delta \tilde{g}^D(z)} \tilde{\mathcal{C}}^B_{A_1 \cdots A_k}(x_1,\ldots,x_k;y) = - \tilde{\mathcal{C}}^B_{D A_1 \cdots A_k}(z,x_1,\ldots,x_k;y) \\
&\hspace{4em}+ \sum_{m=1}^k \sum_{E\colon [\op_E] \leq [\op_{A_m}]} \tilde{\mathcal{C}}^E_{D A_m}(z,x_m;x_m) \, \tilde{\mathcal{C}}^B_{A_1 \cdots E \cdots A_k}(x_1,\ldots,x_k;y) \\
&\hspace{4em}+ \sum_{C\colon [\op_C] < [\op_B]} \tilde{\mathcal{C}}^B_{DC}(z,y;y) \, \tilde{\mathcal{C}}^C_{A_1 \cdots A_k}(x_1,\ldots,x_k;y) \\
&\hspace{4em}+ \int \sum_{C,E\colon [\op_C] \leq [\op_E]} \tilde{\mathcal{C}}^C_{DE}(z,u;u) \tilde{g}^E(u) \frac{\delta}{\delta \tilde{g}^C(u)} \tilde{\mathcal{C}}^B_{A_1 \cdots A_k}(x_1,\ldots,x_k;y) \total u \eqend{.}
\end{splitequation}
This equation agrees with the one derived in flat space~\cite{hollandhollands2015a,hollandhollands2015b,froebholland2016} except for the addition of the last term. However, this term is of higher order in couplings, and expanding equation~\eqref{sec_ope_var_redef_opeder} in powers of the couplings the right-hand side is of lower order, and the OPE coefficients can be constructed recursively. Expressing also the variational derivatives of renormalised products~\eqref{sec_ope_var_redef_renorm_abl} in terms of the redefined coupling, using the relation~\eqref{sec_ope_var_redef_zgammasum} (in the limit $\Delta \to \infty$) I also obtain
\begin{splitequation}
\label{sec_ope_var_redef_renorm_abl_neu}
&\frac{\delta}{\delta \tilde{g}^D(z)} \mathcal{R}\left[ \tilde{\op}_{A_1}(x_1) \cdots \tilde{\op}_{A_k}(x_k) \exp\left( - L \right) \right] \\
&\quad= - \mathcal{R}\left[ \tilde{\op}_D(z) \tilde{\op}_{A_1}(x_1) \cdots \tilde{\op}_{A_k}(x_k) \exp\left( - L \right) \right] \\
&\qquad+ \sum_{m=1}^k \sum_{F\colon [\op_F] \leq [\op_{A_m}]} \tilde{\mathcal{C}}^F_{D A_m}(z,x_m;x_m) \, \mathcal{R}\left[ \tilde{\op}_{A_1}(x_1) \cdots \tilde{\op}_F(x_m) \cdots \tilde{\op}_{A_k}(x_k) \exp\left( - L \right) \right] \\
&\qquad+ \int \sum_{C,E\colon [\op_C] \leq [\op_E]} \tilde{\mathcal{C}}^C_{DE}(z,y;y) \tilde{g}^E(y) \frac{\delta}{\delta \tilde{g}^C(y)} \mathcal{R}\left[ \tilde{\op}_{A_1}(x_1) \cdots \tilde{\op}_{A_k}(x_k) \exp\left( - L \right) \right] \total y \eqend{,}
\end{splitequation}
which is now a recursive equation that can again be solved order by order in couplings.

In principle, one would like to perform also a redefinition of the couplings $\tilde{g}^A \to \hat{g}^A$ to remove the last term in equations~\eqref{sec_ope_var_redef_opeder} and~\eqref{sec_ope_var_redef_renorm_abl_neu}, and simply have
\begin{equation}
\sum_E \mathcal{Z}_D{}^E(z) \frac{\delta}{\delta g^E(z)} = \frac{\delta}{\delta \hat{g}^D(z)}
\end{equation}
instead of equation~\eqref{sec_ope_var_redef_zdg_in_tildeg}. However, this is impossible in general, as can be seen as follows: for such a redefinition to exist, one must satisfy the integrability conditions
\begin{equation}
0 = \frac{\delta}{\delta \hat{g}^A(x)} \frac{\delta}{\delta \hat{g}^B(y)} - \frac{\delta}{\delta \hat{g}^B(y)} \frac{\delta}{\delta \hat{g}^A(x)} \eqend{,}
\end{equation}
which using equations~\eqref{sec_ope_var_redef_gamma_def} and~\eqref{sec_ope_var_redef_zgammasum} (in the limit $\Delta \to \infty$) reduces to
\begin{splitequation}
\label{sec_ope_var_redef_hatg_commutator}
&\frac{\delta}{\delta \hat{g}^A(x)} \frac{\delta}{\delta \hat{g}^B(y)} - \frac{\delta}{\delta \hat{g}^B(y)} \frac{\delta}{\delta \hat{g}^A(x)} \\
&= \sum_{C,D} \left[ \left( \mathcal{Z}_A{}^C(x) \frac{\delta}{\delta g^C(x)} \mathcal{Z}_B{}^D(y) \right) \frac{\delta}{\delta g^D(y)} - \left( \mathcal{Z}_B{}^D(y) \frac{\delta}{\delta g^D(y)} \mathcal{Z}_A{}^C(x) \right) \frac{\delta}{\delta g^C(x)} \right] \\
&= \sum_{C,D,E} \left[ \mathcal{Z}_A{}^C(x) \Gamma_{C B}{}^E(x,y) \mathcal{Z}_E{}^D(y) \frac{\delta}{\delta g^D(y)} - \mathcal{Z}_B{}^D(y) \Gamma_{D A}{}^E(y,x) \mathcal{Z}_E{}^C(x) \frac{\delta}{\delta g^C(x)} \right] \\
&= \sum_{C,E\colon [\op_E] \leq [\op_B]} \tilde{\mathcal{C}}^E_{A B}(x,y;y) \mathcal{Z}_E{}^C(y) \frac{\delta}{\delta g^C(y)} - \sum_{C,E\colon [\op_E] \leq [\op_A]} \tilde{\mathcal{C}}^E_{B A}(y,x;x) \mathcal{Z}_E{}^C(x) \frac{\delta}{\delta g^C(x)} \\
&= \int \left[ \sum_{E\colon [\op_E] \leq [\op_B]} \tilde{\mathcal{C}}^E_{A B}(x,z;z) \delta(y,z) - \sum_{E\colon [\op_E] \leq [\op_A]} \tilde{\mathcal{C}}^E_{B A}(y,z;z) \delta(x,z) \right] \frac{\delta}{\delta \hat{g}^E(z)} \total z \eqend{.}
\end{splitequation}
Clearly the quantity in brackets will not vanish in general, as it does not vanish even in the free theory and for $\op_A = \op_B = \phi$. Interpreting the result geometrically~\cite{sonoda1992,sonoda1993a,sonoda1993b,dolan1994a,dolan1994b,sonoda1994,paganisonoda2017}, consider the manifold of couplings $\hat{g}^E$, and define a \emph{covariant variational derivative}
\begin{equation}
\frac{\nabla}{\nabla \hat{g}^D(z)} \equiv \sum_E \mathcal{Z}_D{}^E(z) \frac{\delta}{\delta g^E(z)} \eqend{.}
\end{equation}
The result~\eqref{sec_ope_var_redef_hatg_commutator} then tells us that the commutator of two covariant variational derivatives is given by
\begin{equation}
\left[ \frac{\nabla}{\nabla \hat{g}^A(x)}, \frac{\nabla}{\nabla \hat{g}^B(y)} \right] = \sum_E \int \hat{T}_{AB}{}^E(x,y,z) \frac{\nabla}{\nabla \hat{g}^E(z)} \total z \eqend{,}
\end{equation}
where
\begin{equation}
\label{sec_ope_var_redef_torsion}
\hat{T}_{AB}{}^E(x,y,z) \equiv \left.\begin{cases} \tilde{\mathcal{C}}^E_{A B}(x,y;z) \delta(y,z) & [\op_E] \leq [\op_B] \\ 0 & \text{else} \end{cases}\right\} - \begin{cases} \tilde{\mathcal{C}}^E_{B A}(y,x;z) \delta(x,z) & [\op_E] \leq [\op_A] \\ 0 & \text{else} \end{cases}
\end{equation}
is the torsion of the covariant variational derivative. A change of couplings is a change of coordinates (a diffeomorphism) in this manifold, but since torsion is a tensor it cannot vanish in any coordinate system if it doesn't vanish in a given one. Note that the problem doesn't arise from the fact that the couplings are position-dependent: even if one takes the adiabatic limit where the couplings become constant, and consequently has to integrate the torsion~\eqref{sec_ope_var_redef_torsion} over $x$ and $y$, it can still be non-vanishing if $\op_A \neq \op_B$. Accordingly, if there is only one coupling and the adiabatic limit is taken where this coupling becomes constant, an additional redefinition of this coupling exists such that the last line of formula~\eqref{sec_ope_var_redef_opeder} is absent.

\subsection{Existence, factorisation and associativity of the OPE}
\label{sec_ope_exist}

With the recursive formula~\eqref{sec_ope_var_redef_opeder} for the OPE coefficients, it is now easy to show that the OPE is an asymptotic expansion, and that the coefficients factorise and fulfill the associativity condition. Consider thus the remainder of the OPE~\eqref{sec_ope_schwinger_numerator}
\begin{splitequation}
\label{sec_ope_exist_operemainder}
N^{\omega,\Delta}_{A_1 \cdots A_k}(x_1,\ldots,x_k;y) &\equiv \expect{ \mathcal{R}\left[ \tilde{\op}_{A_1}(x_1) \cdots \tilde{\op}_{A_k}(x_k) \exp\left( - L \right) \right] }_\omega \\
&\qquad- \sum_{B\colon [\op_B] < \Delta} \tilde{\mathcal{C}}^B_{A_1 \cdots A_k}(x_1,\ldots,x_k;y) \, \expect{ \mathcal{R}\left[ \tilde{\op}_B(y) \exp\left( - L \right) \right] }_\omega \eqend{,}
\end{splitequation}
where the index $\Delta > 0$ determines how many terms of the OPE are subtracted, and the remainder of the factorisation~\eqref{sec_ope_factorisation}
\begin{splitequation}
\label{sec_ope_exist_factremainder}
N^{B,m,\Delta}_{A_1 \cdots A_k}(x_1,\ldots,x_k;y,z) &\equiv \tilde{\mathcal{C}}^B_{A_1 \cdots A_k}(x_1,\ldots,x_k;z) \\
&- \sum_{C\colon [\op_C] < \Delta} \tilde{\mathcal{C}}^C_{A_1 \cdots A_m}(x_1,\ldots,x_m;y) \, \tilde{\mathcal{C}}^B_{C A_{m+1} \cdots A_k}(y,x_{m+1},\ldots,x_k;z) \eqend{,}
\end{splitequation}
where $\Delta > 0$ has the same meaning as before, while the index $m$ denotes that factorisation occurs between the first $m$ and the last $k-m$ operators.

Using the variational formula~\eqref{sec_ope_var_redef_opeder}, equation~\eqref{sec_ope_var_redef_renorm_abl_neu} and that the (non-interacting) expectation value $\expect{ \cdot }_\omega$ is independent of the couplings, a long but straightforward computation yields
\begin{splitequation}
\label{sec_ope_exist_operemainder_der}
&\frac{\delta}{\delta \tilde{g}^D(z)} N^{\omega,\Delta}_{A_1 \cdots A_k}(x_1,\ldots,x_k;y) = - N^{\omega,\Delta}_{D A_1 \cdots A_k}(z,x_1,\ldots,x_k;y) \\
&\hspace{4em}+ \sum_{i=1}^k \sum_{C\colon [\op_C] \leq [\op_{A_i}]} \tilde{\mathcal{C}}^C_{D A_i}(z,x_i;x_i) \, N^{\omega,\Delta}_{A_1 \cdots C \cdots A_k}(x_1,\ldots,x_k;y) \\
&\hspace{4em}+ \sum_{B\colon [\op_B] < \Delta} \tilde{\mathcal{C}}^B_{A_1 \cdots A_k}(x_1,\ldots,x_k;y) \, N^{\omega,\Delta}_{DB}(z,y;y) \\
&\hspace{4em}+ \int \sum_{C,E\colon [\op_C] \leq [\op_E]} \tilde{\mathcal{C}}^C_{DE}(z,u;u) \tilde{g}^E(u) \frac{\delta}{\delta \tilde{g}^C(u)} N^{\omega,\Delta}_{A_1 \cdots A_k}(x_1,\ldots,x_k;y) \total u
\end{splitequation}
and
\begin{splitequation}
\label{sec_ope_exist_factremainder_der}
&\frac{\delta}{\delta \tilde{g}^D(u)} N^{B,m,\Delta}_{A_1 \cdots A_k}(x_1,\ldots,x_k;y,z) = - N^{B,m+1,\Delta}_{D A_1 \cdots A_k}(u,x_1,\ldots,x_k;y,z) \\
&\hspace{4em}+ \sum_{i=1}^k \sum_{E\colon [\op_E] \leq [\op_{A_i}]} \tilde{\mathcal{C}}^E_{D A_i}(u,x_i;x_i) \, N^{B,m,\Delta}_{A_1 \cdots E \cdots A_k}(x_1,\ldots,x_k;y,z) \\
&\hspace{4em}+ \sum_{C\colon [\op_C] < [\op_B]} \tilde{\mathcal{C}}^B_{DC}(u,z;z) \, N^{C,m,\Delta}_{A_1 \cdots A_k}(x_1,\ldots,x_k;y,z) \\
&\hspace{4em}+ \sum_{C\colon [\op_C] < \Delta} \tilde{\mathcal{C}}^C_{A_1 \cdots A_m}(x_1,\ldots,x_m;y) \, N^{B,2,\Delta}_{D C A_{m+1} \cdots A_k}(u,y,x_{m+1},\ldots,x_k;y,z) \\
&\hspace{4em}+ \int \sum_{C,E\colon [\op_C] \leq [\op_E]} \tilde{\mathcal{C}}^C_{DE}(u,v;v) \tilde{g}^E(v) \frac{\delta}{\delta \tilde{g}^C(v)} N^{B,m,\Delta}_{A_1 \cdots A_k}(x_1,\ldots,x_k;y,z) \eqend{.}
\end{splitequation}
Together with the variational formula~\eqref{sec_ope_var_redef_opeder} for the OPE coefficients, with these equations it is now easy to show inductively that as the insertion points $x_i$ scale towards the expansion point $y$, the OPE coefficients scale in the appropriate way, that the OPE is an asymptotic expansion, and that the OPE coefficients factorise properly. The scaling of the points is defined in RNC as for the renormalised products: given $x = \exp_y(\xi)$ with $\xi$ the tangent vector at the geodesic from $x$ to $y$, the rescaled point reads $x^\tau \equiv \exp_y(\tau \xi)$, such that $x^1 = x$ and $x^\tau \to y$ as $\tau \to 0$. Furthermore, instead of the fixed expansion points $y$ and $z$ one also needs rescaled expansion points $\tilde{y}^\tau = \exp_y(\tau \tilde{\xi})$ and $\tilde{z}^\tau = \exp_z(\tau \tilde{\zeta})$ for some tangent vectors $\tilde{\xi}$ and $\tilde{\zeta}$ (which may be zero, in which case we have again fixed expansion points). At order $\ell$ in perturbation theory, I then want to show inductively that
\begin{equations}[sec_ope_exist_scaling]
\tilde{\mathcal{C}}^{B,(\ell)}_{A_1 \cdots A_k}(x_1^\tau,\ldots,x_k^\tau;\tilde{y}^\tau) &= \bigo{ \tau^{[\op_B] - \sum_{i=1}^k [\op_{A_i}] - \ell [L]} } \eqend{,} \label{sec_ope_exist_scaling_coeff} \\
N^{\omega,\Delta,(\ell)}_{A_1 \cdots A_k}(x_1^\tau,\ldots,x_k^\tau;\tilde{y}^\tau) &= \bigo{ \tau^{\Delta - \sum_{i=1}^k [\op_{A_i}] - \ell [L]} } \eqend{,} \label{sec_ope_exist_scaling_operemainder} \\
N^{B,m,\Delta,(\ell)}_{A_1 \cdots A_k}(x_1^\tau,\ldots,x_m^\tau,x_{m+1},\ldots,x_k;\tilde{y}^\tau,z) &= \bigo{ \tau^{\Delta - \sum_{i=1}^m [\op_{A_i}] - \ell [L]} } \eqend{,} \label{sec_ope_exist_scaling_factremainder}
\end{equations}
where the order in $\tau$ is defined by~\eqref{sec_aqft_renorm_ordertau}, and where $[L]$ is the dimension of the interaction, defined as $[L] \equiv \max(0, [\op_I] - \operatorname{dim} M)$ where $\op_I$ is the interaction operator with the largest dimension. For (super-)renormalisable interactions, we have $[L] = 0$, while for power-counting non-renormalisable interactions $[L] > 0$. For (super-)renormalisable interactions, both the scaling of the OPE coefficients themselves and the speed at which the OPE remainder vanishes (or the factorisation holds) are thus independent of the order of perturbation theory, while for non-renormalisable interactions the OPE coefficients become more singular at each order, such that one needs to include more terms in the OPE for a given accuracy.\footnote{For a strictly super-renormalisable theory, one could even show the above bounds with $[L] = [\op_I] - \operatorname{dim} M < 0$, such that the OPE coefficients become less singular as the perturbation order increases. However, renormalisable interaction terms are generically generated by the renormalisation procedure anyway.}

To show the scalings~\eqref{sec_ope_exist_scaling}, I start with the free theory, where the OPE can be thought of as a generalised Taylor expansion (with $\mathcal{D}^B$ as the derivative operator), as will be clear from the following. Using Wick's theorem~\eqref{sec_aqft_free_hadamard_wick}, one computes for even $n$
\begin{equation}
\expect{ \normord{ \phi(x_1) \cdots \phi(x_n) }_H }_\omega = \sum_{p \in P_n^2} \prod_{\{j,k\} \in p} W(x_j,x_k) \eqend{,}
\end{equation}
where the sum is over all (unordered) pairings $p$ of $\{ 1, \ldots, n \}$, and $\expect{ \normord{ \phi(x_1) \cdots \phi(x_n) }_H }_\omega = 0$ for odd $n$. Since $W(x,y)$ is a smooth function of the points $x$ and $y$, one can expand this in a Taylor series with remainder around a point $y$, assuming all points $x_i = \exp_y(\xi_i)$ (with the tangent vectors $\xi_i$ along the geodesic from $x_i$ to $y$) lie in a common normal geodesic neighbourhood. The Taylor expansion reads
\begin{splitequation}
\expect{ \normord{ \phi(x_1) \cdots \phi(x_n) }_H }_\omega &= \sum_{k=0}^N \frac{1}{k!} \left[ \frac{\partial^k}{\partial \tau^k} \expect{ \normord{ \phi(\exp_y(\tau \xi_1)) \cdots \phi(\exp_y(\tau \xi_n)) }_H }_\omega \right]_{\tau = 0} + \bigo{\vec{\xi}^{N+1}} \\
&\quad= \sum_{k=0}^N \frac{1}{k!} \sum_{\abs{\vec{k}} = k} \frac{\abs{\vec{k}}!}{\vec{k}!} \vec{\xi}^{\otimes \vec{k}} \expect{ \normord{ \nabla^{k_1} \phi(y) \cdots \nabla^{k_n} \phi(y) }_H }_\omega + \bigo{\vec{\xi}^{N+1}} \\
&\quad= \sum_{\vec{k}\colon \abs{\vec{k}} \leq N} \frac{1}{\vec{k}!} \vec{\xi}^{\otimes \vec{k}} \expect{ \normord{ \op_\vec{k}(y) }_H }_\omega + \bigo{\vec{\xi}^{N+1}} \eqend{,} \raisetag{2.6em}
\end{splitequation}
where $\vec{\xi} = (\xi_1,\ldots,\xi_n)$, $\vec{k} = (k_1,\ldots,k_n)$ is a standard multi-index of length $n$, and where I used that derivatives of fields in composite operators are exactly defined using derivatives in Riemann normal coordinates~\eqref{sec_aqft_ops_normord_op_def}. On the other hand, we have~\eqref{sec_aqft_ops_dx_def}
\begin{equation}
\mathcal{D}^B_y \normord{ \phi(x_1) \cdots \phi(x_n) }_H = \frac{\delta_{n,\abs{B}}}{b_1! \cdots b_n!} \mathcal{P}_B \, \mathbb{S}_{\{b_i\}} \, \xi_1^{\otimes b_1} \cdots \xi_n^{\otimes b_n} \eqend{,}
\end{equation}
and thus
\begin{equation}
\mathbb{S}_{\{k_i\}} \frac{1}{\vec{k}!} \vec{\xi}^{\otimes \vec{k}} = \frac{1}{\mathcal{P}_\vec{k}} \mathcal{D}^\vec{k}_y \normord{ \phi(x_1) \cdots \phi(x_n) }_H \eqend{.}
\end{equation}
Combining this with the fact that
\begin{equation}
\frac{1}{\vec{k}!} \vec{\xi}^{\otimes \vec{k}} \expect{ \normord{ \op_\vec{k}(y) }_H }_\omega = \mathbb{S}_{\{k_i\}} \frac{1}{\vec{k}!} \vec{\xi}^{\otimes \vec{k}} \expect{ \normord{ \op_\vec{k}(y) }_H }_\omega
\end{equation}
since $\op_\vec{k}$ does not depend on the order of the $k_i$ in the multi-index $\vec{k}$, and that the sum over all multi-indices $\vec{k}$ can then be replaced by a sum over all operators $B$ by including an additional factor $\mathcal{P}_B$ that counts how many permutations of indices give the same operator, it follows that
\begin{splitequation}
\expect{ \normord{ \phi(x_1) \cdots \phi(x_n) }_H }_\omega &= \sum_{\vec{k}\colon \abs{\vec{k}} \leq N} \frac{1}{\mathcal{P}_\vec{k}} \mathcal{D}^\vec{k}_y \normord{ \phi(x_1) \cdots \phi(x_n) }_H \, \expect{ \normord{ \op_\vec{k}(y) }_H }_\omega + \bigo{\vec{\xi}^{N+1}} \\
&= \sum_{B\colon[\op_B] \leq N + n [\phi]} \mathcal{D}^B_y \normord{ \phi(x_1) \cdots \phi(x_n) }_H \, \expect{ \normord{ \op_B(y) }_H }_\omega + \bigo{\vec{\xi}^{N+1}} \eqend{.}
\end{splitequation}

The analoguous expression for composite operators is obtained by acting with derivatives (using that $\mathcal{D}^B$ commutes with derivatives), and setting some of the $x_i$ equal according to equation~\eqref{sec_aqft_ops_normord_op_def}. In this way, one obtains
\begin{splitequation}
&\expect{ \normord{ \op_{A_1}(x_1) \cdots \op_{A_n}(x_n) }_H }_\omega = \hspace{-0.5em} \sum_{B\colon[\op_B] \leq N + \sum_{i=1}^n \abs{\op_{A_i}} [\phi]} \hspace{-1em} \mathcal{D}^B_y \normord{ \op_{A_1}(x_1) \cdots \op_{A_n}(x_n) }_H \, \expect{ \normord{ \op_B(y) }_H }_\omega \\
&\hspace{14em}+ \bigo{\vec{\xi}^{N+1 - \sum_{i=1}^n \left( [\op_{A_i}] - \abs{\op_{A_i}} [\phi] \right)}} \\
&\quad= \sum_{B\colon[\op_B] \leq \Delta-1} \mathcal{D}^B_y \normord{ \op_{A_1}(x_1) \cdots \op_{A_n}(x_n) }_H \, \expect{ \normord{ \op_B(y) }_H }_\omega + \bigo{\vec{\xi}^{\Delta - \sum_{i=1}^n [\op_{A_i}]}} \eqend{,} \raisetag{2.6em}
\end{splitequation}
where the remainder has a lower order than before because derivatives have been taken. Using now the local Wick expansion~\eqref{sec_aqft_renorm_r_wick} of the renormalised products, it follows that
\begin{splitequation}
&\expect{ \mathcal{R}\left[ \op_{A_1}(x_1) \cdots \op_{A_n}(x_n) \right] }_\omega \\
&\quad= \sum_{B_1,\ldots,B_n} r\left[ \partial^{B_1} \op_{A_1}(x_1) \cdots \partial^{B_n} \op_{A_n}(x_n) \right] \expect{ \normord{ \op_{B_1}(x_1) \cdots \op_{B_n}(x_n) }_H }_\omega \\
&\quad= \sum_{B_1,\ldots,B_n} r\left[ \partial^{B_1} \op_{A_1}(x_1) \cdots \partial^{B_n} \op_{A_n}(x_n) \right] \\
&\qquad\times \left[ \sum_{C\colon[\op_C] < \Delta} \mathcal{D}^C_y \normord{ \op_{B_1}(x_1) \cdots \op_{B_n}(x_n) }_H \, \expect{ \normord{ \op_C(y) }_H }_\omega + \bigo{\vec{\xi}^{\Delta - \sum_{i=1}^n [\op_{B_i}]}} \right] \eqend{.}
\end{splitequation}
Since the sums over the $B_i$ are finite and $\mathcal{D}^C$ is linear and only acts on normal-ordered products, the sums over $C$ and the $B_i$ can be interchanged and the local Wick expansion used in reverse, such that the first term reads
\begin{splitequation}
&\sum_{C\colon [\op_C] < \Delta} \mathcal{D}^C_y \mathcal{R}\left[ \op_{A_1}(x_1) \cdots \op_{A_n}(x_n) \right] \, \expect{ \normord{ \op_C(y) }_H }_\omega \\
&\quad= \sum_{C\colon[\op_C] < \Delta} \mathcal{C}^{C,(0)}_{A_1 \cdots A_n}(x_1,\ldots,x_n;y) \, \expect{ \mathcal{R}\left[ \op_C(y) \right] }_\omega \eqend{,}
\end{splitequation}
using the definition of the free OPE coefficients~\eqref{sec_ope_coeffs_redef_recursion} and the renormalised products of a single argument~\eqref{sec_aqft_renorm_r_1arg}. On the other hand, using the scaling condition~\eqref{sec_aqft_renorm_r_scaling} for the $r$ distributions and that $[ \partial^B \op_A ] = [\op_A] - [\op_B]$, the remainder is seen to be of order $\bigo{ \vec{\xi}^{\Delta - \sum_{i=1}^n [\op_{A_i}]} }$, and the scaling~\eqref{sec_ope_exist_scaling_operemainder} for the remainder of the OPE holds in the free theory (also for a rescaled expansion point $\tilde{y}^\tau$, since the difference between $y$ and $\tilde{y}^\tau$ is of higher order in $\tau$).

Note that this result is really a statement about the smoothness of the state-dependent part $W(x,y)$ of the two-point function. In fact, the same proof applies to any linear functional that is smooth and commutes with derivatives, and is not specific to the expectation value $\expect{ \cdot }_\omega$. That is, given any linear functional $F$ such that $F\left( \normord{ \op_{A_1}(x_1) \cdots \op_{A_n}(x_n) }_H \right)$ is smooth in the positions $x_i$ and commutes with derivatives with respect to the $x_i$, we have by the above argument that
\begin{splitequation}
F\left( \mathcal{R}\left[ \op_{A_1}(x_1) \cdots \op_{A_n}(x_n) \right] \right) &= \sum_{C\colon [\op_C] < \Delta} \mathcal{D}^C_y \mathcal{R}\left[ \op_{A_1}(x_1) \cdots \op_{A_n}(x_n) \right] F\left( \mathcal{R}\left[ \op_C(y) \right] \right) \\
&\qquad+ \bigo{\vec{\xi}^{\Delta - \sum_{i=1}^n [\op_{A_i}]}} \eqend{.}
\end{splitequation}
In particular, taking $F = \mathcal{D}^B_y$ (which commutes with derivatives by definition) and $\Delta = [\op_B]$ and using that $\mathcal{D}^B_y \mathcal{R}\left[ \op_C(y) \right] = \mathcal{D}^B_y \normord{ \op_C(y) }_H = \delta^B_C$~\eqref{sec_aqft_ops_dx_delta}, it follows that
\begin{equation}
\mathcal{C}^{B,(0)}_{A_1 \cdots A_n}(x_1,\ldots,x_n;y) = \mathcal{D}^B_y \mathcal{R}\left[ \op_{A_1}(x_1) \cdots \op_{A_n}(x_n) \right] = \bigo{\vec{\xi}^{[\op_B] - \sum_{i=1}^n [\op_{A_i}]}} \eqend{,}
\end{equation}
which is the scaling~\eqref{sec_ope_exist_scaling_coeff} of the OPE coefficients. On the other hand, taking
\begin{equation}
F A = \mathcal{D}^B_z\left( A \mathcal{R}\left[ \op_{A_{m+1}}(x_{m+1}) \cdots \op_{A_n}(x_n) \right] \right) \eqend{,}
\end{equation}
one computes under the assumption that $x_i \neq x_j$ for all $1 \leq i \leq m < j \leq n$ that
\begin{splitequation}
&\mathcal{C}^{B,(0)}_{A_1 \cdots A_n}(x_1,\ldots,x_n;z) = \mathcal{D}^B_z\left( \mathcal{R}\left[ \op_{A_1}(x_1) \cdots \op_{A_m}(x_m) \right] \mathcal{R}\left[ \op_{A_{m+1}}(x_{m+1}) \cdots \op_{A_n}(x_n) \right] \right) \\
&\quad= F \mathcal{R}\left[ \op_{A_1}(x_1) \cdots \op_{A_m}(x_m) \right] \\
&\quad= \sum_{C\colon [\op_C] < \Delta} \mathcal{D}^C_y \mathcal{R}\left[ \op_{A_1}(x_1) \cdots \op_{A_m}(x_m) \right] F\left( \mathcal{R}\left[ \op_C(y) \right] \right) + \bigo{\vec{\xi}^{\Delta - \sum_{i=1}^m [\op_{A_i}]}} \\
&\quad= \sum_{C\colon [\op_C] < \Delta} \mathcal{C}^{C,(0)}_{A_1 \cdots A_m}(x_1,\ldots,x_m;y) \mathcal{D}^B_z\left( \mathcal{R}\left[ \op_C(y) \right] \mathcal{R}\left[ \op_{A_{m+1}}(x_{m+1}) \cdots \op_{A_n}(x_n) \right] \right) \\
&\qquad\quad+ \bigo{\vec{\xi}^{\Delta - \sum_{i=1}^m [\op_{A_i}]}} \\
&\quad= \sum_{C\colon [\op_C] < \Delta} \mathcal{C}^{C,(0)}_{A_1 \cdots A_m}(x_1,\ldots,x_m;y) \, \mathcal{C}^{B,(0)}_{C A_{m+1} \cdots A_n}(y,x_{m+1},\ldots,x_n;z) + \bigo{\vec{\xi}^{\Delta - \sum_{i=1}^m [\op_{A_i}]}} \eqend{,}
\end{splitequation}
using the factorisation~\eqref{sec_aqft_renorm_r_factor} of the renormalised products when the points are distinct. Note that here the tangent vectors in the remainder are only $\vec{\xi} = (\xi_1,\ldots,\xi_m)$ corresponding to the operators on which $F$ acted, and the scaling~\eqref{sec_ope_exist_scaling_factremainder} is obtained.

These results can now be extended to the interacting theory. The terms of order $\ell$ in perturbation theory are extracted using the functional Euler operator
\begin{equation}
\sum_E \int \tilde{g}^E(z) \frac{\delta}{\delta \tilde{g}^E(z)} \total z
\end{equation}
as the terms with eigenvalue $\ell$. Performing a change of integration variable $z \to z^\tau$ and $u \to u^\tau$, equation~\eqref{sec_ope_var_redef_opeder} then gives (replacing $y$ by $\tilde{y}^\tau$)
\begin{splitequation}
&\tilde{\mathcal{C}}^{B,(\ell+1)}_{A_1 \cdots A_k}(x_1^\tau,\ldots,x_k^\tau;\tilde{y}^\tau) = \sum_D \int \tilde{g}^D(z^\tau) \biggg[ - \tilde{\mathcal{C}}^{B,(\ell)}_{D A_1 \cdots A_k}(z^\tau,x_1^\tau,\ldots,x_k^\tau;\tilde{y}^\tau) \\
&\hspace{2em}+ \sum_{\ell'=0}^{\ell} \frac{\ell!}{\ell'! (\ell-\ell')!} \biggg[ \sum_{m=1}^k \sum_{E\colon [\op_E] \leq [\op_{A_m}]} \tilde{\mathcal{C}}^{E,(\ell')}_{D A_m}(z^\tau,x_m^\tau;x_m^\tau) \, \tilde{\mathcal{C}}^{B,(\ell-\ell')}_{A_1 \cdots E \cdots A_k}(x_1^\tau,\ldots,x_k^\tau;\tilde{y}^\tau) \\
&\hspace{4em}- \int \sum_{C,E\colon [\op_C] \leq [\op_E]} \tilde{\mathcal{C}}^{C,(\ell')}_{DE}(z^\tau,u^\tau;u^\tau) \tilde{g}^E(u^\tau) \frac{\delta}{\delta \tilde{g}^C(u^\tau)} \tilde{\mathcal{C}}^{B,(\ell-\ell')}_{A_1 \cdots A_k}(x_1^\tau,\ldots,x_k^\tau;\tilde{y}^\tau) \total u^\tau \\
&\hspace{4em}- \sum_{C\colon [\op_C] < [\op_B]} \tilde{\mathcal{C}}^{B,(\ell')}_{DC}(z^\tau,\tilde{y}^\tau;\tilde{y}^\tau) \, \tilde{\mathcal{C}}^{C,(\ell-\ell')}_{A_1 \cdots A_k}(x_1^\tau,\ldots,x_k^\tau;\tilde{y}^\tau) \biggg] \biggg] \total z^\tau \eqend{.} \raisetag{2.7em}
\end{splitequation}
Assume now that inductively the scaling~\eqref{sec_ope_exist_scaling_coeff} has been shown for all orders up to order $\ell$, such that it can be used for the terms on the right-hand side. Since the couplings $\tilde{g}^D(z^\tau)$ and $\tilde{g}^E(u^\tau)$ are smooth functions and do not scale at all, it follows that all but the terms involving $u^\tau$ have the scaling $\bigo{ \tau^{[\op_B] - [\op_D] - \sum_{i=1}^k [\op_{A_i}] - \ell [L]} }$. These other terms instead scale as $\bigo{ \tau^{[\op_B] - [\op_D] - [\op_E] - \sum_{i=1}^k [\op_{A_i}] - (\ell-1) [L]} }$, where the $(\ell-1) [L]$ and the fact that $[\op_C]$ does not appear result from the fact that $\delta \tilde{\mathcal{C}}^{B,(\ell-\ell')}_{A_1 \cdots A_k}(x_1^\tau,\ldots,x_k^\tau;\tilde{y}^\tau) / \delta \tilde{g}^C(u^\tau)$ is of perturbative order $\ell-\ell'-1$ because of the additional functional derivative with respect to the coupling, but has an additional insertion of $[\op_C]$ according to equation~\eqref{sec_ope_var_redef_opeder}. Finally, in RNC centered at $y$ and with $n = \operatorname{dim} M$ we have~\cite{cheegerebin1975}
\begin{equation}
\total z = \sqrt{\det g(z)} \, \total^n \exp_y(\xi) = \total^n \xi + \bigo{\xi^{n+1}}
\end{equation}
and thus $\total z^\tau = \bigo{\tau^n}$, such that with the estimate $\tau^{n - [\op_D]} \leq \tau^{-[L]}$ the required scaling\linebreak $\bigo{ \tau^{[\op_B] - \sum_{i=1}^k [\op_{A_i}] - (\ell+1) [L]} }$ for the left-hand side follows. In exactly the same way, the other scalings~\eqref{sec_ope_exist_scaling_operemainder} and~\eqref{sec_ope_exist_scaling_factremainder} follow from equations~\eqref{sec_ope_exist_operemainder_der} and~\eqref{sec_ope_exist_factremainder_der}.

\section{Examples}
\label{sec_example}

It remains to compute some examples to show how to use the recursive formula~\eqref{sec_ope_var_redef_opeder} in practice. I consider for simplicity a massless, conformally coupled scalar field on hyperbolic space, the Euclidean version of Anti-de Sitter space (EAdS). $(d+1)$-dimensional EAdS of radius $\ell$ has the metric
\begin{equation}
\total s^2 = g_{\mu\nu} \total x^\mu \total x^\nu = \frac{\ell^2}{z^2} \left[ \total z^2 + \delta_{ab} \total \vec{x}^a \total \vec{x}^b \right]
\end{equation}
in Poincaré coordinates, which (in contrast to Lorentzian AdS) cover the whole space. Hyperbolic space is maximally symmetric, and the geodesic distance $d(x,y)$ between two points $x$ and $y$ can be expressed as
\begin{splitequation}
\label{sec_example_d_u}
d(x,y) &= 2 \ell \operatorname{arcsinh}\left( \sqrt{ \frac{u(x,y)}{4} } \right) = 2 \ell \ln\left( \frac{\sqrt{u(x,y)} + \sqrt{4+u(x,y)}}{2} \right) \eqend{,} \\
u(x,y) &= 4 \sinh^2\left( \frac{d(x,y)}{2 \ell} \right) = 2 \cosh\left( \frac{d(x,y)}{\ell} \right) - 2
\end{splitequation}
in terms of the chordal distance
\begin{equation}
u(x,y) = \frac{(z_x-z_y)^2 + ( \vec{x} - \vec{y} )^2}{z_x z_y} \geq 0 \eqend{.}
\end{equation}
In a maximally symmetric space, all tensors depending on two or more points and transforming covariantly under the isometries of that space can be expressed using the geodesic distance between the given points, the tangent vectors along the geodesic connecting them and the tensor of parallel transport along the geodesic~\cite{peters1969,allenjacobson1986}. In turn, these can be expressed using the chordal distance and derivatives thereof. In particular, the tangent vector $\xi(x,y)^\mu$ at $x$ along the geodesic from $x$ to $y$ is given by a multiple of $\nabla_x^\mu u(x,y)$, which is determined by imposing the normalisation $\xi(x,y)^\mu \xi(x,y)^\nu g_{\mu\nu}(x) = d(x,y)^2$. Using
\begin{equation}
\label{sec_example_urel_1}
g_{\mu\nu}(x) \nabla_x^\mu u(x,y) \nabla_x^\nu u(x,y) = \frac{u(x,y) (4+u(x,y))}{\ell^2} = g_{\mu\nu}(y) \nabla_y^\mu u(x,y) \nabla_y^\nu u(x,y)
\end{equation}
(which is easily obtained from the above expressions), the tangent vector thus reads
\begin{splitequation}
\label{sec_example_xidef}
\xi(x,y)^\mu &= - \frac{\ell d(x,y)}{\sqrt{ u(x,y) (4+u(x,y)) }} \nabla_x^\mu u(x,y) = - 2 \ell^2 \frac{\ln\left( \frac{\sqrt{u(x,y)} + \sqrt{4+u(x,y)}}{2} \right)}{\sqrt{ u(x,y) (4+u(x,y)) }} \nabla_x^\mu u(x,y) \\
&= - \frac{\ell d(x,y)}{2 \sinh\left( \frac{d(x,y)}{\ell} \right)} \nabla_x^\mu u(x,y) = - d(x,y) \nabla_x^\mu d(x,y) \eqend{,} \raisetag{2.6em}
\end{splitequation}
which has the right flat-space limit $\xi(x,y)^\mu \to (y-x)^\mu$ for $d(x,y) \to \sqrt{(x-y)^2}$. Similarly, the tensor of parallel transport $g_{\mu\nu}(x,y)$ can be defined by $g_{\mu\nu}(x,y) \xi(y,x)^\nu = - \xi(x,y)^\mu$ and the coincidence limit $\lim_{y \to x} g_{\mu\nu}(x,y) = g_{\mu\nu}(x)$~\cite{allenjacobson1986}. From the second condition, it follows that $g_{\mu\nu}(x,y) g^{\mu\nu}(x,y) = d+1$, and one computes
\begin{equation}
g_{\mu\nu}(x,y) = - \frac{\ell^2}{2} \left[ \nabla^x_\mu \nabla^y_\nu u(x,y) - \frac{1}{4+u(x,y)} \nabla^x_\mu u(x,y) \nabla^y_\nu u(x,y) \right] \eqend{,}
\end{equation}
using also that
\begin{equation}
\label{sec_example_urel_2}
\nabla^x_\mu \nabla^x_\nu u(x,y) = \frac{2+u(x,y)}{\ell^2} g_{\mu\nu}(x) \eqend{,}
% \ell^2 \nabla^2 f(u) = u (u+4) f''(u) + (d+1) (u+2) f'(u)
% \nabla^2 f(d) = f''(d) + \frac{D}{\ell} \coth\left( \frac{d}{\ell} \right) f'(d)
\end{equation}
and that covariant derivatives with respect to different points commute. In the flat-space limit we have $g_{\mu\nu}(x,y) \to - \frac{1}{2} \partial^x_\mu \partial^y_\nu d(x,y)^2 = \eta_{\mu\nu}$ for $d(x,y) \to \sqrt{(x-y)^2}$ as required.

The free propagator of a scalar of mass $m$ satisfies the equation of motion
\begin{equation}
\label{sec_example_g_eom}
\left[ \nabla^2 + \frac{\Delta (d-\Delta)}{\ell^2} \right] G_\Delta(x,y) = - \delta(x,y) = - \frac{\delta^{d+1}(x-y)}{\sqrt{-g}} \eqend{,}
\end{equation}
where $m^2 = \ell^{-2} \Delta (\Delta-d)$. Conformal coupling is obtained for $\Delta = (d+1)/2$, and since the maximally symmetric solution for the propagator only depends on the chordal distance $u(x,y)$, outside of coincidence one obtains the equation
\begin{equation}
u (u+4) G''(u) + (d+1) ( u + 2 ) G'(u) + \frac{d^2-1}{4} G(u) = 0 \eqend{.}
\end{equation}
A one-parameter family of Hadamard states is given by $G^{(\alpha)}(u) = H(u) + W^{(\alpha)}(u)$, where the correctly normalised Hadamard parametrix (singular as $u \to 0$) is given by
\begin{equation}
H(u) = \frac{\ell^{1-d} \, \Gamma\left( \frac{d+1}{2} \right)}{2 (d-1) \pi^\frac{d+1}{2}} u^{-\frac{d-1}{2}}
\end{equation}
(the logarithmic term that appears for even dimensions is absent for conformal coupling), while the state-dependent regular term reads
\begin{equation}
W^{(\alpha)}(u) = \alpha (u+4)^{-\frac{d-1}{2}}
\end{equation}
for $\alpha \in \mathbb{R}$. The usual choice of state is given by requiring fast fall-off towards the conformal boundary as $u \to \infty$~\cite{burgessluetken1984,allenjacobson1986}, which is obtained for
\begin{equation}
\alpha = - \frac{\ell^{1-d} \, \Gamma\left( \frac{d+1}{2} \right)}{2 (d-1) \pi^\frac{d+1}{2}} \eqend{.}
\end{equation}
For concreteness, in the following I work in $d+1 = 5$ dimensions, and have thus
\begin{equation}
\label{sec_example_hw_5d}
G(u) = H(u) + W(u) \eqend{,} \qquad H(u) = \frac{\ell^{-3}}{8 \pi^2} u^{-\frac{3}{2}} \eqend{,} \qquad W(u) = - \frac{\ell^{-3}}{8 \pi^2} (u+4)^{-\frac{3}{2}} \eqend{,}
\end{equation}
consistent with the dimension of the scalar $[\phi] = \frac{3}{2}$.

\subsection{Two-point OPE coefficients in the free theory}
\label{sec_example_free2}

The free OPE coefficients $\mathcal{C}^{B,(0)}_{A_1 \cdots A_k}$ can be obtained directly from the defining equation~\eqref{sec_ope_coeffs_def_eq}
\begin{equation}
\mathcal{C}^{B,(0)}_{A_1 \cdots A_k}(x_1,\ldots,x_k;y) = \mathcal{D}^B_y \mathcal{R}\left[ \op_{A_1}(x_1) \cdots \op_{A_k}(x_k) \right] \eqend{,}
\end{equation}
and one first needs to determine the renormalised products. Again for simplicity, I choose the expansion point to be $y = x_k$, which gives short expressions for the OPE coefficients. For one argument, the renormalised product is just the normal-ordered expression~\eqref{sec_aqft_renorm_r_1arg}
\begin{equation}
\mathcal{R}\left[ \op_A(x) \right] = \normord{ \op_A(x) }_H \eqend{.}
\end{equation}
For more arguments, one has to proceed inductively both in the number of arguments and the dimension of the composite operators. For non-coinciding points $x \neq y$, by the factorisation condition~\eqref{sec_aqft_renorm_r_factor} one obtains
\begin{equation}
\mathcal{R}\left[ \phi(y) \phi(x) \right] = \mathcal{R}\left[ \phi(y) \right] \mathcal{R}\left[ \phi(x) \right] = \normord{ \phi(y) }_H \normord{ \phi(x) }_H = \phi(y) \phi(x) \eqend{,}
\end{equation}
and since $[\phi] + [\phi] = 3 < 5 = \dim M$, the renormalisation (extension to coinciding points $x = y$) is unique. Using the formula~\eqref{sec_aqft_free_hadamard_product} for the product of two normal-ordered expressions, one obtains
\begin{equation}
\phi(y) \phi(x) = \normord{ \phi(y) \phi(x) }_H + H(y,x) \1
\end{equation}
and the local Wick expansion
\begin{equation}
\mathcal{R}\left[ \phi(y) \phi(x) \right] = \sum_{A,B} r\left[ \partial^A \phi(y) \partial^B \phi(x) \right] \normord{ \op_A(y) \op_B(x) }_H
\end{equation}
with $r\left[ \op_A(x) \1 \right] = r\left[ \op_A(x) \right]$ and
\begin{equation}
\label{sec_example_free2_r_1}
r\left[ \phi(y) \phi(x) \right] = H(y,x) \eqend{,} \quad r\left[ \phi(x) \right] = 0 \eqend{,} \quad r\left[ \1 \right] = 1 \eqend{.}
\end{equation}
Admittedly, this is quite trivial, but serves to illustrate all steps (which I will abbreviate for the second, more complicated example).

Using the definition of $\mathcal{D}^B_x$~\eqref{sec_aqft_ops_dx_def}, I then compute
\begin{equation}
\mathcal{D}^\1_x \1 = 1 \eqend{,} \qquad \mathcal{D}^{\phi \nabla^k \phi}_x \normord{ \phi(y) \phi(x) }_H = \frac{1}{k!} \xi(x,y)^{\otimes k} \eqend{,}
\end{equation}
with the tangent vector $\xi(x,y)^\mu$~\eqref{sec_example_xidef}, and all other $\mathcal{D}^B_x$ acting on $\1$ or $\normord{ \phi(y) \phi(x) }_H$ give zero. It follows immediately that
\begin{equation}
\label{sec_example_free2_phi_phi_coeffs}
\mathcal{C}^{\1,(0)}_{\phi\phi}(y,x;x) = H(y,x) \eqend{,} \qquad \mathcal{C}^{\phi \nabla^{\mu_1 \cdots \mu_k} \phi,(0)}_{\phi\phi}(y,x;x) = \frac{1}{k!} \xi(x,y)^{\mu_1} \cdots \xi(x,y)^{\mu_k} \eqend{,}
\end{equation}
and all other $\mathcal{C}^B_{\phi\phi}(y,x;x)$ vanish, where I recall the notation $\nabla^{\mu_1 \cdots \mu_k} \equiv \nabla^{(\mu_1} \cdots \nabla^{\mu_k)}$. Instead of explicitly using $\mathcal{D}^B_x$, one can also perform a Taylor expansion of the Hadamard normal-ordered product to obtain
\begin{equation}
\phi(y) \phi(x) = H(y,x) \1 + \sum_{k=0}^\infty \frac{1}{k!} \xi^{\otimes k} \normord{ \nabla^k \phi(x) \phi(x) }_H \eqend{,}
\end{equation}
and comparing with the OPE
\begin{equation}
\phi(y) \phi(x) = \mathcal{R}\left[ \phi(y) \phi(x) \right] = \sum_B \mathcal{C}^B_{\phi\phi}(y,x;x) \mathcal{R}\left[ \op_B(x) \right] = \sum_B \mathcal{C}^B_{\phi\phi}(y,x;x) \, \normord{ \op_B(x) }_H \eqend{,}
\end{equation}
the same OPE coefficients are obtained.

Furthermore I need the (almost trivial) OPE between $\phi^\ell$ for $\ell \geq 1$ and $\1$, given by
\begin{equation}
\label{sec_example_free2_phi1_1phi_ope}
\normord{ \phi^\ell(y) }_H \1(x) = \normord{ \phi^\ell(y) }_H = \sum_{k=0}^\infty \frac{1}{k!} \xi^{\otimes k} \normord{ \nabla^k \phi^\ell(x) }_H \eqend{,} \qquad \1(y) \normord{ \phi^\ell(x) }_H = \normord{ \phi^\ell(x) }_H \eqend{.}
\end{equation}
By the generalised Leibniz rule, we have
\begin{equation}
\nabla^k \phi^\ell = \sum_{i_1 + \cdots + i_\ell = k} \frac{k!}{i_1! \cdots i_\ell!} \nabla^{i_1} \phi \cdots \nabla^{i_\ell} \phi = \sum_{i_1 + \cdots + i_\ell = k} \frac{k!}{i_1! \cdots i_\ell!} \op_{(i_1,\ldots,i_\ell)} \eqend{,}
\end{equation}
where the explicit symmetrisation of covariant derivatives is unnecessary because of the factor $\xi^{\otimes k}$ in~\eqref{sec_example_free2_phi1_1phi_ope} which is already symmetric. By definition, the factor $\mathcal{P}_{(i_1,\ldots,i_\ell)}$~\eqref{sec_aqft_ops_calp_def} counts how many permutations of the $i_j$ give the same composite operator, such that
\begin{splitequation}
\normord{ \phi^\ell(y) }_H \1(x) &= \sum_{k=0}^\infty \sum_{i_1 + \cdots + i_\ell = k} \frac{1}{i_1! \cdots i_\ell!} \xi^{\otimes (i_1 + \cdots + i_\ell)} \normord{ \op_{(i_1,\ldots,i_\ell)}(x) }_H \\
&= \sum_{B\colon \op_B = \op_{(i_1,\ldots,i_\ell)}} \frac{\mathcal{P}_{(i_1,\ldots,i_\ell)}}{i_1! \cdots i_\ell!} \xi^{\otimes (i_1 + \cdots + i_\ell)} \normord{ \op_B(x) }_H
\end{splitequation}
and the non-vanishing OPE coefficients are given by
\begin{equation}
\label{sec_example_free2_1_phi_coeffs}
\mathcal{C}^{\nabla^{i_1} \phi \cdots \nabla^{i_\ell} \phi,(0)}_{\phi^\ell \1}(y,x;x) = \frac{\mathcal{P}_{(i_1,\ldots,i_\ell)}}{i_1! \cdots i_\ell!} \xi(x,y)^{\otimes (i_1 + \cdots + i_\ell)} \eqend{,} \qquad \mathcal{C}^{\phi^\ell,(0)}_{\1 \phi^\ell}(y,x;x) = 1 \eqend{.}
\end{equation}

I furthermore need the coefficients of the OPE of two composite operators which include each more than one basic field $\phi$. In this case, the renormalised products must be non-trivially extended to coinciding points, which I recall corresponds to renormalisation. Concretely, I look at the OPE of $\phi^4$ with $\phi$, $\phi^2$ and $\phi \nabla^\rho \phi$, and thus first determine the corresponding renormalised products. For distinct points $x \neq y$, we use the factorisation condition for the renormalised products and the formula~\eqref{sec_aqft_free_hadamard_product} for the product of two normal-ordered expressions, resulting in
\begin{equations}[sec_example_free2_phi4renorm]
\mathcal{R}\left[ \phi^4(y) \otimes \nabla^k \phi(x) \right] &= \normord{ \phi^4(y) }_H \nabla^k \phi(x) = \normord{ \phi^4(y) \nabla^k \phi(x) }_H + 4 \nabla_x^k H(y,x) \normord{ \phi^3(y) }_H \eqend{,} \\
\begin{split}
\mathcal{R}\left[ \phi^4(y) \otimes ( \phi \nabla^k \phi )(x) \right] &= \normord{ \phi^4(y) }_H \normord{ \phi(x) \nabla^k \phi(x) }_H \\
&= \normord{ \phi^4(y) \phi(x) \nabla^k \phi(x) }_H + 4 H(y,x) \normord{ \phi^3(y) \nabla^k \phi(x) }_H \\
&\quad+ 4 \nabla_x^k H(y,x) \normord{ \phi^3(y) \phi(x) }_H + 12 H(y,x) \nabla_x^k H(y,x) \normord{ \phi^2(y) }_H \eqend{.}
\end{split}
\end{equations}
Comparing with the local Wick expansion of the renormalised products
\begin{equation}
\label{sec_example_free2_wick_ab}
\mathcal{R}\left[ \op_A(y) \otimes \op_B(x) \right] = \sum_{C,D} r\left[ \partial^C \op_A(y) \otimes \partial^D \op_B(x) \right] \normord{ \op_C(y) \op_D(x) }_H \eqend{,}
\end{equation}
and imposing that the $r$ commute with covariant derivatives, we obtain in addition to~\eqref{sec_example_free2_r_1}
\begin{equations}[sec_example_free2_r_2]
r\left[ \phi^\ell(x) \right] &= 0 \quad\text{ for } \ell = 2,3,4 \eqend{,} \\
r\left[ \phi^\ell(y) \otimes \phi(x) \right] &= 0 \quad\text{ for } \ell = 2,3,4 \eqend{,} \\
r\left[ \phi^\ell(y) \otimes ( \phi \nabla^k \phi )(x) \right] &= 0 \quad\text{ for } \ell = 1,3,4 \eqend{,} \\
r\left[ \phi^2(y) \otimes ( \phi \nabla^k \phi )(x) \right] &= 2 H(y,x) \nabla_x^k H(y,x) \eqend{.}
\end{equations}
\iffalse

+ \sum_{D} 4 r\left[ \phi^3(y) \otimes \partial^D ( \phi \nabla^k \phi )(x) \right] \normord{ \phi(y) \op_D(x) }_H
+ \sum_{D} 6 r\left[ \phi^2(y) \otimes \partial^D ( \phi \nabla^k \phi )(x) \right] \normord{ \phi^2(y) \op_D(x) }_H
+ \sum_{D} 4 r\left[ \phi(y) \otimes \partial^D ( \phi \nabla^k \phi )(x) \right] \normord{ \phi^3(y) \op_D(x) }_H
+ \sum_{D} r\left[ \partial^D ( \phi \nabla^k \phi )(x) \right] \normord{ \phi^4(y) \op_D(x) }_H

=
+ \normord{ \phi^4(y) \phi(x) \nabla^k \phi(x) }_H
+ 4 H(y,x) \normord{ \phi^3(y) \nabla^k \phi(x) }_H
+ 4 \nabla_x^k H(y,x) \normord{ \phi^3(y) \phi(x) }_H
+ 12 H(y,x) \nabla_x^k H(y,x) \normord{ \phi^2(y) }_H
\fi
The extension of the second term to coinciding points $x = y$ is unique whenever the degree of divergence is not integer, which is the case for even $\ell$. For $\ell = 3$ the degree of divergence of $r\left[ \phi^3(y) \otimes \phi(x) \right]$ is given by $4 [\phi] = 6 > 5$, which is integer and greater than the space dimension. Therefore, a renormalisation ambiguity exists, which is a local term ($\delta$ and its derivatives), multiplied by the metric or curvature tensors of the correct dimension. However, the only possible term in this case would be $\nabla^\mu \delta(x,y)$, which does not have the correct tensor structure. Also for the third and last term, the degree of divergence is not integer and there is a unique (vanishing) extension to coinciding points. For the fourth term, we first have to define some extension, since the degree of divergence is $4[\phi]+k = 6+k > 5$. Using the explicit form
\begin{equation}
H(x,y) = \frac{\ell^{-3}}{8 \pi^2} u(x,y)^{-\frac{3}{2}} \eqend{,}
\end{equation}
this can be easily done by computing the naive product for $x \neq y$, and then extracting derivatives until the remainder has degree of divergence $< 5$. Using the relations~\eqref{sec_example_urel_1}, \eqref{sec_example_urel_2} and
\begin{equation}
\label{sec_example_urel_3}
\ell^2 \nabla^2 f(u) = u (u+4) f''(u) + 5 (u+2) f'(u) \eqend{,}
% \nabla_\mu \nabla_\nu f(u) = f''(u) \nabla_\mu u \nabla_\nu u + f'(u) \frac{u+2}{\ell^2} g_{\mu\nu}
\end{equation}
we obtain for $x \neq y$
\begin{equations}
H(y,x)^2 &= \frac{\ell^{-6}}{64 \pi^4} u(x,y)^{-3} = \frac{\ell^{-6}}{(4\pi)^4} \left( \ell^2 \nabla_x^2 + 4 \right) u(x,y)^{-2} \eqend{,} \\
H(y,x) \nabla_x^\mu H(y,x) &= - \frac{3 \ell^{-6}}{128 \pi^4} u(x,y)^{-4} \nabla_x^\mu u(x,y) = \frac{1}{2} \nabla_x^\mu H(y,x)^2 \eqend{,} \\
\begin{split}
H(y,x) \nabla_x^{\mu\nu} H(y,x) &= \frac{15 \ell^{-6}}{256 \pi^4} u(x,y)^{-5} \nabla_x^\mu u(x,y) \nabla_x^\nu u(x,y) - \frac{3 \ell^{-6}}{128 \pi^4} u(x,y)^{-4} \frac{u(x,y)+2}{\ell^2} g^{\mu\nu} \\
&= \frac{\ell^{-2}}{16} \left[ 5 \ell^2 \nabla_x^{\mu\nu} - g^{\mu\nu} \left( \ell^2 \nabla_x^2 + 12 \right) \right] H(y,x)^2 \eqend{,}
\end{split} \\
\begin{split}
H(y,x) \nabla_x^{\mu\nu\rho} H(y,x) &= - \frac{105 \ell^{-6}}{512 \pi^4} u(x,y)^{-6} \nabla_x^\mu u(x,y) \nabla_x^\nu u(x,y) \nabla_x^\rho u(x,y) \\
&\quad+ \frac{3 \ell^{-6}}{256 \pi^4} u(x,y)^{-5} \frac{13 u(x,y) + 30}{\ell^2} g^{(\mu\nu} \nabla_x^{\rho)} u(x,y) \\
&= \frac{\ell^{-2}}{32} \left[ 7 \ell^2 \nabla_x^{\mu\nu\rho} - 3 g^{(\mu\nu} \nabla_x^{\rho)} \left( \ell^2 \nabla_x^2 + 12 \right) \right] H(y,x)^2 \eqend{.}
\end{split}
\end{equations}
Since $u(x,y)^{-2}$ has degree of divergence $4 < 5$, it has a unique extension to coinciding points $x = y$ as a well-defined distribution. Since derivatives acting on a distribution are also always defined, we have found an extension (renormalisation) of all the above expressions. We now have to determine renormalisation ambiguities. For $H(y,x)^2$, which has degree of divergence $6$, the only possible ambiguity is given by $\nabla^\mu \delta(x,y)$, which however has not the correct tensor structure, such that the renormalisation of $H(y,x)^2$ is unique. The same happens for all the other terms, and we thus have the unique renormalisations
\begin{equations}[sec_example_free2_hdh_renorm]
\left[ H(y,x)^2 \right]^\text{ren} &= \frac{\ell^{-6}}{(4\pi)^4} \left( \ell^2 \nabla_x^2 + 4 \right) u(x,y)^{-2} \eqend{,} \\
\left[ H(y,x) \nabla_x^\mu H(y,x) \right]^\text{ren} &= \frac{1}{2} \nabla_x^\mu \left[ H(y,x)^2 \right]^\text{ren} \eqend{,} \\
\left[ H(y,x) \nabla_x^{\mu\nu} H(y,x) \right]^\text{ren} &= \frac{\ell^{-2}}{16} \left[ 5 \ell^2 \nabla_x^{\mu\nu} - g^{\mu\nu} \left( \ell^2 \nabla_x^2 + 12 \right) \right] \left[ H(y,x)^2 \right]^\text{ren} \eqend{,} \\
\left[ H(y,x) \nabla_x^{\mu\nu\rho} H(y,x) \right]^\text{ren} &= \frac{\ell^{-2}}{32} \left[ 7 \ell^2 \nabla_x^{\mu\nu\rho} - 3 g^{(\mu\nu} \nabla_x^{\rho)} \left( \ell^2 \nabla_x^2 + 12 \right) \right] \left[ H(y,x)^2 \right]^\text{ren} \eqend{.}
\end{equations}
At this point, it was important that we consider locally covariant renormalised products, where the ambiguities are strongly restricted to only contain polynomials of curvature tensors and their covariant derivatives. Otherwise, a possible ambiguity term for $H(y,x)^2$ would be $\ell^{-1} \delta(x,y)$, but this is non-polynomial in the Ricci scalar. Expanding the normal-ordered products in~\eqref{sec_example_free2_phi4renorm} around $y = x$, we can then read off the needed OPE coefficients
\begin{equations}[sec_example_free2_phi4_phi2_coeffs]
\mathcal{C}^{\phi^2,(0)}_{\phi^4 \phi^2}(y,x;x) &= 12 \left[ H(y,x)^2 \right]^\text{ren} \eqend{,} \\
\mathcal{C}^{\phi \nabla^\mu \phi,(0)}_{\phi^4 \phi^2}(y,x;x) &= 24 \left[ H(y,x)^2 \right]^\text{ren} \xi(x,y)^\mu \eqend{,} \\
\mathcal{C}^{\nabla^\mu \phi \nabla^\nu \phi,(0)}_{\phi^4 \phi^2}(y,x;x) &= 12 \left[ H(y,x)^2 \right]^\text{ren} \xi(x,y)^\mu \xi(x,y)^\nu \eqend{,} \\
\mathcal{C}^{\phi \nabla^{\mu\nu} \phi,(0)}_{\phi^4 \phi^2}(y,x;x) &= 12 \left[ H(y,x)^2 \right]^\text{ren} \xi(x,y)^\mu \xi(x,y)^\nu \eqend{,} \\
% \mathcal{C}^{\phi^2,(0)}_{\phi^4 ( \phi \nabla^\rho \phi)}(y,x;x) &= 12 \left[ H(y,x) \nabla_x^\rho H(y,x) \right]^\text{ren} \eqend{,} \\
% \mathcal{C}^{\phi \nabla^\mu \phi,(0)}_{\phi^4 ( \phi \nabla^\rho \phi)}(y,x;x) &= 24 \left[ H(y,x) \nabla_x^\rho H(y,x) \right]^\text{ren} \xi(x,y)^\mu \eqend{,} \\
\mathcal{C}^{\nabla^\mu \phi \nabla^\nu \phi,(0)}_{\phi^4 (\phi \nabla^\rho \phi)}(y,x;x) &= \mathcal{C}^{\phi \nabla^{\mu\nu} \phi,(0)}_{\phi^4 (\phi \nabla^\rho \phi)}(y,x;x) = 12 \left[ H(y,x) \nabla_x^\rho H(y,x) \right]^\text{ren} \xi(x,y)^\mu \xi(x,y)^\nu \eqend{.}
\end{equations}

\subsection{Three-point OPE coefficients in the free theory}
\label{sec_example_free3}

To determine the first-order corrections, we also need the three-point OPE coefficients $\mathcal{C}^{B,(0)}_{\phi^4 \phi \phi}$ and $\mathcal{C}^{B,(0)}_{\phi^4 \phi^2 \phi^2}$. Again, we first construct the renormalised products, and then perform a local Wick expansion and compare with the OPE in the free theory. Thus, for $y \neq x \neq z$ we have factorisation of the renormalised products and compute
\begin{splitequation}
\mathcal{R}\left[ \phi^4(z) \otimes \phi(y) \otimes \phi(x) \right] &= \left[ \normord{ \phi^4(z) \phi(y) }_H + 4 H(z,y) \normord{ \phi^3(z) }_H \right] \phi(x) \\
&= \normord{ \phi^4(z) \phi(y) \phi(x) }_H + H(y,x) \normord{ \phi^4(z) }_H + 4 H(z,x) \normord{ \phi^3(z) \phi(y) }_H \\
&\quad+ 4 H(z,y) \normord{ \phi^3(z) \phi(x) }_H + 12 H(z,y) H(z,x) \normord{ \phi^2(z) }_H \eqend{,} \raisetag{1.4em}
\end{splitequation}
where we used the result~\eqref{sec_example_free2_phi4renorm} for the renormalised product of two composite operators. Using the local Wick expansion, we obtain
\begin{equations}[sec_example_free3_r_1]
r\left[ \phi^4(z) \otimes \phi(y) \otimes \phi(x) \right] &= 0 \eqend{,} \\
r\left[ \phi^2(z) \otimes \phi(y) \otimes \phi(x) \right] &= 2 H(z,y) H(z,x)
\end{equations}
in addition to the results~\eqref{sec_example_free2_r_1} and~\eqref{sec_example_free2_r_2}. Since the degree of divergence of the product $H(z,y) H(z,x)$ is $6 < 2\cdot5$, it is already well-defined as a distribution in 3 points. Expanding the normal-ordered products around $x$, we thus obtain directly the OPE coefficients
\begin{equations}[sec_example_free3_phi4phiphi_coeffs]
\mathcal{C}^{\phi^2,(0)}_{\phi^4 \phi \phi}(z,y,x;x) &= 12 H(z,y) H(z,x) \eqend{,} \\
\mathcal{C}^{\phi \nabla^\mu \phi,(0)}_{\phi^4 \phi \phi}(z,y,x;x) &= 24 H(z,y) H(z,x) \xi(x,z)^\mu \eqend{,} \\
\mathcal{C}^{\phi \nabla^{\mu\nu} \phi,(0)}_{\phi^4 \phi \phi}(z,y,x;x) &= \mathcal{C}^{\nabla^\mu \phi \nabla^\nu \phi,(0)}_{\phi^4 \phi \phi}(z,y,x;x) = 12 H(z,y) H(z,x) \xi(x,z)^\mu \xi(x,z)^\nu \eqend{,}
\end{equations}
with all other coefficients $\mathcal{C}^{B,(0)}_{\phi^4 \phi \phi}$ with $[\op_B] < 6$ vanishing.

\subsection{The OPE at first order}
\label{sec_example_pert}

I now would like to compute the OPE of $\phi(x) \phi(y)$ up to terms vanishing as $d(x,y)^2$, where $d(x,y)$ is the geodesic distance between $x$ and $y$, to first order in the interaction $g \phi^4$. Since the theory is massive, one would like to take the adiabatic limit (of constant coupling $g$), but we will see that this is actually not possible. Nevertheless, as long as $x$ and $y$ are close together and within a region where $g$ is constant, we will see that the OPE coefficients do not depend on the concrete form of the cutoff.

In $5$ dimensions, the field $\phi$ has dimension $\frac{3}{2}$ and the interaction has therefore dimension $[L] = 4 [\phi] - 5 = 1$. Using the result~\eqref{sec_ope_exist_scaling_operemainder} for the scaling behaviour of the remainder, one needs to take into account the terms up to $\Delta = 2 + 2 [\phi] + [L] = 6$. To shorten notation, let me denote in this section the interacting composite fields by capital letters, such that $\mathcal{R}\left[ \phi(x) \phi(y) \exp\left( - L \right) \right] = \Phi(x) \Phi(y)$ etc. Then one has
\begin{splitequation}
\label{sec_example_pert_ope_phiphi}
\Phi(y) \Phi(x) &= \sum_{B\colon [\op_B] < 6} \mathcal{C}^B_{\phi \phi}(y,x;x) \, \op_B(x) + \bigo{ d(y,x)^2 } + \bigo{ g^2 } \\
&= \mathcal{C}^\1_{\phi \phi}(y,x;x) \, \1 + \mathcal{C}^{\phi^2}_{\phi \phi}(y,x;x) \, \Phi^2(x) + \mathcal{C}^{\phi \nabla^\mu \phi}_{\phi \phi}(y,x;x) \, \left( \Phi \nabla^\mu \Phi \right)(x) \\
&\quad+ \mathcal{C}^{\phi \nabla^{\mu\nu} \phi}_{\phi \phi}(y,x;x) \, \left( \Phi \nabla^{\mu\nu} \Phi \right)(x) + \mathcal{C}^{\nabla^\mu \phi \nabla^\nu \phi}_{\phi \phi}(y,x;x) \, \left( \nabla^\mu \Phi \nabla^\nu \Phi \right)(x) \\
&\quad+ \bigo{ d(y,x)^2 } + \bigo{ g^2 }
\end{splitequation}
in any Hadamard state $\omega$, where only those OPE coefficients are displayed that do not vanish because of the $\mathbb{Z}_2$ symmetry $\phi \to -\phi$ (of both the free theory and the interaction).

Integrating the recursive formula~\eqref{sec_ope_var_redef_opeder} over $z$ with a cutoff function $f(z)$, one obtains at first order
\begin{splitequation}
\label{sec_example_opeder_g_AB}
\mathcal{C}^{C,(1)}_{A B}(y,x;x) &= \int f(z) \Bigg[ - \mathcal{C}^{C,(0)}_{\phi^4 A B}(z,y,x;x) + \sum_{E\colon [\op_E] \leq [\op_A]} \mathcal{C}^{E,(0)}_{\phi^4 A}(z,y;y) \, \mathcal{C}^{C,(0)}_{E B}(y,x;x) \\
&\hspace{4em}+ \sum_{E\colon [\op_E] \leq [\op_B]} \mathcal{C}^{E,(0)}_{\phi^4 B}(z,x;x) \, \mathcal{C}^{C,(0)}_{A E}(y,x;x) \\
&\hspace{4em}+ \sum_{E\colon [\op_E] < [\op_C]} \mathcal{C}^{C,(0)}_{\phi^4 E}(z,x;x) \, \mathcal{C}^{E,(0)}_{A B}(y,x;x) \Bigg] \total z \eqend{,}
\end{splitequation}
where the order in $g$ is denoted by the index in parentheses. For the OPE of $\phi$ with itself, this reduces to
\begin{equation}
\mathcal{C}^{C,(1)}_{\phi \phi}(y,x;x) = \int f(z) \Bigg[ - \mathcal{C}^{C,(0)}_{\phi^4 \phi \phi}(z,y,x;x) + \sum_{E\colon [\op_E] < [\op_C]} \mathcal{C}^{C,(0)}_{\phi^4 E}(z,x;x) \, \mathcal{C}^{E,(0)}_{\phi \phi}(y,x;x) \Bigg] \total z \eqend{,}
\end{equation}
where all other contributions vanished either because $\mathcal{C}^{\phi,(0)}_{\phi^4 \phi} = 0$ or by the $\mathbb{Z}_2$ symmetry. We see that for these coefficients, only IR subtractions need to be made, and no UV subtractions. For the coefficients in the OPE~\eqref{sec_example_pert_ope_phiphi}, one then obtains concretely
\begin{equations}
\mathcal{C}^{\1,(1)}_{\phi \phi}(y,x;x) &= - \int f(z) \, \mathcal{C}^{\1,(0)}_{\phi^4 \phi \phi}(z,y,x;x) \total z \eqend{,} \\
\mathcal{C}^{\phi^2,(1)}_{\phi \phi}(y,x;x) &= - \int f(z) \, \mathcal{C}^{\phi^2,(0)}_{\phi^4 \phi \phi}(z,y,x;x) \total z \eqend{,} \\
\mathcal{C}^{\phi \nabla^\mu \phi,(1)}_{\phi \phi}(y,x;x) &= \int f(z) \Bigg[ - \mathcal{C}^{\phi \nabla^\mu \phi,(0)}_{\phi^4 \phi \phi}(z,y,x;x) + \mathcal{C}^{\phi \nabla^\mu \phi,(0)}_{\phi^4 \phi^2}(z,x;x) \, \mathcal{C}^{\phi^2,(0)}_{\phi \phi}(y,x;x) \Bigg] \total z \eqend{,} \\
\begin{split}
\mathcal{C}^{\phi \nabla^{\mu\nu} \phi,(1)}_{\phi \phi}(y,x;x) &= \int f(z) \Bigg[ - \mathcal{C}^{\phi \nabla^{\mu\nu} \phi,(0)}_{\phi^4 \phi \phi}(z,y,x;x) + \mathcal{C}^{\phi \nabla^{\mu\nu} \phi,(0)}_{\phi^4 \phi^2}(z,x;x) \, \mathcal{C}^{\phi^2,(0)}_{\phi \phi}(y,x;x) \\
&\hspace{6em}+ \mathcal{C}^{\phi \nabla^{\mu\nu} \phi,(0)}_{\phi^4 (\phi \nabla^\rho \phi)}(z,x;x) \, \mathcal{C}^{\phi \nabla^\rho \phi,(0)}_{\phi \phi}(y,x;x) \Bigg] \total z \eqend{,} \raisetag{2.4em}
\end{split} \\
\begin{split}
\mathcal{C}^{\nabla^\mu \phi \nabla^\nu \phi,(1)}_{\phi \phi}(y,x;x) &= \int f(z) \Bigg[ - \mathcal{C}^{\nabla^\mu \phi \nabla^\nu \phi,(0)}_{\phi^4 \phi \phi}(z,y,x;x) + \mathcal{C}^{\nabla^\mu \phi \nabla^\nu \phi,(0)}_{\phi^4 \phi^2}(z,x;x) \, \mathcal{C}^{\phi^2,(0)}_{\phi \phi}(y,x;x) \\
&\hspace{6em}+ \mathcal{C}^{\nabla^\mu \phi \nabla^\nu \phi,(0)}_{\phi^4 (\phi \nabla^\rho \phi)}(z,x;x) \, \mathcal{C}^{\phi \nabla^\rho \phi,(0)}_{\phi \phi}(y,x;x) \Bigg] \total z \eqend{,} \raisetag{2.4em}
\end{split}
\end{equations}
where I again used that many of the two-point OPE coefficients vanish in the free theory. Since the coefficient $\mathcal{C}^{\1,(0)}_{\phi^4 \phi \phi}$ vanishes~\eqref{sec_example_free3_phi4phiphi_coeffs}, we also have $\mathcal{C}^{\1,(1)}_{\phi \phi}(y,x;x) = 0$. This is in fact not surprising, since $\mathcal{C}^\1_{\phi \phi}$ is the singular part of the two-point function of $\phi$, whose leading-order (one-loop) correction is of order $g^2$.

For the first non-vanishing correction $\mathcal{C}^{\phi^2,(1)}_{\phi \phi}$ one obtains using~\eqref{sec_example_free3_phi4phiphi_coeffs} and~\eqref{sec_example_hw_5d}
\begin{splitequation}
\label{sec_example_pert_coeff_phi2_phiphi}
&\mathcal{C}^{\phi^2,(1)}_{\phi \phi}(y,x;x) = - 12 \int f(z) \, H(z,y) H(z,x) \total z = - \frac{3 \ell^{-6}}{(2\pi)^4} \int f(z) \, u(z,y)^{-\frac{3}{2}} u(z,x)^{-\frac{3}{2}} \total z \\
&\quad= - \frac{3 \ell^{-1}}{(2\pi)^4} \int \int_0^\infty f(\zeta,\vec{z}) \, \left[ \frac{z_x \zeta}{(z_x-\zeta)^2 + ( \vec{x} - \vec{z} )^2} \frac{z_y \zeta}{(z_y-\zeta)^2 + ( \vec{y} - \vec{z} )^2} \right]^\frac{3}{2} \zeta^{-5} \total \zeta \total^4 \vec{z} \eqend{.}
\end{splitequation}
In principle, one would like to take the adiabatic limit $f \to 1$, but this would result in an IR divergence close to the boundary $\zeta \to 0$ since the integrand diverges there $\sim \zeta^{-2}$. One possibility would be to consider normal ordering not with respect to the Hadamard parametrix, but the full propagator; call the corresponding OPE coefficient $\hat{\mathcal{C}}^{\phi^2,(1)}_{\phi \phi}$. For $f = 1$, the integral is then clearly EAdS-invariant and therefore only depends on the chordal distance $u(x,y)$, such that one can set $\vec{x} = \vec{y} = \vec{0}$ to simplify the actual evaluation. With the propagator~\eqref{sec_example_hw_5d}, I compute
\begin{splitequation}
\label{sec_example_pert_coeffhat_phi2_phiphi}
\hat{\mathcal{C}}^{\phi^2,(1)}_{\phi \phi} &= - 12 \int G(z,y) G(z,x) \total z \\
&= - \frac{3 \ell^{-1}}{(2\pi)^4} \int \int_0^\infty \left[ \left( \frac{z_x \zeta}{(z_x-\zeta)^2 + \vec{z}^2} \right)^\frac{3}{2} - \left( \frac{z_x \zeta}{(z_x+\zeta)^2 + \vec{z}^2} \right)^\frac{3}{2} \right] \\
&\hspace{8em}\times \left[ \left( \frac{z_y \zeta}{(z_y-\zeta)^2 + \vec{z}^2} \right)^\frac{3}{2} - \left( \frac{z_y \zeta}{(z_y+\zeta)^2 + \vec{z}^2} \right)^\frac{3}{2} \right] \zeta^{-5} \total \zeta \total^4 \vec{z} \\
&= - \frac{3 \ell^{-1}}{8 \pi^2} ( z_x z_y )^\frac{3}{2} \int_0^\infty \bigg[ \frac{1}{( z_x + z_y + 2 \zeta )^2} - \frac{1}{( z_x + \zeta + \abs{z_y-\zeta} )^2} \\
&\hspace{8em}- \frac{1}{( z_y + \zeta + \abs{z_x-\zeta} )^2} + \frac{1}{( \abs{z_x-\zeta} + \abs{z_y-\zeta} )^2} \bigg] \zeta^{-2} \total \zeta
\end{splitequation}
using spherical coordinates $\total^4 \vec{z} = 2 \pi^2 \abs{\vec{z}}^3 \total \abs{\vec{z}}$. The integral over $\zeta$ is done by splitting the integration range into varions regions depending on whether $\zeta$ is larger or smaller than $z_x$ and $z_y$, and which of $z_x$ and $z_y$ is larger. This results in
\begin{equation}
\hat{\mathcal{C}}^{\phi^2,(1)}_{\phi \phi} = \frac{3 \ell^{-1}}{2 \pi^2} ( z_+ z_- )^\frac{3}{2} \left[ \left( \frac{1}{(z_+ + z_-)^3} - \frac{1}{(z_+ - z_-)^3} \right) \ln\left( \frac{z_+}{z_-} \right) + \frac{4 z_-}{\left( z_+^2 - z_-^2 \right)^2} \right]
\end{equation}
with $z_+ = \max(z_x,z_y)$ and $z_- = \min(z_x,z_y)$. Expressing those in terms of $u = u(x,y)$ according to
\begin{equation}
z_+ = \frac{u+2 + \sqrt{u(u+4)}}{2} z_-
\end{equation}
(recall that $\vec{x} = \vec{y} = 0$), one finally obtains
\begin{splitequation}
\label{sec_example_pert_coeff_phi2_phiphi_ing}
\hat{\mathcal{C}}^{\phi^2,(1)}_{\phi \phi} &= \frac{3 \ell^{-1}}{\pi^2} \left[ \frac{4}{u(u+4) \left( \sqrt{u} + \sqrt{u+4} \right)} + \left[ (u+4)^{-\frac{3}{2}} - u^{-\frac{3}{2}} \right] \ln\left( \frac{\sqrt{u} + \sqrt{u+4}}{2} \right) \right] \\
&= - \frac{3 \ell^{-1}}{4 \pi^2} \left[ u(x,y)^{-\frac{1}{2}} + \frac{1}{6} - \frac{1}{2} \sqrt{u(x,y)} \right] + \bigo{d(x,y)^2} \eqend{,} \raisetag{2em}
\end{splitequation}
where it was used that $u(x,y) \sim d(x,y)^2$ as $x \to y$. In terms of the geodesic distance $d(x,y)$ itself, related to $u$ by~\eqref{sec_example_d_u}, we have
\begin{splitequation}
\hat{\mathcal{C}}^{\phi^2,(1)}_{\phi \phi} &= - \frac{3}{\pi^2} \exp\left( \frac{3 d}{2 \ell} \right) \frac{d \left[ 1 + 3 \exp\left( \frac{2 d}{\ell} \right) \right] - 2 \ell \left[ \exp\left( \frac{2 d}{\ell} \right) - 1 \right]}{\left[ \exp\left( \frac{2 d}{\ell} \right) - 1 \right]^3 \ell^2} \\
&= - \frac{3}{4 \pi^2} \left[ \frac{1}{d(x,y)} + \frac{1}{6 \ell} - \frac{13 d(x,y)}{24 \ell^2} \right] + \bigo{d(x,y)^2} \eqend{.}
\end{splitequation}
For a covariant definition, one would expect a covariant expansion whose coefficients are proportional to the curvature, hence quadratic in the hyperbolic radius $\ell$. However, since normal-ordering was performed with respect to the full propagator (which is not covariant), this does not hold. This is easily seen by performing an expansion of the propagator~\eqref{sec_example_hw_5d} itself in terms of the geodesic distance:
\begin{equation}
G(x,y) = \frac{1}{8\pi^2} \left[ \frac{1}{d(x,y)^3} - \frac{1}{8 \ell^2 d(x,y)} - \frac{1}{8 \ell^3} \right] + \bigo{d(x,y)} \eqend{,}
\end{equation}
where also odd powers of $\ell$ appear.

However, since local covariance is important to restrict the renormalisation ambiguity, this is not the most useful solution. I thus return to the original coefficient~\eqref{sec_example_pert_coeff_phi2_phiphi}, and observe the following: since the propagator $G(x,y)$ fulfills the equation of motion~\eqref{sec_example_g_eom}, the parametrix fulfills
\begin{equation}
\label{sec_example_pert_h_eom}
\left( \nabla_x^2 + \frac{15}{4 \ell^2} \right) H(x,y) = - \delta(x,y) - \left( \nabla_x^2 + \frac{15}{4 \ell^2} \right) W(u(x,y)) = - \delta(x,y) \eqend{,}
\end{equation}
since the state-dependent part $W(u)$~\eqref{sec_example_hw_5d} is a homogeneous solution. From equation~\eqref{sec_example_pert_coeff_phi2_phiphi} I then obtain
\begin{equation}
\label{sec_example_pert_d2_coeff_phi2_phiphi}
\left( \nabla_y^2 + \frac{15}{4 \ell^2} \right) \mathcal{C}^{\phi^2,(1)}_{\phi \phi}(y,x;x) = 12 f(y) H(y,x) = 12 H(y,x) \eqend{,}
\end{equation}
assuming that $y$ lies in the region where the coupling is constant: $f = 1$. For $y$ sufficiently close to $x$ (and thus sufficiently far from the region where the effect of the cutoff $f \neq 1$ becomes important), one can assume that the OPE coefficient $\mathcal{C}^{\phi^2,(1)}_{\phi \phi}(y,x;x)$ is a function of the chordal distance $u(x,y)$ only: $\mathcal{C}^{\phi^2,(1)}_{\phi \phi}(y,x;x) = C(u(x,y))$. Equation~\eqref{sec_example_pert_d2_coeff_phi2_phiphi} then gives the second-order differential equation
\begin{equation}
\label{sec_example_pert_d2_coeff_phi2_phiphi_inu}
u (u+4) C''(u) + 4 (u+2) C'(u) + \frac{15}{4} C(u) = \frac{3 \ell^{-1}}{2 \pi^2} u^{-\frac{3}{2}} \eqend{,}
\end{equation}
which can be solved for small $u$ to any desired asymptotic order.\footnote{The same trick can be used to evaluate higher $n$-point functions in the AdS/CFT correspondence~\cite{dhokerfreedmanrastelli1999}.} From the scaling~\eqref{sec_ope_exist_scaling_coeff}, the most singular term of $\mathcal{C}^{\phi^2,(1)}_{\phi \phi}(y,x;x)$ scales as $d(x,y)^{[\phi^2] - 2 [\phi] - [\phi^4] + \dim M} = d(x,y)^{-1} \sim u(x,y)^{-\frac{1}{2}}$, and I thus make the ansatz
\begin{equation}
C(u) = \frac{c}{\sqrt{u}} \left[ 1 + a_1 u + a_2 u^2 + \bigo{u^3} \right]
\end{equation}
with constants $c$ and $a_i$. Inserting this into the differential equation~\eqref{sec_example_pert_d2_coeff_phi2_phiphi_inu}, it follows that
\begin{equation}
c = - \frac{3 \ell^{-1}}{4 \pi^2} \eqend{,} \qquad a_1 = - \frac{1}{2} \eqend{,} \qquad a_2 = \frac{1}{6} \eqend{,}
\end{equation}
which agrees with the singular (geometrically defined) part of the result~\eqref{sec_example_pert_coeff_phi2_phiphi_ing} for the OPE coefficient normal-ordered with respect to the full propagator. In fact, equation~\eqref{sec_example_pert_d2_coeff_phi2_phiphi_inu} can be solved explicitly to give
\begin{equation}
C(u) = - \frac{3 \ell^{-1}}{\pi^2 (u+4)^\frac{3}{2}} \left[ \sqrt{ \frac{u+4}{u} } - \ln\left( \frac{\sqrt{u} + \sqrt{u+4}}{2} \right) \right] + \frac{c'}{u^\frac{3}{2}} + \frac{c''}{(u+4)^\frac{3}{2}} \eqend{.}
\end{equation}
While the term proportional to $c'$ has the wrong scaling as $u \to 0$ such that $c' = 0$, the term proportional to $c''$ is analytic around $u = 0$ and thus does not contribute to the singular part. It follows that
\begin{splitequation}
\label{sec_example_pert_coeff_phi2_phiphi_sol}
\mathcal{C}^{\phi^2,(1)}_{\phi \phi}(y,x;x) &= - \frac{3 \ell^{-1}}{\pi^2 (u+4)^\frac{3}{2}} \left[ \sqrt{ \frac{u+4}{u} } - \ln\left( \frac{\sqrt{u} + \sqrt{u+4}}{2} \right) \right] \\
&\quad+ \text{ cutoff-dependent terms analytic at } y = x \eqend{,}
\end{splitequation}
where the cutoff-dependent terms could be computed from the integral~\eqref{sec_example_pert_coeff_phi2_phiphi}. In fact, also the coefficient normal-ordered with respect to the full propagator $G$ can be obtained in this way: the analogue of equation~\eqref{sec_example_pert_d2_coeff_phi2_phiphi} reads
\begin{equation}
\left( \nabla_y^2 + \frac{15}{4 \ell^2} \right) \hat{\mathcal{C}}^{\phi^2,(1)}_{\phi \phi}(y,x;x) = 12 G(y,x) \eqend{,}
\end{equation}
which is solved by
\begin{splitequation}
\label{sec_example_pert_coeff_phi2_phiphi_ing2}
\hat{\mathcal{C}}^{\phi^2,(1)}_{\phi \phi}(y,x;x) &= \frac{3 \ell^{-1}}{\pi^2} \left[ (u+4)^{-\frac{3}{2}} - u^{-\frac{3}{2}} \right] \ln\left( \frac{\sqrt{u} + \sqrt{u+4}}{2} \right) \\
&\quad+ \frac{3 \ell^{-1}}{\pi^2} \frac{\sqrt{u+4} - \sqrt{u}}{u (u+4)} + c_1 H(x,y) + c_2 W(x,y) \eqend{.}
\end{splitequation}
Since the propagator $G(x,y)$ scales $\sim u(x,y)^{-\frac{5}{2}}$ in the IR (as $u \to \infty$), from the integral~\eqref{sec_example_pert_coeffhat_phi2_phiphi} that includes one integration over $z$ one obtains that $\hat{\mathcal{C}}^{\phi^2,(1)}_{\phi \phi}(y,x;x)$ scales $\sim u(x,y)^{-5+\frac{5}{2}} = u(x,y)^{-\frac{5}{2}}$ in the IR. Expanding the coefficient~\eqref{sec_example_pert_coeff_phi2_phiphi_ing2} for large $u$, the term of order $u^{-\frac{3}{2}}$ is cancelled by choosing $c_1 = c_2$. On the other hand, the UV scaling is the same as for the normal-ordered coefficient $\hat{\mathcal{C}}^{\phi^2,(1)}_{\phi \phi}(y,x;x) \sim u(x,y)^{-\frac{1}{2}}$ as $u \to 0$, which fixes $c_1 = 0$, and the previous result~\eqref{sec_example_pert_coeff_phi2_phiphi_ing} is recovered.

Similarly, using~\eqref{sec_example_free3_phi4phiphi_coeffs} and~\eqref{sec_example_free2_phi4_phi2_coeffs} for the next OPE coefficient one obtains
\begin{equation}
\label{sec_example_pert_coeff_phidphi_phiphi}
\mathcal{C}^{\phi \nabla^\mu \phi,(1)}_{\phi \phi}(y,x;x) = - 24 \int f(z) \left[ H(z,y) H(z,x) - \left[ H(z,x)^2 \right]^\text{ren} \right] \xi(x,z)^\mu \total z
\end{equation}
and computes
\begin{equation}
\left( \nabla_y^2 + \frac{15}{4 \ell^2} \right) \mathcal{C}^{\phi \nabla^\mu \phi,(1)}_{\phi \phi}(y,x;x) = 24 H(y,x) \xi(x,y)^\mu
\end{equation}
for $x$ and $y$ in the region where $f = 1$. Note that the second contribution to the integral~\eqref{sec_example_pert_coeff_phidphi_phiphi} is completely irrelevant for this result and thus the singular terms of the OPE coefficient $\mathcal{C}^{\phi \nabla^\mu \phi,(1)}_{\phi \phi}$. Instead, it ensures that the integral is convergent in the IR if one considers normal-ordering with respect to the full propagator and takes the adiabatic limit $f \to 1$ (without this term, a logarithmic divergence would appear). Making the ansatz
\begin{equation}
\mathcal{C}^{\phi \nabla^\mu \phi,(1)}_{\phi \phi}(y,x;x) = C(u(x,y)) \xi(x,y)^\mu \eqend{,}
\end{equation}
and using the explicit form~\eqref{sec_example_xidef} of the tangent vector $\xi(x,y)^\mu$, I obtain the differential equation
\begin{equation}
\label{sec_example_pert_d2_coeff_phidphi_phiphi_inu}
u (u+4) C''(u) + \left[ 5 (u+2) + 4 K(u) \right] C'(u) + \left[ \frac{15}{4} + 8 \frac{(u+2) K(u) - 2}{u (u+4)} \right] C(u) = \frac{3 \ell^{-1}}{\pi^2} u^{-\frac{3}{2}}
\end{equation}
with
\begin{equation}
\label{sec_example_pert_k_def}
K(u) \equiv \frac{\sqrt{u (u+4)}}{4 \ln\left( \frac{\sqrt{u} + \sqrt{4+u}}{2} \right)} = 1 + \frac{u}{6} - \frac{u^2}{180} + \bigo{u^3} \eqend{.}
\end{equation}
Equation~\eqref{sec_example_pert_d2_coeff_phidphi_phiphi_inu} can still be solved exactly, and using that the most singular term of $\mathcal{C}^{\phi \nabla^\mu \phi,(1)}_{\phi \phi}(y,x;x)$ scales as $u(x,y)^0$ by the scaling~\eqref{sec_ope_exist_scaling_coeff}, such that the most singular term of $C(u)$ scales as $u^{-\frac{1}{2}}$ since the tangent vector itself scales like $\sqrt{u}$, I obtain
\begin{splitequation}
C(u) &= - \frac{3 \ell^{-1}}{5 \pi^2 (u+4)^\frac{5}{2}} \Bigg[ 16 K(u) \operatorname{Li}_2\left( \frac{u+2 - \sqrt{u(u+4)}}{2} \right) - 4 \sqrt{u (u+4)} \ln u - \frac{16 \pi^2}{6 K(u)} \\
&\hspace{6em}- \frac{16 \sqrt{u+4}}{u^\frac{3}{2}} K(u) + \frac{(32 + 32 u + 5 u^2) \sqrt{u+4}}{u^\frac{3}{2}} - \frac{(u+4)^\frac{7}{2}}{4 u^\frac{3}{2} K(u)} + \frac{u (u+4)}{K(u)} \Bigg] \\
&= - \frac{3 \ell^{-1}}{4 \pi^2 \sqrt{u}} \left[ 1 - \frac{13}{18} u + \frac{77}{24} u^2 + \bigo{u^3} \right] \raisetag{1.8em}
\end{splitequation}
up to cutoff-dependent terms analytic at coincidence.

For the two remaining coefficients $\mathcal{C}^{\phi \nabla^{\mu\nu} \phi,(1)}_{\phi \phi}$ and $\mathcal{C}^{\nabla^\mu \phi \nabla^\nu \phi,(1)}_{\phi \phi}$, one obtains
\begin{splitequation}
\mathcal{C}^{\phi \nabla^{\mu\nu} \phi,(1)}_{\phi \phi}(y,x;x) &= \mathcal{C}^{\nabla^\mu \phi \nabla^\nu \phi,(1)}_{\phi \phi}(y,x;x) \\
&= - 12 \int f(z) \bigg[ H(z,y) H(z,x) - \left[ H(z,x)^2 \right]^\text{ren} \\
&\qquad\qquad- \left[ H(z,x) \nabla^x_\rho H(z,x) \right]^\text{ren} \xi(x,y)^\rho \bigg] \xi(x,z)^\mu \xi(x,z)^\nu \total z \eqend{,}
\end{splitequation}
where the combination in brackets is recognised as the remainder of the covariant and properly renormalised Taylor expansion of $H(z,y) H(z,x)$ around $y = x$. Again, these subtraction terms serve to ensure the IR convergence of the corresponding coefficient normal-ordered with respect to the full propagator in the adiabatic limit. Applying as before the equation-of-motion operator at $y$, it follows that
\begin{splitequation}
\label{sec_example_pert_d2_coeff_phid2phi_phiphi}
&\left( \nabla_y^2 + \frac{15}{4 \ell^2} \right) \mathcal{C}^{\phi \nabla^{\mu\nu} \phi,(1)}_{\phi \phi}(y,x;x) = 12 H(y,x) \xi(x,y)^\mu \xi(x,y)^\nu \\
&\hspace{6em}+ 12 \left( \nabla_y^2 + \frac{15}{4 \ell^2} \right) \xi(x,y)^\rho \int f(z) \left[ H(z,x) \nabla^x_\rho H(z,x) \right]^\text{ren} \xi(x,z)^\mu \xi(x,z)^\nu \total z \eqend{.}
\end{splitequation}
While the first term on the right-hand side is singular as $u \to 0$, the second one is analytic at coincidence. This is seen easiest by using the geodesic distance $d(x,y)$, in terms of which the tangent vector is given by~\eqref{sec_example_xidef}, and computing
\begin{equation}
\label{sec_example_pert_d2xi}
\left( \nabla_y^2 + \frac{15}{4 \ell^2} \right) \xi(x,y)^\rho = \frac{47 d(x,y) - 15 d(x,y) \cosh\left( \frac{2 d(x,y)}{\ell} \right) - 16 \ell \sinh\left( \frac{2 d(x,y)}{\ell} \right)}{16 d(x,y) \sinh^2\left( \frac{d(x,y)}{\ell} \right) \ell^2} \nabla_x^\rho d(x,y)^2
\end{equation}
using that
\begin{equation}
\nabla^2 f(d(x,y)) = f''(d(x,y)) + \frac{4}{\ell} \coth\left( \frac{d(x,y)}{\ell} \right) f'(d(x,y)) \eqend{.}
\end{equation}
Since the right-hand side of equation~\eqref{sec_example_pert_d2xi} is invariant under $d \to -d$ and thus only depends on $d^2$, the numerator vanishes like $d(x,y)^3$ for small $d$, and the relation between $d^2$ and $u$~\eqref{sec_example_d_u} is analytic, it follows that $\left( \nabla_y^2 + \frac{15}{4 \ell^2} \right) \xi(x,y)^\rho$ is analytic at coincidence. Therefore, the second term in equation~\eqref{sec_example_pert_d2_coeff_phid2phi_phiphi} does not contribute to the singular part of the OPE coefficient and can be ignored in the following. Making the ansatz
\begin{equation}
\mathcal{C}^{\phi \nabla^{\mu\nu} \phi,(1)}_{\phi \phi}(y,x;x) = C(u(x,y)) g^{\mu\nu}(x) + \tilde{C}(u(x,y)) \xi(x,y)^\mu \xi(x,y)^\nu \eqend{,}
\end{equation}
up to terms analytic at coincidence, from equation~\eqref{sec_example_pert_d2_coeff_phid2phi_phiphi} the differential equations
\begin{equations}[sec_example_pert_d2_coeff_phid2phi_phiphi_inu]
&u(u+4) C''(u) + 5 (u+2) C'(u) + \frac{15}{4} C(u) = 0 \eqend{,} \\
\begin{split}
&u (u+4) \tilde{C}''(u) + \left[ 5 (u+2) + 8 K(u) \right] \tilde{C}'(u) \\
&\qquad+ \left[ \frac{7}{4} + 8 \frac{K(u)^2 + 2 (u+2) K(u) - 5}{u (u+4)} \right] \tilde{C}(u) = \frac{3 \ell^{-1}}{2 \pi^2} u^{-\frac{3}{2}}
\end{split}
\end{equations}
(with the same function $K(u)$~\eqref{sec_example_pert_k_def} as before) are obtained. The coefficient $\mathcal{C}^{\phi \nabla^{\mu\nu} \phi,(1)}_{\phi \phi}(y,x;x)$ scales as $d(x,y) \sim \sqrt{u(x,y)}$, such that $C(u)$ scales as $\sqrt{u}$ and $\tilde{C}(u)$ scales as $u^{-\frac{1}{2}}$. The only solution of the equation for $C(u)$ with this bound on the scaling is then analytic at coincidence, such that the singular part of $C(u)$ vanishes. On the other hand, the equation for $\tilde{C}(u)$ cannot be solved anymore in closed form, but with the known scaling an asymptotic solution to any required order is easily found:
\begin{equation}
\tilde{C}(u) = - \frac{\ell^{-1}}{4 \pi^2 \sqrt{u}} \left[ 1 - \frac{2}{3} u + \frac{119}{450} u^2 + \bigo{u^3} \right] \eqend{.}
\end{equation}

Inserting the results in equation~\eqref{sec_example_pert_ope_phiphi}, the OPE with $\phi$ with itself reads
\begin{splitequation}
&\Phi(y) \Phi(x) = H(y,x) \1 + \left[ 1 - \frac{3 g \ell^{-1}}{4 \pi^2 \sqrt{u}} \left( 1 - \frac{u}{2} \right) \right] \Phi^2(x) + \left[ 1 - \frac{3 g \ell^{-1}}{4 \pi^2 \sqrt{u}} \right] \xi(x,y)^\mu \left( \Phi \nabla_\mu \Phi \right)(x) \\
&\qquad+ \left[ \frac{1}{2} - \frac{g \ell^{-1}}{4 \pi^2 \sqrt{u}} \right] \xi(x,y)^\mu \xi(x,y)^\nu \, \left( \Phi \nabla_{\mu\nu} \Phi \right)(x) - \frac{g \ell^{-1}}{4 \pi^2 \sqrt{u}} \xi(x,y)^\mu \xi(x,y)^\nu \, \left( \nabla_\mu \Phi \nabla_\nu \Phi \right)(x) \\
&\qquad+ \bigo{ d(y,x)^2 } + \bigo{ g^2 } + \text{terms analytic at coincidence $y = x$} \eqend{,} \raisetag{1.6em}
\end{splitequation}
or expressing the chordal distance $u$ in terms of the geodesic distance $d$ according to~\eqref{sec_example_d_u}
\begin{splitequation}
&\Phi(y) \Phi(x) = H(y,x) \1 + \left[ 1 - \frac{3 g}{4 \pi^2 d} \left( 1 - \frac{13}{24} \frac{d^2}{\ell^2} \right) \right] \Phi^2(x) + \left( 1 - \frac{3 g}{4 \pi^2 d} \right) \xi(x,y)^\mu \left( \Phi \nabla_\mu \Phi \right)(x) \\
&\qquad+ \left( \frac{1}{2} - \frac{g}{4 \pi^2 d} \right) \xi(x,y)^\mu \xi(x,y)^\nu \, \left( \Phi \nabla_{\mu\nu} \Phi \right)(x) - \frac{g}{4 \pi^2 d} \xi(x,y)^\mu \xi(x,y)^\nu \, \left( \nabla_\mu \Phi \nabla_\nu \Phi \right)(x) \\
&\qquad+ \bigo{ d(y,x)^2 } + \bigo{ g^2 } + \text{terms analytic at coincidence $y = x$} \eqend{.} \raisetag{1.6em}
\end{splitequation}

\subsection{An OPE coefficient at second order}
\label{sec_example_pert2}

In this subsection, I only want to compute the OPE coefficient $\mathcal{C}^{\1,(2)}_{\phi \phi}(y,x;x)$, both for normal-ordering with respect to the Hadamard parametrix and the full propagator. In addition to the coefficients~\eqref{sec_example_free2_phi_phi_coeffs}, \eqref{sec_example_free2_1_phi_coeffs}, \eqref{sec_example_free2_phi4_phi2_coeffs} and \eqref{sec_example_free3_phi4phiphi_coeffs}, the free coefficients for the OPE of $\phi^4$ with itself, the OPE of $\phi \nabla^k \phi$ with $\phi$ and some three-point coefficients are needed. For the OPE of $\phi^4$ with itself, I compute using the formula~\eqref{sec_aqft_free_hadamard_product}
\begin{splitequation}
\normord{ \phi^4(y) }_H \, \normord{ \phi^4(x) }_H &= \normord{ \phi^4(y) \phi^4(x) }_H + 16 H(x,y) \normord{ \phi^3(y) \phi^3(x) }_G + 72 H(x,y)^2 \normord{ \phi^2(y) \phi^2(x) }_G \\
&\quad+ 96 H(x,y)^3 \normord{ \phi(y) \phi(x) }_G + 24 H(x,y)^4 \eqend{.} \raisetag{1.8em}
\end{splitequation}
Expanding the normal-ordered products around $y = x$, the required OPE coefficients are obtained as
\begin{equations}
\mathcal{C}^{\1,(0)}_{\phi^4 \phi^4}(y,x;x) &= 24 H(x,y)^4 \eqend{,} \\
\mathcal{C}^{\phi^2,(0)}_{\phi^4 \phi^4}(y,x;x) &= 96 H(x,y)^3 \eqend{,} \\
\mathcal{C}^{\phi \nabla^\mu \phi,(0)}_{\phi^4 \phi^4}(y,x;x) &= 96 H(x,y)^3 \xi(x,y)^\mu \eqend{,} \\
\mathcal{C}^{\phi \nabla^{\mu\nu} \phi,(0)}_{\phi^4 \phi^4}(y,x;x) &= 48 H(x,y)^3 \xi(x,y)^\mu \xi(x,y)^\nu \eqend{,} \\
\mathcal{C}^{\phi \nabla^{\mu\nu\rho} \phi,(0)}_{\phi^4 \phi^4}(y,x;x) &= 16 H(x,y)^3 \xi(x,y)^\mu \xi(x,y)^\nu \xi(x,y)^\rho \eqend{,}
\end{equations}
with $\mathcal{C}^{\nabla^\mu \phi \nabla^\nu \phi,(0)}_{\phi^4 \phi^4}$ and $\mathcal{C}^{\nabla^\mu \phi \nabla^{\nu\rho} \phi,(0)}_{\phi^4 \phi^4}$ vanishing. Note that $H(x,y)^4$ etc.\ are not well-defined as a distribution because they are too singular at coincidence, but using the same trick as before this will actually be unproblematic. Moreover, the coefficient of $\phi$ in the OPE of $\phi \nabla^k \phi$ with $\phi$ for $k = 0,1,2,3$ is needed, which is obtained from
\begin{equation}
\normord{ \phi(y) \nabla^k \phi(y) }_H \, \phi(x) = \normord{ \phi(y) \nabla^k \phi(y) \phi(x) }_H + H(y,x) \nabla^k \phi(y) + \nabla_y^k H(y,x) \phi(y)
\end{equation}
as
\begin{equation}
\mathcal{C}^{\phi,(0)}_{\phi^2 \phi}(y,x;x) = 2 H(y,x) \eqend{,} \qquad \mathcal{C}^{\phi,(0)}_{(\phi \nabla^k \phi) \phi}(y,x;x) = \nabla_y^k H(y,x) \quad\text{for } k > 0 \eqend{.}
\end{equation}

The required three-point coefficients come from the OPE of $\phi \nabla^k \phi$ with $\phi$ and $\phi$ for $k = 0,1,2,3$, as well as the OPE of $\phi^4$ with $\phi^4$ and $\phi$. For the first, only the coefficient of the unit operator is required, which in the normal-ordered case is just the expectation value (with the propagator $G$ replaced by the parametrix $H$)
\begin{equation}
\mathcal{C}^{\1,(0)}_{(\phi \nabla^k \phi) \phi \phi}(z,y,x;x) = H(z,y) \nabla_z^k H(z,x) + H(z,x) \nabla_z^k H(z,y) \eqend{.}
\end{equation}
For the second, we need the coefficient of $\phi$, which comes from
\begin{splitequation}
\normord{ \phi^4(z) }_H \, \normord{ \phi^4(y) }_H \, \phi(x) &= \normord{ \phi^4(z) }_H \left[ \normord{ \phi^4(y) \phi(x) }_H + 4 H(y,x) \normord{ \phi^3(y) }_H \right] \\
&= 24 H(y,z)^4 \phi(x) + 96 H(x,z) H(y,z)^3 \phi(y) + 96 H(x,y) H(y,z)^3 \phi(z) \\
&\qquad+ \text{terms involving normal-ordered products with more fields} \eqend{,}
\end{splitequation}
such that expanding around $z = x = y$ one obtains
\begin{equation}
\mathcal{C}^{\phi,(0)}_{\phi^4 \phi^4 \phi}(z,y,x;x) = 24 H(y,z)^4 + 96 H(x,z) H(y,z)^3 + 96 H(x,y) H(y,z)^3 \eqend{.}
\end{equation}
Lastly a four-point coefficient is needed, which is again just the expectation value
\begin{equation}
\mathcal{C}^{\1,(0)}_{\phi^4 \phi^4 \phi \phi}(u,z,y,x;x) = 24 H(x,y) H(u,z)^4 + 96 H(u,z)^3 \left[ H(x,u) H(y,z) + H(x,z) H(y,u) \right] \eqend{.}
\end{equation}

Integrating the recursive formula~\eqref{sec_ope_var_redef_opeder} over $z$ with a cutoff function $f(z)$, one obtains
% \mathcal{C}^{B} = \mathcal{C}^{B,(0)} + g \mathcal{C}^{B,(1)} + 1/2 g^2 \mathcal{C}^{B,(2)}
\begin{splitequation}
&\mathcal{C}^{\1,(2)}_{\phi \phi}(y,x;x) = \int f(z) \Bigg[ - \mathcal{C}^{\1,(1)}_{\phi^4 \phi \phi}(z,y,x;x) + \mathcal{C}^{\phi,(1)}_{\phi^4 \phi}(z,y;y) \, \mathcal{C}^{\1,(0)}_{\phi \phi}(y,x;x) \\
&\qquad+ \mathcal{C}^{\phi,(1)}_{\phi^4 \phi}(z,x;x) \, \mathcal{C}^{\1,(0)}_{\phi \phi}(y,x;x) - \int f(u) \, \mathcal{C}^{\phi^4,(0)}_{\phi^4 \phi^4}(z,u;u) \, \mathcal{C}^{\1,(0)}_{\phi^4 \phi \phi}(u,y,x;x) \total u \Bigg] \total z \eqend{,}
\end{splitequation}
where the formula~\eqref{sec_ope_var_redef_opeder} was used again to substitute the last term, as well as the $\mathbb{Z}_2$ symmetry and that $\mathcal{C}^{\phi,(0)}_{\phi^4 \phi} = 0$. In fact, since also $\mathcal{C}^{\1,(0)}_{\phi^4 \phi \phi} = 0$, the last term does not contribute at this order either. For the appearing first-order coefficients, we have from the formula~\eqref{sec_ope_var_redef_opeder} that (using again the $\mathbb{Z}_2$ symmetry)
\begin{splitequation}
&\mathcal{C}^{\1,(1)}_{\phi^4 \phi \phi}(z,y,x;x) = \int f(u) \bigg[ - \mathcal{C}^{\1,(0)}_{\phi^4 \phi^4 \phi \phi}(u,z,y,x;x) + \mathcal{C}^{\1,(0)}_{\phi^4 \phi^4}(u,z;z) \, \mathcal{C}^{\1,(0)}_{\1 \phi \phi}(z,y,x;x) \\
&\qquad+ \mathcal{C}^{\phi^2,(0)}_{\phi^4 \phi^4}(u,z;z) \, \mathcal{C}^{\1,(0)}_{\phi^2 \phi \phi}(z,y,x;x) + \mathcal{C}^{\phi \nabla^\mu \phi,(0)}_{\phi^4 \phi^4}(u,z;z) \, \mathcal{C}^{\1,(0)}_{(\phi \nabla^\mu \phi) \phi \phi}(z,y,x;x) \\
&\qquad+ \mathcal{C}^{\phi \nabla^{\mu\nu} \phi,(0)}_{\phi^4 \phi^4}(u,z;z) \, \mathcal{C}^{\1,(0)}_{(\phi \nabla^{\mu\nu} \phi) \phi \phi}(z,y,x;x) \\
&\qquad+ \mathcal{C}^{\phi \nabla^{\mu\nu\rho} \phi,(0)}_{\phi^4 \phi^4}(u,z;z) \, \mathcal{C}^{\1,(0)}_{(\phi \nabla^{\mu\nu\rho} \phi) \phi \phi}(z,y,x;x) \bigg] \total u \eqend{,}
\end{splitequation}
and
\begin{splitequation}
&\mathcal{C}^{\phi,(1)}_{\phi^4 \phi}(y,x;x) = \int f(u) \bigg[ - \mathcal{C}^{\phi,(0)}_{\phi^4 \phi^4 \phi}(u,y,x;x) + \mathcal{C}^{\1,(0)}_{\phi^4 \phi^4}(u,y;y) \, \mathcal{C}^{\phi,(0)}_{\1 \phi}(y,x;x) \\
&\qquad+ \mathcal{C}^{\phi^2,(0)}_{\phi^4 \phi^4}(u,y;y) \, \mathcal{C}^{\phi,(0)}_{\phi^2 \phi}(y,x;x) + \mathcal{C}^{\phi \nabla^\mu \phi,(0)}_{\phi^4 \phi^4}(u,y;y) \, \mathcal{C}^{\phi,(0)}_{(\phi \nabla^\mu \phi) \phi}(y,x;x) \\
&\qquad+ \mathcal{C}^{\phi \nabla^{\mu\nu} \phi,(0)}_{\phi^4 \phi^4}(u,y;y) \, \mathcal{C}^{\phi,(0)}_{(\phi \nabla^{\mu\nu} \phi) \phi}(y,x;x) \\
&\qquad+ \mathcal{C}^{\phi \nabla^{\mu\nu\rho} \phi,(0)}_{\phi^4 \phi^4}(u,y;y) \, \mathcal{C}^{\phi,(0)}_{(\phi \nabla^{\mu\nu\rho} \phi) \phi}(y,x;x) \bigg] \total u \eqend{.}
\end{splitequation}
Putting all together and using the OPE coefficients computed above, it follows that
\begin{splitequation}
\label{sec_example_pert2_c1_phi_phi}
&\mathcal{C}^{\1,(2)}_{\phi \phi}(y,x;x) = 16 \iint f(z) f(u) H(z,u)^3 \Bigg[ 12 H(x,u) H(y,z) - 12 H(x,z) H(y,z) \\
&\qquad+ 6 \xi(z,u)^\mu \Big[ \left( H(x,y) - H(x,z) \right) \nabla^z_\mu H(y,z) + (x \leftrightarrow y) \Big] \\
&\qquad+ 3 \xi(z,u)^\mu \xi(z,u)^\nu \Big[ \left( H(x,y) - H(x,z) \right) \nabla^z_{\mu\nu} H(y,z) + (x \leftrightarrow y) \Big] \\
&\qquad+ \xi(z,u)^\mu \xi(z,u)^\nu \xi(z,u)^\rho \Big[ \left( H(x,y) - H(x,z) \right) \nabla^z_{\mu\nu\rho} H(y,z) + (x \leftrightarrow y) \Big] \Bigg] \total z \total u \eqend{,} \raisetag{6.6em}
\end{splitequation}
where some terms have cancelled because of the symmetric integration over $z$ and $u$.

Now the same trick as before can be used: if $x$ is close (but not equal) to $y$, $f(x) = f(y) = 1$ and both $x$ and $y$ are far away from the region where $f \neq 1$, one can assume that the OPE coefficient is a function of the chordal distance $u(x,y)$ and that the effects of the cutoff function $f$ are negligible. Applying the equation-of-motion operator $\left( \nabla^2 + \frac{15}{4 \ell^2} \right)$ at both $x$ and $y$ to equation~\eqref{sec_example_pert2_c1_phi_phi} and using that the Hadamard parametrix is a fundamental solution~\eqref{sec_example_pert_h_eom}, it follows that
\begin{splitequation}
\left( \nabla_x^2 + \frac{15}{4 \ell^2} \right) \left( \nabla_y^2 + \frac{15}{4 \ell^2} \right) \mathcal{C}^{\1,(2)}_{\phi \phi}(y,x;x) = 192 H(x,y)^3 \qquad (x \neq y) \eqend{.}
\end{splitequation}
This can in fact be exactly solved with the solution
\begin{splitequation}
\label{sec_example_pert2_c1_phi_phi_sol}
\mathcal{C}^{\1,(2)}_{\phi \phi}(y,x;x) &= \frac{\ell^{-5}}{2240 \pi^6 u(x,y)^\frac{5}{2}} \left[ 3 + \ln u(x,y) \right] + c \, \mathcal{C}^{\1,(0)}_{\phi\phi}(y,x;x) + c' \mathcal{C}^{\phi^2,(1)}_{\phi\phi}(y,x;x) \\
&\quad+ \text{terms analytic at coincidence $y = x$} \eqend{,}
\end{splitequation}
where $c$ and $c'$ are two arbitrary constants. In the specific scheme that is used to obtain the recursive formula~\eqref{sec_ope_var_redef_opeder}, these coefficients can be fixed by computing in the same way the OPE coefficient $\hat{\mathcal{C}}^{\1,(2)}_{\phi \phi}$ normal-ordered with respect to the full propagator. This fulfills the analogue of equation~\eqref{sec_example_pert2_c1_phi_phi}:
\begin{splitequation}
\left( \nabla_x^2 + \frac{15}{4 \ell^2} \right) \left( \nabla_y^2 + \frac{15}{4 \ell^2} \right) \hat{\mathcal{C}}^{\1,(2)}_{\phi \phi}(y,x;x) = 192 G(x,y)^3 \qquad (x \neq y) \eqend{,}
\end{splitequation}
where the only difference is that on the right-hand side the full propagator appears instead of the Hadamard parametrix; its exact solution is
\begin{splitequation}
\hat{\mathcal{C}}^{\1,(2)}_{\phi \phi}(y,x;x) &= \frac{\ell^{-5}}{2240 \pi^6} \left( u^{-\frac{3}{2}} - 105 (u+4)^{-\frac{3}{2}} \right) \ln u + \frac{3 - 78 u}{560 \pi^6 \ell^5 u^\frac{5}{2} (u+4)} \\
&\quad+ \frac{\ell^{-5}}{2240 \pi^6} \left( (u+4)^{-\frac{3}{2}} - 105 u^{-\frac{3}{2}} \right) \ln(u+4) + \frac{3 (105 + 26 u)}{560 \pi^6 \ell^5 u (u+4)^\frac{5}{2}} \\
&\quad+ c_1 u^{-\frac{3}{2}} + c_2 (u+4)^{-\frac{3}{2}} + \frac{c_3}{u^\frac{3}{2}} \left[ \frac{4 u^\frac{1}{2} (u+3)}{3 (u+4)^\frac{3}{2}} - \ln\left( \frac{\sqrt{u} + \sqrt{4+u}}{2} \right) \right] \\
&\quad+ \frac{c_4}{(u+4)^\frac{3}{2}} \left[ \sqrt{ \frac{u+4}{u}} - \ln\left( \frac{\sqrt{u} + \sqrt{4+u}}{2} \right) \right] \eqend{.}
\end{splitequation}
To determine the unknown constants $c_i$ in this solution, consider the integral formula for this coefficient, which is given by equation~\eqref{sec_example_pert2_c1_phi_phi} with the Hadamard parametrix $H$ replaced by the propagator $G$, and taking the adiabatic limit $f \to 1$ (which does exist for the propagator but not the parametrix). Looking at the IR behaviour $u \to \infty$ of the integral, I note that the IR fall-off of the propagator $G$ as $u \to \infty$ is $\sim u^{-\frac{5}{2}}$, and the formula has the fifth power of the propagator and two integrations, such that I expect an IR fall-off of the OPE coefficient $\hat{\mathcal{C}}^{\1,(2)}_{\phi \phi}(y,x;x) \sim u^{-\frac{25}{2}+5} = u^{-\frac{15}{2}}$. This is achieved by choosing
\begin{equation}
c_1 = \frac{3}{4} c_2 = - c_3 = - c_4 = \frac{13 \ell^{-5}}{140 \pi^6} \eqend{,}
\end{equation}
which uniquely determines $\hat{\mathcal{C}}^{\1,(2)}_{\phi \phi}$. Since with arbitrary constants $\hat{\mathcal{C}}^{\1,(2)}_{\phi \phi}$ only falls of $\sim u^{-\frac{3}{2}}$ as $u \to \infty$, non-trivial cancellations have taken place to obtain this fast fall-off behaviour. Comparing now the singular behaviour of $\hat{\mathcal{C}}^{\1,(2)}_{\phi \phi}$ and $\mathcal{C}^{\1,(2)}_{\phi \phi}$ as $u \to 0$ then fixes the constants in~\eqref{sec_example_pert2_c1_phi_phi_sol} to be
\begin{equation}
c = \frac{517 - 840 \ln 2}{1120 \pi^4} \ell^{-2} \eqend{,} \qquad c' = \frac{937}{26880 \pi^4} \ell^{-4} \eqend{.}
\end{equation}
Note however that the difference between $\hat{\mathcal{C}}^{\1,(2)}_{\phi \phi}$ and $\mathcal{C}^{\1,(2)}_{\phi \phi}$ contains a term more singular than $\mathcal{C}^{\phi^2,(1)}_{\phi\phi}$, i.e., $c'$ multiplies a subleading singularity. This is in fact to be expected, since the value of $c'$ is scheme-dependent, and can be changed by another redefinition at first order: $\Phi^2 \to \Phi^2 + \tilde{c} g \1$, which results in $c' \to c' + \tilde{c}$. Only the most singular part of the OPE coefficient $\mathcal{C}^{\1,(2)}_{\phi \phi}$~\eqref{sec_example_pert2_c1_phi_phi_sol} is scheme-independent und uniquely determined.

\section{Outlook}
\label{sec_outlook}

While the formula~\eqref{sec_ope_var_redef_opeder} gives in principle a straightforward way to compute OPE coefficients to arbitrary order in perturbation theory, its practical application is hampered by the difficulties of evaluating integrals in curved space. Even if one considers normal ordering with respect to the full propagator, where the adiabatic limit of constant couplings can be taken, a direct evaluation of the integals is seldomly possible. This is not surprising, since in that case (for example) the coefficient of the unit operator is nothing else but the expectation value of the fields appearing in the OPE, and so their evaluation cannot be any simpler than the computation of correlation functions. Even in the simplest case of a maximally symmetric space (Euclidean AdS, or hyperbolic space), computations are very difficult (see e.g. the recent works~\cite{aharonyetal2017,giombisleighttaronna2018,bertansachs2018,bertansachsskvortsov2019,carmidipietrokomatsu2019,liuetal2019,ponomarev2020,carmi2020} for loop calculations in AdS). On the other hand, for a renormalisable interaction the UV counterterms in this formula are sufficient to remove all non-integrable divergences, such that it is possible to evaluate the OPE coefficients numerically. If a further redefinition of composite operators can be determined such that all non-integrable divergences are removed also for non-renormalisable interactions is currently under investigation.

Nevertheless, the most interesting application would be in the context of the AdS/CFT correspondence with all external points on the AdS boundary. In this case, one can use the Symanzik formula~\cite{symanzik1972} for the AdS integrations, which is a huge simplification compared to the general case with external points in the AdS bulk where no such formula exists. Since quite generally a local bulk field theory with small coupling corresponds to a large-$N$ dual CFT with a gap on the boundary~\cite{heemskerketal2009,aldaybissiperlmutter2017}, one thus obtains a way to systematically compute $1/N$ corrections to the planar limit $N \to \infty$ of such theories. The flat-space analogue of formula~\eqref{sec_ope_var_redef_opeder} has already been applied directly to a conformal field theory~\cite{hollands2018} (see also~\cite{behan2017} and~\cite{bashmakovbertoliniraj2017}), and allows one to compute corrections in conformal perturbation theory where an existing CFT is perturbed by a strictly marginal operator. Using conformal symmetry the integrals can be performed explicitly, and one obtains a coupled system of ODEs describing the change of the conformal data with the coupling, which can in principle even be solved numerically to obtain non-perturbative results. However, the obvious application where the unperturbed CFT is free (Gaussian) is hampered by the fact that the spectrum of such a CFT is highly degenerate, which violates a non-degeneracy condition that is necessary for the derivation of the ODEs~\cite{hollands2018}. It is quite probable that this non-degeneracy condition can be more easily fulfilled in the planar limit $N \to \infty$, which is already a non-trivial interacting theory, and that a formula for $1/N$ corrections could be solved numerically. One issue that needs to be adressed to fulfill this program is that even if the external points are on the AdS boundary, the integration that one needs to obtain corrections to the OPE coefficients from~\eqref{sec_ope_var_redef_opeder} still involves lower-order OPE coefficients with points in the bulk. Since the form of the bulk OPE coefficients is quite involved, it seems difficult to directly obtain a formula for the conformal data. I believe that a further redefinition of composite operators is needed to change the cutoff function in this integral in such a way as to involve only boundary points, similarly to the redefinition needed in the flat-space formula of Holland and Hollands~\cite{hollandhollands2015a} to pass from the massive to the massless theory. This is currently under investigation, as well as the extension of formula~\eqref{sec_ope_var_redef_opeder} to gauge theories and fermions.

\begin{acknowledgments}
It is a pleasure to thank Chris Fewster, Igor Khavkine and Mojtaba Taslimi Tehrani for discussions, and in particular Pawe{\l} Duch for extensive discussions on (algebraic) quantum field theory.

This work has been funded by the Deutsche Forschungsgemeinschaft (DFG, German Research Foundation) --- project nos. 415803368 and 406116891 within the Research Training Group RTG 2522/1.
\end{acknowledgments}

\appendix

\section{Local Wick expansion}
\label{appendix_wick}

Here I show that the existence of the local Wick expansion~\eqref{sec_aqft_renorm_r_wick}
\begin{equation}
\mathcal{R}\left[ \op_{A_1}(x_1) \cdots \op_{A_k}(x_k) \right] = \sum_{B_1,\ldots,B_k} r\left[ \partial^{B_1} \op_{A_1}(x_1) \cdots \partial^{B_k} \op_{A_k}(x_k) \right] \normord{ \op_{B_1}(x_1) \cdots \op_{B_k}(x_k) }_H
\end{equation}
follows from field independence and local covariance. The Wick expansion is proven by induction in $\sum_{i=1}^k [\op_{A_i}]$. Namely, it holds with $r\left[ \unitmatrix \right] = 1$ in the case where $k = 1$ and $\op_{A_1} = \unitmatrix$. Assume that it has been proven for all $\sum_{i=1}^k [\op_{A_i}] < D$, and consider the difference
\begin{splitequation}
\Delta(x_1,\ldots,x_k) &\equiv \mathcal{R}\left[ \op_{A_1}(x_1) \cdots \op_{A_k}(x_k) \right] - \sum_{B_1,\ldots,B_k, \sum_{i=1}^k [\op_{B_i}] > 0} \\
&\quad\times r\left[ \partial^{B_1} \op_{A_1}(x_1) \cdots \partial^{B_k} \op_{A_k}(x_k) \right] \normord{ \op_{B_1}(x_1) \cdots \op_{B_k}(x_k) }_H \eqend{,}
\end{splitequation}
where all the appearing $r$'s are known by induction since $\sum_{i=1}^k [\partial^{B_i} \op_{A_i}] < D$ because the sum is restricted to the case where at least one of the $\op_{B_i}$ is not the unit operator. By field independence of the renormalised products, one obtains
\begin{equation}
\frac{\delta}{\delta \phi(y)} \Delta(x_1,\ldots,x_k) = \sum_{i=1}^k \Delta_i(x_1,\ldots,x_k,y)
\end{equation}
with
\begin{splitequation}
&\Delta_i(x_1,\ldots,x_k,y) = \mathcal{R}\left[ \op_{A_1}(x_1) \cdots \frac{\delta \op_{A_i}(x_i)}{\delta \phi(y)} \cdots \op_{A_k}(x_k) \right] \\
&\qquad- \sum_{B_1,\ldots,B_k} r\left[ \partial^{B_1} \op_{A_1}(x_1) \cdots \partial^{B_k} \op_{A_k}(x_k) \right] \normord{ \op_{B_1}(x_1) \cdots \frac{\delta \op_{B_i}(x_i)}{\delta \phi(y)} \cdots \op_{B_k}(x_k) }_H \eqend{,}
\end{splitequation}
where the restriction on the sum over the $B_j$ could be removed because $\op_{B_i} \neq \unitmatrix$ (otherwise the functional derivative vanishes). Since the renormalised product that appears on the right-hand side has a lower total dimension by induction we know that it has a local Wick expansion. Using that
\begin{equation}
\frac{\delta \op_A(x)}{\delta \phi(y)} = \sum_{i=0}^\infty \left( \frac{\partial}{\partial \nabla^i \phi} \op_A(x) \right) \nabla^i \delta(x,y) \eqend{,}
\end{equation}
it follows that
\begin{splitequation}
&\mathcal{R}\left[ \op_{A_1}(x_1) \cdots \frac{\delta \op_{A_i}(x_i)}{\delta \phi(y)} \cdots \op_{A_k}(x_k) \right] \\
&\quad= \sum_{j=0}^\infty \nabla^j \delta(x_i,y) \mathcal{R}\left[ \op_{A_1}(x_1) \cdots \left( \frac{\partial}{\partial \nabla^j \phi} \op_{A_i}(x_i) \right) \cdots \op_{A_k}(x_k) \right] \\
&\quad= \sum_{j=0}^\infty \nabla^j \delta(x_i,y) \sum_{B_1,\ldots,B_k} r\left[ \partial^{B_1} \op_{A_1}(x_1) \cdots \partial^{B_i} \left( \frac{\partial}{\partial \nabla^j \phi} \op_{A_i}(x_i) \right) \cdots \partial^{B_k} \op_{A_k}(x_k) \right] \\
&\qquad\quad\times \normord{ \op_{B_1}(x_1) \cdots \op_{B_k}(x_k) }_H
\end{splitequation}
and
\begin{splitequation}
&\Delta_i(x_1,\ldots,x_k,y) \\
&\quad= \sum_{j=0}^\infty \nabla^j \delta(x_i,y) \sum_{B_1,\ldots,B_k} r\left[ \partial^{B_1} \op_{A_1}(x_1) \cdots \partial^{B_i} \left( \frac{\partial}{\partial \nabla^j \phi} \op_{A_i}(x_i) \right) \cdots \partial^{B_k} \op_{A_k}(x_k) \right] \\
&\qquad\qquad\times \normord{ \op_{B_1}(x_1) \cdots \op_{B_k}(x_k) }_H \\
&\qquad- \sum_{j=0}^\infty \nabla^j \delta(x_i,y) \sum_{B_1,\ldots,B_k} r\left[ \partial^{B_1} \op_{A_1}(x_1) \cdots \partial^{B_k} \op_{A_k}(x_k) \right] \\
&\qquad\qquad\times \normord{ \op_{B_1}(x_1) \cdots \left( \frac{\partial}{\partial \nabla^j \phi} \op_{B_i}(x_i) \right) \cdots \op_{B_k}(x_k) }_H \eqend{.}
\end{splitequation}
In the second sum, instead of the sum over all $B_k$, the distribution $r$ depending on $\partial^{B_i} \op_{A_i}$ and then considering $\partial/\partial \left( \nabla^j \phi \right) \op_{B_i}(x_i)$ in the Hadamard-normal ordered product, one can also sum over all $B_k$, the distribution $r$ depending on $\partial^{B_i} \left( \frac{\partial}{\partial \nabla^j \phi} \op_{A_i} \right)$ and consider $\op_{B_i}(x_i)$ in the Hadamard-normal ordered product. Therefore, all terms cancel and it follows that
\begin{equation}
\frac{\delta}{\delta \phi(y)} \Delta(x_1,\ldots,x_k) = 0 \eqend{.}
\end{equation}
Hence, $\Delta$ must be proportional to the unit operator, and one can simply define the distribution $r$ to be the proportionality coefficient:
\begin{equation}
r\left[ \op_{A_1}(x_1) \otimes \cdots \otimes \op_{A_k}(x_k) \right] \unitmatrix \equiv \Delta(x_1,\ldots,x_k) \eqend{.}
\end{equation}

\bibliography{literature}

\end{document}